\documentclass[12pt]{article}
\usepackage[mathscr]{eucal}
\usepackage{epsfig,amsfonts}
\usepackage{amsmath}
\usepackage{amsthm,amssymb}
\usepackage{graphicx}
\usepackage{hhline}
\usepackage{cite}
\usepackage{psfrag}
\usepackage{enumerate}
\makeatletter
\@addtoreset{equation}{section}
\makeatother

\def\be{\begin{equation}}
\def\ee{\end{equation}}
\def\bdm{\begin{displaymath}}
\def\edm{\end{displaymath}}
\def\bea{\begin{eqnarray}}
\def\eea{\end{eqnarray}}

\def\zb{{\bar z}}

\def\ri{{\rm i}}

\def\XXint#1#2#3{{\setbox0=\hbox{$#1{#2#3}{\int}$}
    \vcenter{\hbox{$#2#3$}}\kern-.5\wd0}}

\def\hf{{\frac{1}{2}}}

\def\wr{{\tt W}}
\def\be{\begin{equation}}
\def\ee{\end{equation}}
\def\wh{\widehat}
\def\beq{\begin{equation}}
\def\eeq{\end{equation}}
\def\sh{\textstyle \frac{1}{2}}

\newcommand{\rd}{\mbox{d}}
\newcommand{\re}{\mbox{e}}

\begin{document}

\begin{titlepage}
\begin{flushright}
RUNHETC-2013-18 \\
\end{flushright}

\vspace{1.cm}

\begin{center}
\begin{LARGE}

{\bf Integrable structure of Quantum Field Theory:\\
\bigskip
Classical flat connections versus
quantum\\
\bigskip
 stationary states}
\end{LARGE}
\vspace{1.3cm}

\begin{large}

{\bf Vladimir V. Bazhanov$^{1,2}$ and  Sergei  L. Lukyanov}$^{3,4}$

\end{large}

\vspace{1.cm}
$^1$Department of Theoretical Physics,\\
         Research School of Physics and Engineering,\\
    Australian National University, Canberra, ACT 0200, Australia\\\ \\
$^2$Mathematical Sciences Institute,\\
      Australian National University, Canberra, ACT 0200,
      Australia\\\ \\
${}^{3}$NHETC, Department of Physics and Astronomy\\
     Rutgers University\\
     Piscataway, NJ 08855-0849, USA\\
\vspace{.2cm}
and\\
\vspace{.2cm}
${}^{4}$L.D. Landau Institute for Theoretical Physics\\
  Chernogolovka, 142432, Russia\\
\vspace{1.0cm}

\end{center}

\vspace{0.7cm}

\begin{center}
\centerline{\bf Abstract} \vspace{.8cm}

\parbox{13cm}{
We establish a  correspondence between an infinite set of
special solutions of the  (classical) modified  sinh-Gordon equation and a
set of stationary states in the finite-volume Hilbert space 
of the integrable 2D QFT invented by V.A. Fateev.
The modified sinh-Gordon equation arise in this
case as a zero-curvature condition for a class of multivalued
connections on the punctured Riemann sphere, similarly to Hitchin's
self-duality equations. 
The proposed correspondence between the classical and
quantum integrable systems provides a powerful tool for deriving
functional and 
integral equations which determine the full spectrum of local integrals
of motion for massive QFT in a finite volume. 
Potential applications of our results to the problem
of non-perturbative quantization of classically integrable non-linear
sigma models are briefly discussed.}
\end{center}
\vspace{.8cm}


\vfill

\end{titlepage}
\newpage

\tableofcontents

\section{Introduction and summary}

It is difficult to assign a precise mathematical meaning for the
concept of integrability in Quantum Field Theory.  A naive intuition
goes back to Liouville of the $19^{\rm th}$ century and suggests an
existence of a sufficiently large set of mutually commuting operators
whose joint spectra fully specify stationary states of the quantum
system.  For deeper insights, it is useful to consider 2D Conformal
Field Theory (CFT), where significant simplifications occur due to the
presence of an infinite dimensional algebra of (extended) conformal
symmetry \cite{Zamolodchikov:1989zs}. 
For a finite-size 2D CFT (with
the spatial coordinate compactified on a circle of the circumference
$R$), a mathematically satisfactory construction of an infinite set of
mutually commuting local Integrals of Motion (IM) can be given and the
simultaneous diagonalization of these operators turns out to be a
well-defined problem within  the representation theory of the associated
conformal algebra.

Different conformal algebras, as well as different sets of mutually
commuting local IM yield a variety of integrable structures in CFT.
The series of works
\cite{Bazhanov:1994ft,Bazhanov:1996dr,Bazhanov:1998dq} was dedicated
to the simplest of these structures, associated with the
diagonalization of the local IM from the quantum KdV hierarchy
\cite{Sasaki:1987mm,Eguchi:1989hs,Kupershmidt:1989bf,FF5}.  Subsequent
studies of this problem culminated in a rather surprising
link between the integrable structures of CFT and spectral
theory of Ordinary Differential Equations (ODE)
\cite{Dorey:1998pt,Bazhanov:1998wj,Bazhanov:2003ni}.  In particular,
in \cite{Bazhanov:2003ni} a one-to-one correspondence was conjectured
between the joint eigenbasis of the IM from the quantum KdV hierarchy
and a certain class of differential operators of the second order
$-\partial_z^2+V_L(z)$, with singular potentials
$V_L(z)$ (``monster'' potentials in terminology of
\cite{Bazhanov:2003ni}). Apart from a regular singularity at $z=0$ and
an irregular singular point at $z=\infty$, the monster potentials
possess $L$ regular singular points $\{x_a\}_{a=1}^L$. These
potentials are not of much immediate interest in quantum mechanics,
but arise rather naturally in the context of the theory of
isomonodromic deformations.  Solutions of the corresponding
Schr\"odinger equations are single valued (monodromy-free) at $z=x_a$
and their monodromy properties turn out to be similar to that of the
radial wave functions for the three-dimensional isotropic anharmonic
oscillator.  The monodromy-free condition was formulated in a form of
the system of $L$ algebraic equations imposed on the set
$\{x_a\}_{a=1}^L$.\footnote{An alternative, but equivalent form of the
  monodromy-free condition was given in
\cite{Fioravanti:2004cz, Feigin:2007mr}.}  The correspondence  
proposed in \cite{Bazhanov:2003ni} precisely relates
the set of monster potentials $V_L(z)$ and the joint eigenbasis for all
quantum KdV integrals of motion in the level $L$ subspace of the highest weight
representation of the Virasoro algebra. In particular, 
this implies that a number of the potentials 
$V_L(z)$ for a given value of $L$ exactly coincides with a number of
partitions ${\tt p}_1(L)$ of the integer $L$ into parts of one kind.

Since 1998, the link to the  spectral theory of ODE have been
extended to a large variety of  integrable CFT structures (for
review, see \cite{Dorey:2007zx}), so that a natural question has
emerged on whether a similar  relation exist for {\em massive} integrable
QFT. This question  remained more or less dormant until 
the work \cite{Gaiotto:2008cd}, after which 
the so-called thermodynamic Bethe
Ansatz equations have started to appear in different contexts of SUSY gauge
theories
\cite{Gaiotto:2009hg,Alday:2009yn, Alday:2009dv, Alday:2010vh}. 
These remarkable developments have led to the work
\cite{Lukyanov:2010rn}, which  
established a link between eigenvalues of IM in the  vacuum
sector of  the massive sine/sinh-Gordon model and some new spectral
problem generalizing the one from \cite{Dorey:1998pt,Bazhanov:1998wj}. 

This work is aimed to extend the results of
\cite{Bazhanov:2003ni,Lukyanov:2010rn}    
and provide an explicit example of the correspondence 
between stationary states of massive integrable QFT in a finite volume and
singular  differential operators of a certain class.
At first glance, the best candidate for this purpose
should be the sine-Gordon model,
which always served as a basis for the development of integrable QFT.
However, in spite of some technical complexity, a more general model
introduced by Fateev \cite{Fateev:1996ea} (which contains
the sine-Gordon model as a particular case) turned out to be more
appropriate for this task.
The situation here is analogous to that in the  Painlev\'e  theory.
Even though the Painlev\'e VI is the most complicated
and general  equation  in the  Painlev\'e classification, 
geometric structures behind this  equation are much more transparent 
than those related to its degenerations. From this point of view, the fact that
the sine-Gordon model is a certain degeneration of the Fateev model,
could be understood as a QFT version of the relationship between  the
Painlev\'e VI and a particular case of
Painlev\'e III.

The organization of this paper is as follows. 
In Section\,\ref{sec21}  we introduce the notion 
of  Generalized 
Hypergeometric Opers (GHO's) ---  
a special class of Fuchsian differential operators of the second order
\bea\label{assaisoa}
{\cal D}=-\partial_z^2+T_L(z)
\eea
with $3+L$ regular singular points at
$z=z_1,z_2,z_3$ and $z=x_1,\ldots x_L$.
The variable $z$ can be regarded as a complex coordinate 
on the Riemann sphere with $3+L$ punctures.
Projective transformations of $z$ allows one to send the three points
$z_i$ to any designated positions. 
At the same time other parameters of GHO are chosen in such a way that
the remaining $L$  regular singular points satisfy the monodromy-free
condition. Therefore, the monodromy  properties of GHO for $L>0$
turn out to be similar to those for $L=0$ (i.e. the ordinary
hypergeometric  differential operator of the second order). 
The complex numbers $\{x_a\}$ can be thought as local coordinates in
the $L$-dimensional moduli space of GHO's.

In Section\,\ref{sectwo}  we consider more general differential
operators, which inherit the 
monodromy-free property of GHO's.
We call them the Perturbed Generalized
Hypergeometric Opers (PGHO's).  These operators have the form 
\bea\label{oaissua}
{\cal D}{(\lambda)}=-\partial_z^2
+T_L(z)+\lambda^2\ {\cal P}(z)\ , 
\eea
where
\bea\label{oasioaq}
{\cal P}(z)=\frac{(z_3-z_2)^{a_1}\,(z_1-z_3)^{a_2}\,(z_2-z_1)^{a_3}}
{(z-z_1)^{2-a_1}(z-z_2)^{2-a_2}(z-z_3)^{2-a_3}}
\eea
and the parameters $0<a_i<2$ satisfy the constraint 
\bea\label{iasoaisa}
a_1+a_2+a_3=2\ .
\eea
Due to the last relation the quantity 
${\cal P}(z)(\rd z)^2$ transforms as a
quadratic differential under $\mathbb{PSL}(2,\mathbb{C})$
transformations and the points  $z_1,z_2,z_3$ on the Riemann sphere
can still be sent to any given positions.
In the presence of ``perturbation'' the monodromy properties of the
operators \eqref{oaissua}
are changed dramatically in comparison with $\lambda=0$ case. 
However, one can still find positions of 
the punctures $x_1,\ldots x_L$, so that they remain 
monodromy-free singular points for any values of $\lambda$. 
In this case the coordinates 
$\{x_i\}_{i=1}^L$ obey a system of $L$ algebraic equations 
similar to that from \cite{Bazhanov:2003ni,Fioravanti:2004cz,
  Feigin:2007mr}. Therefore, the  moduli space of the PGHO's constitute a 
{\it finite discrete subset} ${\cal A}^{(L)}$ in the moduli space  
of GHO's.\footnote{To the
best of our knowledge,  the PGHO for $L=0$ was originally introduced
(up to change of variables) in the unpublished  
work (2001) of A.~B.~Zamolodchikov and the second  author 
(see also \cite{Lukyanov:2012wq}). Its particular cases 
were studied in a series of works on integrable models of boundary
interactions \cite{Lukyanov:2003rt,Lukyanov:2003nj,Lukyanov:2006gv}. 
For $L>0$, the  PGHO's appeared  in  the work \cite{Feigin:2007mr}.}
It  appears that, for a given $L$,  the cardinality  of ${\cal A}^{(L)}$
coincides with a number of partitions ${\tt p}_3(L)$ of the integer
$L$ into parts of three kinds. 
In Sections\,\ref{jassysa}-\ref{sec6}   
we  interpret this  fact in the spirit of
\cite{Bazhanov:2003ni}  and present arguments in support of existence 
of 
a one-to-one correspondence
between  elements of ${\cal A}^{(L)}$
and   the level-$L$  common eigenbasis 
of  the local IM  in the integrable hierarchy 
introduced by Fateev in \cite{Fateev:1996ea}.
The  arguments  closely follow the line of 
\cite{Bazhanov:1994ft,Bazhanov:1996dr,Bazhanov:1998dq,
Bazhanov:1998wj, Bazhanov:2003ni} adapted to the algebra of  
extended conformal symmetry, which can be
regarded   as a quantum 
Hamiltonian reduction of the exceptional affine superalgebra ${\hat
  D}(2,1;\alpha)$ \cite{Feigin:2001yq} (the ``corner-brane''    
$W$-algebra, in terminology of \cite{Lukyanov:2012wq}). 

The generalization of the above constructions  
to the case of  massive QFT is given in Sections\,\ref{sec7}-\ref{ninesec}.
It is based on the idea from \cite{Lukyanov:2010rn},
which was inspired by the works \cite{Gaiotto:2009hg,Alday:2009yn,
  Alday:2009dv, Alday:2010vh}.
As far as our attention has been confined to the case of CFT, 
there was no need to separately consider the antiholomorphic PGHO, 
${\bar {\cal D}}{({\bar \lambda})}=
-\partial_{\bar z}^2+{\bar T}_{\bar L}({\bar z})
+{\bar \lambda}^2\ {\bar {\cal P}}({\bar z})$,
since there is only a nomenclature 
difference between the holomorphic and antiholomorphic cases.
In massive QFT,
following  \cite{Lukyanov:2010rn},
one should substitute the pair of   
PGHO's
$({\cal D}{(\lambda)},\,{\bar
  {\cal D}}{({\bar \lambda})})$ 
by a pair of $(2\times 2)$-matrix valued differential operators
\bea\label{asopssaopasopq}
{\boldsymbol { D}}{(\lambda)}=\partial_z-{\boldsymbol
  A}_z\ ,\ \ \ \ \  {\boldsymbol {\bar  D}}{({\bar \lambda})}= 
\partial_{\bar z}-{ {\boldsymbol A}}_{\bar z}
\eea
with
\beq\label{ystopsso}
\begin{array}{rcl}
{\boldsymbol
  A}_z&=&-{\textstyle\frac{1}{2}}\ \partial_z\eta\,\sigma_3+ \sigma_+\,\re^{+\eta}+ 
\sigma_-\,\lambda^2\, {\cal P}(z)\, \re^{-\eta}\\[.3cm]
{ {\boldsymbol A}}_{\bar z}&=&+
{\textstyle\frac{1}{2}}\ \partial_{\bar
  z}\eta\,\sigma_3+ \sigma_-\, 
\re^{+\eta}+\sigma_+\,\bar{\lambda}^2 \ {\bar {\cal P}}({\bar z})\,\re^{-\eta}\ ,
\end{array}
\eeq
where $\sigma_3,\sigma_\pm=(\sigma_1\pm \ri \sigma_2)/2$ are the
standard Pauli matrices. 
In fact,  $({\boldsymbol A}_z,\,  {\boldsymbol A}_{\bar z})$ 
forms an
$\mathfrak{sl}(2)$  connection whose flatness 
is a necessary condition for the existence of solution of the  linear problem 
\bea\label{apsosaospa}
{\boldsymbol { D}}{(\lambda)}\, {\boldsymbol \Psi}=0\ ,\ \ \  \ \ \ 
 {\boldsymbol {\bar  D}}{({\bar \lambda})}\,{\boldsymbol \Psi}=0\ .
\eea
The zero-curvature condition leads to 
the Modified Sinh-Gordon equation (MShG):
\bea\label{asospsosap}
\partial_z\partial_{\bar z}\eta-\re^{2\eta}+ \rho^4\ {\cal P}(z){\bar
  {\cal P}}({\bar z})\, \re^{-2\eta}=0\ ,\qquad 
\rho^2={\lambda{\bar\lambda}}\ .
\eea
%
We consider a particular class of singular solutions of this equation, 
distinguished by special monodromy properties of the associated linear
problem \eqref{apsosaospa}.
The set of constraints  imposed on these solutions is discussed in
Section\,\ref{sec7}. 
In summary,  $\re^{-\eta}$ should be a {\it  smooth, single valued
  complex function without zeroes} 
on the Riemann sphere with $3+L+{\bar L}$
punctures. Since
$z=\infty$ is assumed to be a regular point on the  sphere, then
\bea
\label{sospsaosapyst}
\re^{-\eta}\sim |z|^{2}\ \ \ \ \ \ \ \  {\rm as}\ \ \ \ \ \  \ \   |z|\to  \infty\ .
\eea
At the same time,  $\re^{-\eta}$ develops a singular behavior at $z=z_i$,
\bea\label{sospsaosap}
\re^{-\eta}\sim |z-z_i|^{-2m_i}\ \ \ \ \ \ \  {\rm
  as}\ \ \ \ \ \  \ \ |z-z_i|\to  0\ \ \ \ \ \ (i=1,2,3)\ , 
\eea
and  also at
$z=x_a$
and ${\bar z}={\bar y}_b$
\bea\label{sopasopspoa}
\re^{-\eta}&\sim& \frac{{\bar z}-{\bar
    x}_a}{z-x_a}\ \ \ \ \ \ \ \ \ (a=1,\ldots L)\nonumber\\ 
\re^{-\eta}&\sim& \frac{{ z}-{ y}_b}{{\bar z}-{\bar
    y}_b}\ \ \ \ \ \ \ \ \ \ (b=1,\ldots {\bar L})\ . 
\eea
The  arbitrary parameters $m_i$ in  the asymptotic formulae \eqref{sospsaosap} 
should be  restricted to the domains\footnote{At $m_i=
\frac{1}{2},\,\frac{1}{4}(2-a_i)$ the  leading asymptotic
\eqref{sospsaosap} involves logarithms. Here we ignore such
subtleties.} 
\bea\label{asioasiso}
-{\textstyle\frac{1}{2}}\leq m_i\leq -{\textstyle\frac{1}{4}}\ (2-a_i)\ ,
\eea
whereas positions of the punctures \eqref{sopasopspoa}
are  constrained by  a certain   monodromy-free  condition. 
The latter is now understood as a requirement
that $ \re^{\pm \frac{1}{2}\eta\sigma_3}\ {\boldsymbol \Psi}$ (where
${\boldsymbol \Psi}$ is a general solution 
of the  auxiliary linear problem \eqref{apsosaospa}) 
is single-valued in the neighborhood of the punctures
 $z=x_a$ $(a=1,\ldots L)$ and ${\bar z}={\bar y}_b$ $(b=1,\ldots {\bar L})$.
Following the consideration from \cite{Feigin:2007mr}, the
monodromy-free condition can be transformed into a set
of $L+{\bar L}$ constraints imposed on the 
regular part of local  expansions of $(\partial_z\eta,\partial_{\bar
  z}\eta)$  at  the monodromy-free punctures: 
\bea\label{isuposksi}
\partial_z\eta&=&\frac{1}{z-x_a}+\frac{1}{2}\ \gamma_a+o(1)\\
\partial_{\bar z}\eta&=&-\frac{1}{{\bar z}-{\bar
    x}_a}+o(1)\ \ \ \ \ \ \ \ \ \ \ \ \ \ \ \ \ \ \  
\ (a=1,\ldots L)\nonumber
\eea
and
\bea\label{saopsosap}
\partial_{\bar z}\eta&=&\frac{1}{{\bar z}-{\bar y}_b}+\frac{1}{2}\
	{\bar \gamma}_b+o(1)\\ 
\partial_z\eta&=&-\frac{1}{z-y_b}+o(1) \ \ \ \ \ \ \ \ \ \  \ \ \ \ \ \ \ \ \ \ ( b=1,\ldots {\bar L})\ ,\nonumber
\eea
where $\gamma_a=\partial_z\log {\cal P}(z)|_{z=x_a},\ {\bar \gamma}_b=
\partial_{\bar z}\log {\bar {\cal P}}({\bar z})|_{{\bar z}={\bar y}_b}$.
We expect that as far as
positions of the  punctures $z_i$ are fixed, the triple ${\bf
  m}=(m_1,m_2,m_3)$ \eqref{asioasiso} and 
the pair  $(L,{\bar L})$ are chosen,
the MShG equation possesses a {\it finite  set} ${\cal A}^{(L,{\bar L})}_{\bf m}$
of solutions 
satisfying  all  the above requirements. 
We can now define the moduli space ${\cal A}_{\bf m}$ which is the
union of such  finite sets:
\bea\label{oaioasoas}
{\cal A}_{\bf m}=\cup_{L,{\bar L}}^\infty\,{\cal A}^{(L,{\bar L})}_{\bf m}\ .
\eea
Notice that, to a  certain extent, ${\cal A}_{\bf m}$ can be regarded
as a Hitchin moduli space \cite{Hitchin:1986vp}. 

\begin{figure}
\centering
\includegraphics[width=6.  cm]{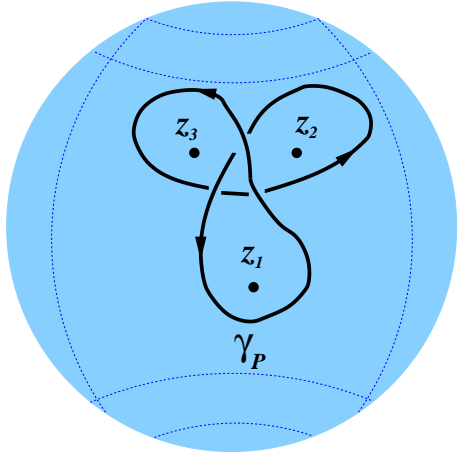}
\caption{The Pochhammer  loop on the  Riemann sphere.}
\label{fig1av}
\end{figure}

An essential ingredient of the formal theory of 
the partial differential equation \eqref{asospsosap}
is the existence of an infinite hierarchy of one-forms, 
which are closed by virtue of the equation \eqref{asospsosap} itself.
In the case under consideration this formal property leads to the  existence
of an infinite set of conserved charges  
$\{{\mathfrak q}_{2n-1},\, {\bar {\mathfrak q}}_{2n-1}\}_{n=1}^\infty$, 
which can be used to characterize the elements of 
the moduli space  ${\cal A}_{\bf m}$. 
The proof of this statement goes along the following lines. 
It easy to see that the flat  connection 
${\boldsymbol A}={\boldsymbol A}_z\,\rd z+{\boldsymbol A}_{\bar z}\,\rd {\bar z}$ \eqref{ystopsso}
associated with an  element of ${\cal A}_{\bf m}$
is not single-valued on the punctured sphere. However, 
it does return to the original value
after a continuation along the Pochhammer loop --- 
the contour $\gamma_P$ depicted in Fig.\ref{fig1av}.
Therefore one  can consider the Wilson loop
\bea\label{sospsaosapo}
W={\rm Tr}\bigg[{\cal P}\exp\Big(\oint_{\gamma_P} {\boldsymbol  A}\,\Big)\,\bigg]\ ,
\eea
whose definition does not depend on the precise
shape of the integration contour. In particular, it
is not sensitive to deformations of $\gamma_P$
which sweep through
the monodromy-free punctures.
By construction the Wilson loop is
an entire function of the spectral parameter $\theta$,
\bea\label{osapsoasp}
\lambda=\rho\ \re^{\theta}\ ,\ \ \ \ \ \ {\bar \lambda}=\rho\ \re^{-\theta}\ .
\eea
Furthermore,
since the shift of the argument $\theta\mapsto\theta+\ri\pi$ does not affect the connection ${\boldsymbol A}$,
the  Wilson
loop $W=W(\theta)$ is a periodic function of the period $\ri\pi$. 
The textbook calculation \cite{Faddeev:1987ph} yields the following asymptotic expansions:
\bea\label{aaosioasao}
\log W\asymp
\begin{cases}
& -  q_0\, \rho\re^{\theta}+\sum_{n=1}^\infty c_n\  {\mathfrak q}_{2n-1}\ \re^{-(2n-1)\theta}\ \ \ 
{\rm as}\ \ \                            
\Re e(\theta)\to+\infty,\ \ |\Im  m(\theta)|<\frac{\pi}{2}\\
& - q_0\, \rho\re^{-\theta}+\sum_{n=1}^\infty c_n\ {\bar {\mathfrak q}}_{2n-1}\ \re^{(2n-1)\theta}\ \ \
{\rm as}\ \ \ 
\Re e(\theta)\to-\infty,\ \ |\Im  m(\theta)|<\frac{\pi}{2}
\end{cases}
\ .
\eea
Here  $q_0=-\frac{4\pi^2}{\prod_{i=1}^3\Gamma(1-\frac{a_i}{2})}$, whereas  
$c_n=\frac{(-1)^n}{2n!}\ \frac{\Gamma(n-\frac{1}{2})}{\sqrt{\pi}}$ 
stand for the
constants that  set a conventional multiplicative normalization (see Eqs.\eqref{osapsospa}-\eqref{aopssapissu} bellow) for
the  conserved charges $\{{\mathfrak q}_{2n-1},\, {\bar {\mathfrak q}}_{2n-1}\}_{n=1}^\infty$. 

The main result of this work is presented in Section\,\ref{sec8}, where
we  conjecture a correspondence  between elements of 
the moduli space  ${\cal A}_{\bf m}$ \eqref{oaioasoas}
and a subset ${\cal H}^{(0)}_{\bf k}$ of the  stationary states
of the Fateev model in a finite volume.
To describe ${\cal H}^{(0)}_{\bf k}$
explicitly, let us  recall some basic facts about the model. 
The Fateev model is governed by
the  following  Lagrangian in $1+1$ Minkowski
space
\bea\label{aposoasio}
{\cal L}&=& \frac{1}{16\pi}\ \sum_{i=1}^3
\big(\, (\partial_t\varphi_i)^2-(\partial_x\varphi_i)^2\,\big)\\
&+&2\mu\
\big(\, \re^{\ri\, \alpha_3\varphi_3}\ \cos(\alpha_1\varphi_1+\alpha_2\varphi_2)+\re^{-\ri \alpha_3\varphi_3}\
\cos(\alpha_1\varphi_1-\alpha_2\varphi_2)\,\big)
\nonumber
\eea
for the three scalar fields $\varphi_i=\varphi_i(x,t)$.
Here $\alpha_i$ are   coupling constants satisfying the constraint
\bea\label{aposapoas}
\alpha_1^2+\alpha_2^2+\alpha_3^2=\frac{1}{2}\ .
\eea
In this work we
focus on the case where $\alpha_i^2>0$.
The parameter $\mu$ in the Lagrangian sets the mass scale, $\mu\sim [\,{\rm mass}\,]$.
We will consider  the theory in a finite-size geometry (where the
spatial coordinate $x$ compactified on a circle of
circumference $R$) with the periodic boundary conditions
\bea\label{sissiaosai}
\varphi_i(x+R,t)=\varphi_i(x,t)\ .
\eea
Due to the periodicity of the potential 
term in  \eqref{aposoasio} in $\varphi_i$,  the space of states ${\cal H}$ splits on the orthogonal
subspaces ${\cal H}_{k_1,k_2,k_3}:={\cal H}_{\bf k}$ characterized by the three  
``quasimomenta'' $k_i\in [-\frac{1}{2},\frac{1}{2}]$:
\bea\label{sapsapo}
\varphi_i\mapsto\varphi_i+2\pi/\alpha_i\ :\ \ \ |\,\Psi_{\bf k}\,\rangle\mapsto\re^{2\pi\ri k_i}\  |\,\Psi_{\bf k}\,\rangle\ ,
\ \ \ \ \ \ \ \ |\,\Psi_{\bf k}\,\rangle\in {\cal H}_{\bf k}\ .
\eea
Similar to the quantum mechanical problem of
a  particle in  a periodic potential,
the subspaces ${\cal H}_{\bf k}$ possess the band structure; they 
split into discrete components labeled by three   integers:
\bea\label{saossasap}
{\cal H}_{\bf k}=\oplus_{n_1,n_2,n_3\in\mathbb{Z}}\,{\cal H}^{(n_1,n_2,n_3)}_{\bf k}\ .
\eea

The QFT \eqref{aposoasio} is integrable, in particular,
it has an infinite set of commuting local integrals of motion 
$\mathbb{I}^{(+)}_{2n-1}$,\ 
$\mathbb{ I}^{(-)}_{2n-1}$, with $2n=2,\, 4,\,6,\,\ldots$
being the Lorentz spins of the associated local densities \cite{Fateev:1996ea}:
\bea\label{isusospsasopas}
\mathbb{I}^{(\pm)}_{2n-1}=
\int_0^R\frac{\rd x}{ 2\pi}\
\ \Big[\, \sum_{i+j+k=n} C^{(n)}_{ijk}\ (\partial_\pm\varphi_1)^{2i}\,
(\partial_\pm\varphi_2)^{2j}\, (\partial_\pm\varphi_3)^{2k} +\ldots\,  \Big]\ ,
\eea
where  $\partial_\pm=\frac{1}{2} (\partial_x\mp \partial_t)$ 
and
$\ldots$ stand for the terms involving higher
derivatives of $\varphi_i$, as well as the terms proportional to
powers of $\mu$.
The constant $C^{(n)}_{ijk}$ was found in \cite{Lukyanov:2012wq}
\bea\label{saosopsaosa}
C^{(n)}_{ijk}=\frac{n!}{i!\ j!\ k!}\  \ 
\frac{
\big(2\alpha_1^2(1-2 n)\big)_{n-i} \big(2\alpha_2^2\,(1-2 n)\big)_{n-j}\,\big(2\alpha_3^2\,(1-2 n)\big)_{n-k}}{
(2n-1)^3\  (4\alpha^2_1)^{1-i}\ (4\alpha^2_2)^{1-j}\ (4\alpha^2_3)^{1-k}}\ , 
\eea
where   $(x)_n$ stands for the Pochhammer symbol.
Note, that
the displayed terms in \eqref{isusospsasopas} with $C^{(n)}_{ijk}$ given by 
\eqref{saosopsaosa} define the normalization of 
$\mathbb{I}^{(\pm)}_{2n-1}$ unambiguously.
Our primary interest concerns   
eigenvalues of $\mathbb{I}^{(\pm)}_{2n-1}$ in the subspaces 
${\cal H}^{(n_1,n_2,n_3)}_{\bf k}$ \eqref{saossasap}, particularly, in the subspace
${\cal H}^{(0)}_{\bf k}:={\cal H}^{(0,0,0)}_{k_1,k_2,k_3}$
corresponding to  the first Brillouin zone:
\bea\label{akssasuaias}
I^{(\pm)}_{2n-1}\ :\ \ \ \ \ \ \ 
\mathbb{I}^{(\pm)}_{2n-1}\ |\,\Psi^{(0)}_{\bf k}\,\rangle=
I^{(\pm)}_{2n-1}\ |\,\Psi^{(0)}_{\bf k}\,\rangle\ ,\ \ \ \ \ \ \  |\,\Psi^{(0)}_{\bf k}\,\rangle\in 
{\cal H}^{(0)}_{\bf k}\ .
\eea
It seems  natural to expect that for
$0\leq k_i\leq\frac{1}{2}$, 
the sets of eigenvalues 
$\{I^{(+)}_{2n-1},I^{(-)}_{2n-1}\}_{n=1}^\infty$ fully specify the
common eigenbasis
of  the  local   IM in ${\cal H}^{(0)}_{\bf k}$.

In the recent paper \cite{Lukyanov:2013wra}  it was  argued  that  the  {\it  vacuum} eigenvalues
$\{I^{(+)}_{2n-1},I^{(-)}_{2n-1}\}_{n=1}^\infty$
(i.e.  those corresponding to the unique state  in ${\cal
  H}^{(0)}_{\bf k}$ with the lowest value of the energy
$E=I^{(+)}_1+I^{(-)}_{1}$) 
are simply related to  the set of    conserved charges
 $\{{\mathfrak q}_{2n-1},\, {\bar {\mathfrak q}}_{2n-1}\}_{n=1}^\infty$
associated with the unique element
${\cal A}^{(0,0)}_{\bf m}$ of  the moduli space  ${\cal A}_{\bf m}$ \eqref{oaioasoas}. Namely:
\bea\label{aspspsapo}
\mu^{-1}\ \big(\,I_{1}-{\textstyle\frac{1}{2}}\, R\,{\cal E}_0\,\big)=d_1\ {\mathfrak q}_{1}\ ,\ \ \ \ \ \ \ \ \ \ 
\mu^{-1}\ \big(\,{\bar I}_{1}-{\textstyle\frac{1}{2}}\, R\,{\cal E}_0\,\big)=d_1\ {\bar {\mathfrak q}}_{1} 
\eea
and
\bea\label{apospaosp}
\mu^{1-2n}\ I^{(+)}_{2n-1}=d_n\ {\mathfrak q}_{2 n-1}\ ,\ \ \ \ \ \ \ \ \ \ \ \ 
\mu^{1-2n}\ I^{(-)}_{2n-1}=d_n\ {\bar {\mathfrak q}}_{2 n-1} \  \ \ \ \ \ \ \ \ \ \ \ \ (n=2,\,3,\,\ldots)\  .
\eea
With the normalization conditions 
described above, the constants $d_n$ and ${\cal E}_0$   reads explicitly as
\bea\label{poapsospsaos}
d_n=(2\pi)^{2n-1}\ 
\ \frac{(-1)^{n-1} }{16\,\pi^2  }\  
\ \prod_{i=1}^3\Gamma\big(\, 2\,(2n-1)\, \alpha^2_i\,\big)
\eea
and
\bea\label{asososa}
{\cal E}_0=-\pi\mu^2\ \prod_{i=1}^3\frac{\Gamma(2\alpha_i^2)}{\Gamma(1-2\alpha_i^2)}\ .
\eea
These relations should be supplemented by the  identification of the
parameters from the 
quantum and classical  integrable problems:
\beq\label{sssaopsa}
\alpha_i^2={\frac{a_i}{4}}\,,\ \ \ \ \ \ \  \qquad
k_i=\frac{1}{a_i}\ (2m_i+1)\ \ \ \ \ \qquad 
(i=1,2,3)\ ,
\eeq
whereas the relation between dimensionless parameter $\mu R$ and $\rho$ is  given by 
\bea\label{saossaops}
\mu R=2 \rho\ .
\eea

In this work we  promote  
Eqs.\eqref{aspspsapo}-\eqref{saossaops} to a general relations
between the joint spectra 
of
the local IM  in the  subspace  ${\cal H}^{(0)}_{\bf k}$  corresponding the first Brillouin zone and
the set of the conserved charges  
associated with
the elements of the moduli space ${\cal A}_{\bf m}$. 
For the values of $k_i$ restricted to the segment
$[\,0,\frac{1}{2}\,]$, this gives a remarkable bijection 
between the joint eigenbasis of the local IM in
${\cal H}^{(0)}_{\bf k}$ and the elements  of ${\cal A}_{\bf m}$.

In Section \ref{ninesec} we demonstrate  that the correspondence
between the classical and quantum integrable systems provides  a
powerful tool for deriving  integral equations which determine the
full spectrum  of  local  IM in the massive QFT.

We conclude this paper with few remarks concerning the QFT
\eqref{aposoasio} in the regime where one of the couplings $\alpha_i$
is pure imaginary.

\section{\label{sec21}Generalized Hypergeometric Oper}

\subsection{Monodromies of the Fuchsian differential equations}

In this preliminary subsection we include some basic concepts and results 
about the Fuchsian differential equations.

Let  $z$ stands for the complex coordinate on $\mathbb{CP}^1\backslash\{z_1,z_2,\ldots z_n\}$,
the Riemann sphere with $n$ punctures.
Consider the  second order
Fuchsian differential operator  $-\partial_z^2+T(z)$,
where $T(z)$ is given by
\bea\label{ssapsapo}
T(z)=-\sum_{i=1}^n\Big(\,\frac{\delta_i}{(z-z_i)^2}+\frac{c_i}{z-z_i}\,
\Big)\ .
\eea
The equation
\bea\label{sopspa}
\big(-\partial_z^2+T(z)\,\big)\,\psi=0
\eea
is a general second-order differential equation with $n$ regular
singular points.  
We will always regard the parameters $\delta_i$ as fixed numbers.
The positions of the singularities $z_i$ and  
the coefficients $c_i$ (which are usually referred to as 
the ``accessory parameters'') will be treated as variables.
The accessory parameters $c_i$ are constrained by the elementary relations
\bea\label{ytosaospa}
\sum_{i=1}^nc_i=0\ ,\ \ \ \ \ \sum_{i=1}^n(z_i\,c_i+\delta_i)=0\ ,\ \ \  \
\sum_{i=1}^n(z_i^2\,c_i+2\,z_i\delta_i)=0\ ,
\eea
ensuring that $T(z)$ has no additional singularity at $z=\infty$. Thus only $n-3$ of these parameters
are independent.
Also, the projective transformations of the variable $z$ allows one to
send three of the points $z_i$,  
say $(z_1,\,z_{2},\,z_3)$, to any designated positions,
usually $(0,\,1,\,\infty)$. 
Therefore,  with fixed $\delta_i$, the differential equation
\eqref{sopspa} essentially depends on $2\, (n-3)$ complex parameters.

The equation \eqref{sopspa} generates a monodromy group --- a 
homomorphism of the fundamental group of the sphere with marked points
into the group $\mathbb{ SL}(2,\mathbb{ C})$,
\bea\label{ksasusa}
{\boldsymbol  M}\ :\ \ 
\pi_1\big(\mathbb{CP}^1\backslash\{z_i\}\big) \mapsto \mathbb{
  SL}(2,\mathbb{ C})\ . 
\eea
Let 
$(\psi_1(z), \psi_2(z))$ is a basis of linearly independent 
solutions of \eqref{sopspa}.  
Then its continuation along any
closed path $\gamma$ defines the
monodromy matrix 
\bea\label{apospaospsap}
{\boldsymbol M(\gamma)}\ :\ \ \ \ \  \big(\psi_1(\gamma\circ z),\psi_2(\gamma\circ z)) = 
(\psi_1(z), \psi_2(z))\,  {\boldsymbol M}(\gamma)\ ,
\eea
which depends only on the homotopy class of $\gamma \in \pi_1
(\mathbb{CP}^{1}\,\backslash\{z_i\})$. Let $\gamma_i \in \pi_1
(\mathbb{CP}^{1}\backslash\{z_i\}),\ i=1,2,\ldots n$ be the elementary paths
around the points $z_i$, and 
\bea\label{apospaospaq}
{\boldsymbol M}^{(i)}:={\boldsymbol M}(\gamma_i) \in \mathbb{ SL}(2,\mathbb{ C}) 
\eea
the associated elements of the monodromy group of \eqref{sopspa}. 
The parameters 
\bea\label{oipsasap}
\delta_i=\delta(p_i)\ \  \ \ \ \ \ \ \ {\rm with}\ \ \ \ \ \ \ \ \ \ \
\delta(p)={\textstyle\frac{1}{4}}-p^2
\eea
determine the conjugacy classes
of ${\boldsymbol M}^{(i)}$ via the equation
\bea\label{oaposaop}
{\rm Tr}\big({\boldsymbol M}^{(i)}\big)=-2\, \cos(2\pi p_i)\ .
\eea

\begin{figure}
\centering
\includegraphics[width=10  cm]{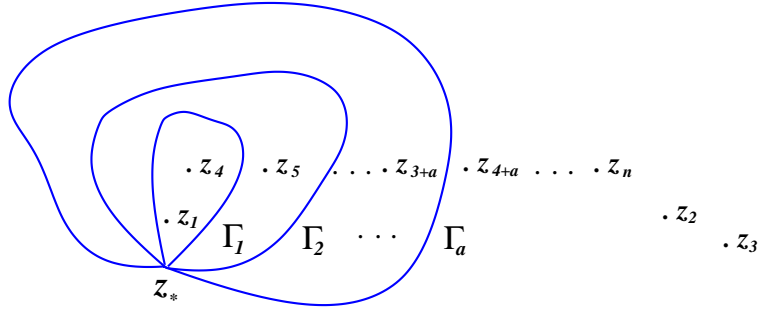}
\caption{The elements $\Gamma_1,\,\Gamma_2,\ldots\Gamma_{n-3}$ of the fundamental
group $\pi_1\big(\mathbb{CP}^1\backslash\{z_i\}\big)$. Choosing the accessory parameters in \eqref{ssapsapo}
according to \eqref{apossosp} fixes the conjugacy classes of the associated elements of the
monodromy group of \eqref{sopspa} as given in \eqref{hsausysau}.}
\label{fig1ab}
\end{figure}

\begin{figure}
\centering
\includegraphics[width=12  cm]{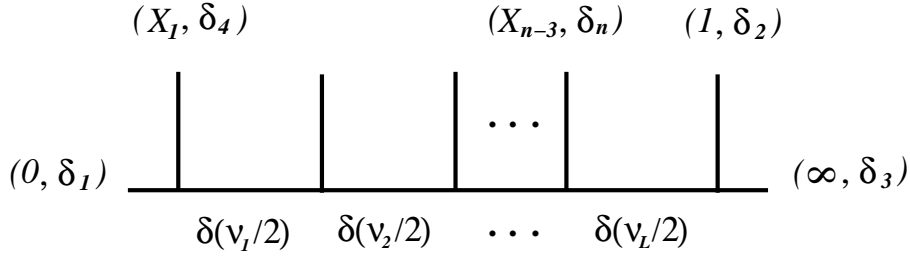}
\caption{Dual diagram for the classical conformal block from Eq.\eqref{apossosp}, $\delta(\nu_a/2)=\frac{1}{4}\,(1-\nu_a^2)$.}
\label{fig1aa}
\end{figure}

Let  $\big\{\Gamma_a\big\}_{a=1}^{n-3}$ be the system of contours    shown in Fig.\,\ref{fig1ab}, such that
$\Gamma_a$  loops around the punctures $z_1,\,z_4,\ldots z_{3+a}$ only;
and let the set  ${\boldsymbol \nu}=(\nu_1,\ldots \nu_{n-3})$ parameterize the conjugacy classes of the corresponding  monodromy matrices
${\boldsymbol M}(\Gamma_a)$,
\bea\label{hsausysau}
{\rm Tr}\big(\, {\boldsymbol M}(\Gamma_a)\,\big)=-2\ \cos(\pi \nu_a)\ \ \ \ \
(a=1,\ldots n-3)\ .\label{aopaposa}
\eea
For given conjugacy classes \eqref{hsausysau}   (i.e., for a given set
${\boldsymbol \nu}$),   
the accessory parameters are determined in terms of the so-called classical conformal block
$f_{\boldsymbol \nu}(X_1,\ldots X_{n-3}) $ corresponding to ``haircomb'' 
diagram  shown in  Fig.\,\ref{fig1aa} (for  details, see e.g.\cite{Nekrasov:2011bc,LLNZ}). Namely,
\bea\label{aopsaps}
c_i=\frac{\partial}{\partial z_i}\ F\ \ \ \ \ \ \ \ \ \ \ \ \ (i=1,\ldots n)\ ,
\eea
where  the shortcut notation $F$ stands for
\bea\label{apossosp}
F=F_0+\delta_3\ \log\Big(\frac{z_{21}}{z_{31}z_{32}}\Big)
+\sum_{i=1\atop i\not= 3}^n\delta_{i}\ \log\Big(\frac{z_{31}z_{32}}{z_{21}z^2_{3i}}\Big)+
f_{\boldsymbol \nu}(X_1,\ldots X_{n-3})\ ,
\eea
with $z_{ij}:=z_i-z_j$ and
the arguments of the conformal block   are  cross ratios  
\bea\label{aoapospaspo}
X_a=\frac{z_{a+3}-z_1}{z_{a+3}-z_3}\ \frac{z_{2}-z_3}{z_2-z_1}\ \ \ \ \ \ \ \ \ (a=1,\ldots n-3)\ .
\eea
A certain  additive normalization of the
classical conformal block is usually assumed.
For this reason,  we reserve the room for an arbitrary coordinate-independent constant $F_0$ in \eqref{apossosp}.

\subsection{\label{sec2}Definition of GHO}

Here we consider a  special class of second order Fuchsian
differential operators. 
We will always assume that the three parameters  $p_i$ 
defining conjugacy classes of the matrices 
${\boldsymbol M}^{(i)}$ in \eqref{apospaospaq}-\eqref{oaposaop},
associated with the elementary  
paths around   the ``fixed'' punctures $z_1,\, z_2,\ z_3$,  
are  positive numbers satisfying the following  constraints
\bea\label{aopsaospap}
0<p_i<{\textstyle\frac{1}{2}}\ \ \ (i=1,\,2,\,3)\ ,\ \  \ \  \
0<p_1+p_2+p_3<{\textstyle\frac{1}{2}}\ . 
\eea
For the remaining $L:=n-3$ punctures 
we require that both linearly independent solutions of \eqref{sopspa}
are single-valued (or {\em monodromy-free}) in the vicinity of these
punctures. It is well known \cite{Duistermaat}
how to reformulate this condition as a set of algebraic relations imposed
on the corresponding parameters
$\delta_{3+a}$, $C_a:=c_{3+a}$ \  and $x_a:=z_{3+a}$, 
$(a=1,\ldots L)$.
Namely, suppose that $T(z)$ has  a Laurent  expansion at 
$z=x_a$ of the form
\bea\label{isiooisasa}
T(z)=-\frac{l_a(l_a+1)}{(z-x_a)^2}-
\frac{C_a}{z-x_a}-\sum_{k=0}^{+\infty} t^{(a)}_k\ (z-x_a)^k \ .
\eea
It is easy to see that  $l_a$ must  be  an integer.
We will focus on the case
$l_a=1$, i.e.
\bea\label{ossasap}
\delta_{3+a}=-2\ \ \ \ \ \ (a=1,\ldots L)\ .
\eea
To ensure that 
solutions of Eq.\eqref{sopspa} are single-valued
in the vicinity of the punctures at $z=x_a\  (a=1,\ldots L)$,
the expansion  coefficients $t^{(a)}_0$ and $t^{(a)}_1$ in \eqref{isiooisasa}
should be constrained as 
\bea\label{opsoa}
(C_a)^3\,  -4\ C_a\  t^{(a)}_0+4 \, t^{(a)}_1=0\ .
\eea
This yields
\bea\label{hsgspossaopa}
&&C_a\ \bigg[\, \frac{1}{4}\ 
(C_a)^2-T_0(x_a)-\sum_{b\not= a}^L\Big(\frac{2}{(x_a-x_b)^2}-
\frac{C_b}{x_a-x_b}\,\Big)
\bigg]\nonumber\\
&&-
T'_0(x_a)+\sum_{b\not= a}^L\Big(\frac{4}{(x_a-x_b)^3}
-\frac{C_b}{(x_a-x_b)^2}\,\Big)
=0\ \ \ \ \ (a=1,\ldots L)\ ,
\eea
where
\bea\label{sopsapos}
T_0(z)=
-\sum_{i=1}^3\Big(\,\frac{\delta_i}{(z-z_i)^2}+\frac{c_i}{z-z_i}\,
\Big)\ .
\eea
The prime in  $T'(z)$ stands for the derivative w.r.t. the variable $z$.
This system of algebraic equations  should be supplemented by 
the three conditions\ \eqref{ytosaospa}, specialized to the case
\eqref{ossasap}: 
\bea\label{osaospa}
\sum_{i=1}^3c_i&=&-\sum_{a=1}^LC_a\ ,\nonumber\\
\sum_{i=1}^3(z_i\,c_i+\delta_i)&=& -\sum_{a=1}^L(x_a\,C_a-2)\ ,\\
\sum_{i=1}^3(z_i^2\,c_i+2x_i\delta_i)&=&- \sum_{a=1}^L(x^2_a\,C_a-4 x_a)\ .\nonumber
\eea

\begin{figure}
\centering
\psfrag{z1}{$(0,\delta_1)$}
\psfrag{z2}{$(X_1,-2)$}
\psfrag{z3}{$\delta(p_1+\epsilon_1)$}
\psfrag{z4}{$(X_2,-2)$}
\psfrag{z5}{$\delta(p_1+\epsilon_1+\epsilon_2)$}
\psfrag{z6}{$(X_{L},-2)$}
\psfrag{z7}{${\textstyle\delta(p_1+\epsilon_1+\ldots+\epsilon_{L})}$}
\psfrag{z8}{$(1,\delta_2)$}
\psfrag{z9}{$(\infty,\delta_3)$}
\psfrag{zz}{$\cdots$}
\includegraphics[width=15  cm]{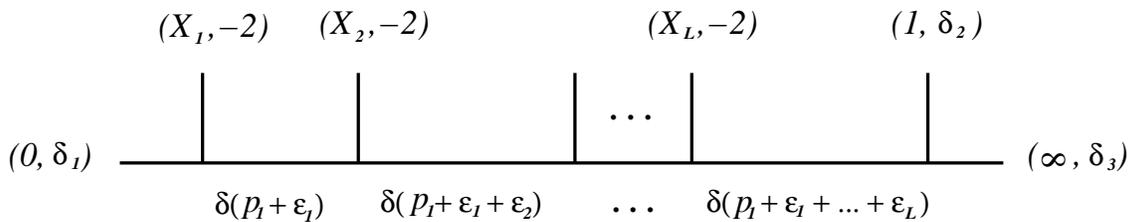}
\caption{Dual diagram for the classical conformal block from  Eq.\eqref{oappospsa}. Here
$\epsilon_a=0,\,\pm 1$\ $(a=1,\ldots L).$}
\label{fig2aac}
\end{figure}

As far as
positions of  the punctures $z_1,z_2,z_3$ and  
corresponding parameters ${\bf p}=(p_1,\,p_2,\,p_3)$ are 
fixed,  
Eqs.\eqref{hsgspossaopa}, \eqref{osaospa} define  an  algebraic 
variety which will be denoted by 
${\cal V}^{(L)}_{\bf p}$. 
If positions of the punctures $(x_1,\ldots x_L)$ 
are used
as local coordinates on ${\cal V}^{(L)}_{\bf p}$, then  a system of
$L$  locally  defined functions 
$C_a(x_1,\ldots x_L)$ 
satisfy the integrability conditions
\bea\label{opsaopsaop}
\frac{\partial}{\partial x_b}\, C_a=
\frac{\partial}{\partial x_a}\, C_b\  .
\eea
These  relations can be verified by
the brute-force calculation  using Eqs.\eqref{hsgspossaopa} and
\eqref{osaospa}, but, of course, 
they follows immediately from the general relation \eqref{aopsaps}.
In this  particular case the classical conformal block in Eq.\eqref{apossosp}
is related to the classical limit of the $(3+L)$-point
correlator involving three generic {\it chiral} vertex operators $V_{i}$
with conformal dimensions $\Delta_{i}\ (i=1,2,3)$
and $L$ degenerate vertices $V_{(3,1)}$ with dimensions $\Delta_{(3,1)}$:
\bea\label{oappospsa}
\langle\, V_{1}(0)\,V_{2}(1)\,V_{3}(\infty)\,
V_{(3,1)}(X_1)\,\ldots\,V_{(3,1)}(X_L)\,\rangle
\sim
\exp\Big(\,\frac{1}{b^2}\ f_{\boldsymbol \nu}(X_1,\ldots X_L)\,\Big),
\quad b^2\to0\,,
\eea
where the parameter $b^2$ and other conventional
notations are inherited from the quantum Liouville theory (see Ref.\cite{Zamolodchikov:1995aa} for details)
\bea\label{oaopsp}
\Delta_i\to\frac{\delta_i}{b^2}\ ,\ \ \ \ \ \Delta_{(3,1)}\to-\frac{2}{b^2}\ \ \ \ {\rm as}
\ \ \ \ \ b^2\to 0\ .
\eea
Due to the well known fusion rule for the degenerate vertex
$V_{(3,1)}$ \cite{Belavin:1984vu}, only a 
discrete set of the parameters ${\boldsymbol \nu}=(\nu_1,\ldots \nu_L)$
is allowed (see Fig.\,\ref{fig2aac}):
\bea\label{sospospa}
\nu_1=2\,(p_1+\epsilon_1)\ ,\ \ \ \ \ \nu_{a}=\nu_{a-1}+2\, \epsilon_a\ \ \ \ \ \ \  (a=2,\ldots L)\ ,
\eea
where the discrete variables  $\epsilon_a$ takes the values $0,\pm 1$ only.
Different configurations $(\epsilon_1,\ldots\epsilon_L)$ correspond to the different locally defined functions
$C^{(\epsilon_1\ldots\epsilon_L)}_a(x_1,\ldots x_L) \ (\epsilon_a=0,\pm 1)$, which are  
branches of  the  multivalued algebraic   function  of the complex variables $x_1,\ldots x_L$.
This is illustrated by the simplest  case with $L=1$ in Appendix A.

In what follows we will refer to the differential operators \eqref{assaisoa}, whose moduli space coincides with the 
algebraic variety ${\cal V}^{(L)}_{\bf p}$
as
to the {\it Generalized Hypergeometric Opers} (GHO's).\footnote{A
  general notion of $\mathfrak{g}$-oper for Riemann surfaces 
with punctures was introduced in \cite{Beilinson}.
In the case of the genus zero surface with $n$-marked points
an
$\mathfrak{sl}(2)$-oper is equivalent  to that of the second order
Fuchsian differential operator.}
The  marked points $x_a\ (a=1,\ldots L)$ will be called as 
{\it monodromy-free\/} punctures.

\subsection{\label{sec13}Connection matrices for GHO}

\begin{figure}
\centering
\includegraphics[width=6.  cm]{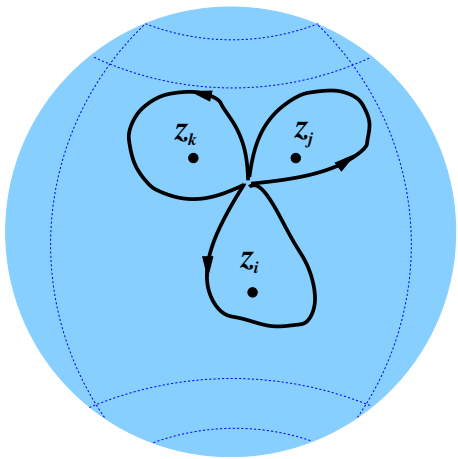}
\caption{The contractible loop $
\gamma_k\circ\gamma_j\circ\gamma_i=\gamma_i\circ\gamma_k\circ\gamma_j=
\gamma_j\circ\gamma_i\circ\gamma_k$ on
the sphere with three punctures.}
\label{fig1ay}
\end{figure}

We have introduced the concept of
GHO because the monodromy group of  such opers 
coincides with the monodromy group of the conventional  hypergeometric
equation. Let us recall some facts about this group.
In the case under consideration
there are only three elementary $\mathbb{SL}(2,{\mathbb C})$-matrices 
${\boldsymbol M}^{(i)}$, ${\boldsymbol M}^{(j)}$ and 
${\boldsymbol M}^{(k)}$  \eqref{oaposaop},
corresponding to the contours $\gamma_i$, $\gamma_j$ and $\gamma_k$, 
shown in Fig.~\ref{fig1ay}. Here $(i,j,k)$ is any {\it cyclic}
permutation of $(1,2,3)$. 
These matrices satisfy an obvious relation
\bea\label{sisioisaosisu}
{\boldsymbol M}^{(i)}{\boldsymbol M}^{(k)}{\boldsymbol M}^{(j)}={\boldsymbol I}\ ,
\eea
because the contour $\gamma_i\circ\gamma_k\circ\gamma_j$ is a
contractible loop.
Further, since $0<p_i<\frac{1}{2}$, one of these matrices, say, 
 ${\boldsymbol M}^{(i)}$ can always be 
chosen diagonal
\bea\label{raospposa}
{\boldsymbol M}^{(i)}= -\re^{-2\pi\ri p_i\sigma_3}\ .
\eea
Here and below we use standard notation for the Pauli matrices
$\sigma_1,\sigma_2, \sigma_3$. 
Then Eqs.\eqref{oaposaop}, \eqref{sisioisaosisu}, \eqref{raospposa}
define ${ {\boldsymbol M}}^{(j)}$ and ${ {\boldsymbol M}}^{(k)}$
up  to  a diagonal  similarity transformation. In particular,
\bea\label{skisosai}
{ {\boldsymbol M}}^{(j)}=\re^{-\omega_i\sigma_3}\ \Bigg[\
\frac{\ri}{ s(2 p_i)}\
\begin{pmatrix}
\re^{2 \pi \ri p_i}\  c(2 p_j)+ \, c(2p_k)&2 \,\Lambda\\
-2\ \Lambda& -\re^{-2 \pi \ri p_i}\, c(2 p_j)-\, c(2p_k)
\end{pmatrix}\ \Bigg]\ \re^{\omega_i\sigma_3}\ .
\eea
The quantity $\omega_i$ is an arbitrary complex number and 
\bea\label{saopsa}
\Lambda=\sqrt{c(p_2 + p_3 - p_1)\, c(p_1 + p_2 - p_3)
c(p_1 - p_2 + p_3)\, c(p_1 + p_2 + p_3)}\ ,
\eea
where we have used the shorthand notations
\bea\label{opsosap}
c(p)=\cos(\pi p)\ ,\ \ \ \ \ \ \ \ \ s(p)=\sin(\pi p)\ .
\eea

We now return to the the equation \eqref{sopspa} corresponding to  the  GHO.
Let $\chi^{(i)}_\sigma(z)$ $(i=1,2,3;\ \sigma=\pm)$  be its solutions  
such that
\bea\label{opaosap}
\chi_\sigma^{(i)}\to\frac{1}{\sqrt{2p_i}}\  (z-z_i)^{\frac{1}{2}+\sigma
  p_i}\,\Big(1+O(z-z_i)\Big)\qquad {\rm as}\ \ \ \ \ \ \ \ \ z\to z_i\ .
\eea
The prefactor here is chosen to satisfy the normalization condition
\bea\label{aoposas}
{\tt W}[\chi_{\sigma'}^{(i)},\,\chi_\sigma^{(i)}]=\sigma\,  \delta_{\sigma+\sigma',0}\ ,
\eea
where ${\tt W}[f,g]=f g'-g  f'$ stands for the Wronskian.
If the constraints \eqref{aopsaospap}
are imposed, the asymptotic conditions \eqref{opaosap}
define\footnote{It is worth noting,
however, that 
\eqref{opaosap} define these solutions only up to phase factors
of the form $\pm \re^{2\pi\ri p_i M}\ (M\in \mathbb{Z})$.} three
different bases (for $i=1,2,3$) in the  two-dimensional  
linear space of solutions of \eqref{sopspa}.
Let us  combine the solutions \eqref{opaosap} for given $i$ into the row
\bea\label{chi-basis}
{\boldsymbol \chi}^{(i)}=(\chi_-^{(i)},\chi_+^{(i)})\,,\qquad i=1,2,3\ .
\eea
Then the two sets of basis vectors  
${\boldsymbol \chi}^{(i)}$ and ${\boldsymbol \chi}^{(j)}$   are related
through a
linear transformation
\bea\label{sksssopsosp}
{\boldsymbol \chi}^{(i)}={\boldsymbol \chi}^{(j)}
\ {\boldsymbol S}^{(j,i)}\ .
\eea
Here ${\boldsymbol S}^{(j,i)}$ stand for $\mathbb{SL}(2,\mathbb{C})$-matrices,
satisfying  the relations
\bea\label{saopsapo}
\det\big({\boldsymbol S}^{(j,i)}\big)=1\,,\qquad
{\boldsymbol S}^{(i,k)}{\boldsymbol S}^{(k,i)}={\boldsymbol I}\,,\qquad
{\boldsymbol S}^{(i,k)}{\boldsymbol S}^{(k,j)}{\boldsymbol S}^{(j,i)}={\boldsymbol I}\,,
\eea
where again $(i,j,k)$ is any cyclic permutation of $(1,2,3)$.
It is easy to see that one needs  six  independent complex numbers to
parameterize the three matrices
${\boldsymbol S}^{(i+1,i)}$ $(i\sim i+3)$ satisfying
\eqref{saopsapo}. Moreover, these connection matrices are subject to
three additional  complex constraints. Indeed, the monodromy matrix 
${\boldsymbol M}^{(j)}$ \eqref{skisosai} can be expressed 
in term of the connection matrix ${\boldsymbol S}^{(j,i)}$:
\bea\label{isaisisoaoaia}
{\boldsymbol M}^{(j)}&=&
-\big({\boldsymbol S}^{(j,i)}\big)^{-1}\ \re^{-2\pi \ri p_j\sigma_3}\
{\boldsymbol S}^{(j,i)}\ .
\eea
This relation combined   with
Eqs.\eqref{saopsapo} leads to
\bea\label{osapsopsasisi}
{\boldsymbol S}^{(j,i)}=-
\re^{-\omega_{j}\sigma_3}\ \re^{\ri\pi p_j\sigma_3}\ \sigma_2\ {\boldsymbol A}^{(k)}\
 \re^{\omega_{i}\sigma_3}\ ,
\eea
with
\bea\label{sosssussa}
{\boldsymbol A}^{(k)}=
\begin{pmatrix}
A^{(k)}_{--}&A^{(k)}_{-+}\\
A^{(k)}_{+-}&A^{(k)}_{++}
\end{pmatrix}\ ,
\eea
where the matrix  elements   read explicitly
\bea\label{ospopsassss}
&&A_{\sigma'\sigma}^{(k)}=\sqrt{\frac{\Lambda}{
s(2 p_i)s (2 p_j)}}\ \ \bigg[\,\frac{c(  p_i +   p_j + p_k)\,c(  p_i +  p_j - p_k)}
{c(  p_i- p_j+  p_k )\, c(   p_i - p_j-p_k)}\,\bigg]^{\frac{\sigma\sigma'}{4}}
\eea
and $\Lambda$ is given by \eqref{saopsa}.
Note that Eq.\eqref{osapsopsasisi} can be equivalently
rewritten as a formula for the  Wronskians:
\bea\label{osapsassasss}
{\tt W}[\chi_{\sigma'}^{(j)},\chi_{\sigma}^{(i)}]
=
-\ri\ \re^{\ri\pi \sigma' p_j}\
 \ A_{\sigma'\sigma}^{(k)}\ \re^{-\sigma'\omega_j-\sigma \omega_i}\ .
\eea

The  complex parameters
$\omega_i\ (i=1,2,3)$, entering the expression \eqref{osapsopsasisi},
remain undetermined.
These parameters do not affect the conjugacy class of
 the representation of $\pi_1\big(\mathbb{CP}^1\backslash\{z_1,z_2,z_3\}\big)$.
Nevertheless, they are important characteristics of the GHO itself.
In the next section we argue 
that, in the case of  GHO,
a coordinate-independent additive normalization of the function
$F$  \eqref{apossosp} 
can be chosen  in such a way that
\bea\label{sopasapsissu}
\exp(\omega_i)=
\exp\Big(-\frac{1}{2}\ \frac{\partial }{\partial p_i}\ F\,\Big)\ \ \ \ \ \ \ \ \ \ 
(i=1,2,3)\ .
\eea
An immediate consequence of this fact is that
\bea\label{asospasoaso}
\exp(\omega_i)\ \Big(\frac{z_{jk}}{z_{ji} z_{ik}}\Big)^{-p_i}
\eea
depends on $L$ projective invariants \eqref{aoapospaspo}  only.
Explicit forms for \eqref{asospasoaso} in the cases $L=0$ and $L=1$ are given by equations
\eqref{saopopsa} and \eqref{soaoposap} from Appendix B, respectively.

Finally, note that  $\exp({\omega_i)}$  is defined up to the phase factor
$\pm \re^{2\pi\ri p M}\ (M\in \mathbb{Z})$. This  ambiguity  is
inherited from the similar 
ambiguity in the definition of $\chi_\sigma^{(i)}$ in Eq.\eqref{opaosap}.

\subsection{\label{Liouv}GHO and complex solutions of the Liouville equation}

Until now we have discussed  holomorphic GHO only. Of course, with 
minor modifications all the above can applied to
the antiholomorphic GHO
\bea\label{usyooapossap}
{\bar {\cal D}}=-\partial_{\bar z}^2-
\sum_{i=1}^3\Big(\,
\frac{{\bar \delta}_i}{({\bar z}-{\bar z}_i)^2}+
\frac{{\bar c}_i}{{\bar z}-{\bar z}_i}\,\Big)-
\sum_{b=1}^{\bar L}\Big(\,\frac{(-2)}{({\bar z}-
{\bar y}_b)^2}+\frac{{\bar C}_b}{{\bar z}-
{\bar y}_b}\,\Big)\ .
\eea
In what follows we assume that the triple
 $\{{\bar z}_i\}_{i=1}^3$   is complex conjugate 
to $\{z_i\}_{i=1}^3$, the corresponding ${\bar \delta}_i=\delta_i$ 
are real and, furthermore,
\bea\label{asioaisosa}
p_i={\bar p}_i\ .
\eea
We will not impose  any relations
between coordinates of  the 
monodromy-free punctures for the holomorphic and antiholomorphic GHO's.
For this reason the coordinates of antiholomorphic monodromy-free 
punctures in \eqref{usyooapossap}
are denoted by 
${\bar y}_b\,,\  b=1,\ldots {\bar L}$
and  ${\bar L}=0,1,2\ldots $ does not necessarily coincide 
with $L$.

Let ${\bar \chi}_{\sigma}^{(i)}$ be the basis solutions of 
${\bar {\cal D}}{\bar\psi}=0$,
which are  defined similarly to Eq.\eqref{opaosap}.
With the same  arguments as above,
one  arrives
to the antiholomorphic analog of  Eq.\eqref{osapsassasss}
\bea\label{osapssss}
{\tt W}[{\bar \chi}_{\sigma'}^{(j)},{\bar \chi}_{\sigma}^{(i)}]
=\ri\ 
\re^{-\ri\pi \sigma' p_j}\
 \ A_{\sigma'\sigma}^{(k)}\ 
\re^{-\sigma' {\bar \omega}_j-\sigma {\bar \omega}_i}\ ,
\eea
where $A_{\sigma'\sigma}^{(k)}$ is the same matrix as
in Eq.\eqref{ospopsassss}.

Consider  a bilinear form
\bea\label{kjjsasajk}
\tau(z,{\bar z})={\boldsymbol \chi}^{(i)}\ 
{\boldsymbol G}^{(i)}\ 
\big(\bar{{\boldsymbol \chi}}^{(i)}\big)^T
\eea
where ${\boldsymbol G}^{(i)}$  is an arbitrary $2\times 2$ matrix and superscript  
$T$ stands for the matrix transposition.
We  specialize  ${\boldsymbol G}^{(i)}$  by the requirement that
$\tau(z,{\bar z})$
is  a single-valued 
function on the punctured sphere.
Imposing this condition in the vicinity
of the  puncture $z_i$, one concludes that 
${\boldsymbol G}^{(i)}$ is  a diagonal matrix.
With the connection formula \eqref{sksssopsosp},
the single-valuedness 
implies that
${\boldsymbol S}^{(ji)}\
{\boldsymbol G}^{(i)}\ \big({\bar{\boldsymbol S}}^{(ji)}\big)^T$
is also a diagonal matrix.
Using  the explicit form of connection 
matrices \eqref{osapsopsasisi}-\eqref{ospopsassss}, 
one finds
\bea\label{aopssaop}
{\boldsymbol G}^{(i)}={\rm const}\
\re^{-(\omega_i+{\bar \omega}_{i})\sigma_3}\ \sigma_3\ .
\eea
If  the undetermined  constant is chosen to be $\pm 1$, then
the complex  function 
\bea\label{apoososa}
\eta\ :\ \ \ \ \ 
\re^{-\eta}:=\tau(z,{\bar z})\ ,
\eea
satisfies  the Liouville equation
\bea\label{iossaiao}
\partial_z\partial_{\bar z}\eta=\re^{2\eta}\ .
\eea 
This fact can be easily verified and it is  well known in the
theory of the classical Liouville equation.
Since we are considering  the complex solution of Eq.\eqref{iossaiao},
the overall sign of the constant in \eqref{aopssaop} it is not
important and we fix it to be $1$. As a result, one has
\bea\label{siisaoiso}
\re^{-\eta}=\re^{-(\omega_i+{\bar \omega}_{i})}\ 
\ \chi_-^{(i)}{\bar \chi}_-^{(i)}-
\re^{\omega_i+{\bar \omega}_{i}}\ \chi_+^{(i)}{\bar \chi}_+^{(i)}\ .
\eea
At the  monodromy-free punctures,
$z=x_a\ (a=1,\ldots L)$ and
${\bar z}={\bar y}_b\ (b=1,\ldots {\bar L})$,  $\re^{-\eta}$
becomes singular,
\bea\label{sopossaposa}
\re^{-\eta}\sim \frac{1}{z-x_a}\ \ \ \ \ \ {\rm  and}\ \ \ \ \  \ \  
\re^{-\eta}\sim \frac{1}{{\bar z}-{\bar y}_b}\ ,
\eea
however it  still  remains single-valued.
Thus, $\re^{-\eta}$ is a complex  single-valued function  on 
the sphere with $3+L+{\bar L}$ punctures.
Notice, that it
does not have any zeroes, as this 
contradicts to the Liouville equation \eqref{iossaiao}.

As it follows from Eq.\eqref{siisaoiso},
the solution $\eta$    satisfies the
asymptotic conditions
\bea\label{aospsao}
\eta&=&- 2\ \log|z|+O(1)\ \ \ \ \ \ \ \ \ \ \  \,\ \ \ \ \ \ \ \  \ \ \ \ \ \ \ \ \ \ \ \ 
{\rm as}\ \ \ \ \ \ |z|\to \infty\nonumber\\
\eta&=&2\,m_i\ \log|z-z_i|+\eta^{\rm (reg)}_i+o(1)\ 
\ \ \ \ \ \ \ \ \ \ \ \ \ \  {\rm as}\ \ \ \  \ \ |z-z_i|\to  0\ ,
\eea
where 
\bea\label{osopposopas}
m_i=p_i-\frac{1}{2}\ ,\ \ \ \ \ \ \ \ \ \ \ \ \ \ \frac{\exp\big(\eta^{\rm (reg)}_i\big)}{2\, p_i}= \re^{\omega_i+{\bar \omega}_i}\ .
\eea
The  constants $\eta^{\rm (reg)}_i$ 
can be regarded as regularized values of the Liouville
field at the punctures $z_i$. They are
be expressed in terms of  the regularized Liouville action \cite{zt,Zamolodchikov:1995aa}.
To explain this important relation we recall  that
the Liouville equation and the asymptotic conditions \eqref{aospsao}
follow from the variational principle for the functional
\bea\label{tystsaooiausais}
{\cal A}_{\rm Liouv}[ \eta]&=&\lim_{\epsilon_i\to 0\atop
R\to \infty}
\, \bigg[\, \frac{1}{\pi}\ \int_{|z-z_i|>\epsilon\atop
|z|<R}\rd^2 z\ 
\big(\, \partial_z{\eta}\partial_{\bar z}{\eta}+\re^{2\eta}
\,\big)
\\
&+&2\ \sum_{i=1}^3 \big(\,m_i\,\eta_i- m_i^2\,\log(\epsilon_i)\,\big)
+2\,\eta_\infty+2\,\log R\, \bigg]\ .\nonumber
\eea
Since the fields configuration is singular at $z\to z_i$, we cut out
a small disk of radius $\epsilon_i$ around the point $z_i$ and add
the boundary terms
with 
\bea\label{papssapo}
\eta_i=\frac{1}{2\pi\epsilon_i}\ \oint_{|z-z_i|=\epsilon_i} \rd \ell \, \eta
\eea
to ensure the behavior \eqref{aospsao} near $z_i$.
To control the large $|z|$-behavior
we regularize the action for large values of $z$ and add the boundary
term with
\bea\label{saisisaos}
\eta_\infty=\frac{1}{2\pi R}\ \oint_{|z|=R}\rd \ell\,\eta\ .
\eea
In addition, we 
include some field independent terms such that  ${\cal A}_{\rm Liouv}$ is
finite and independent on $\epsilon_i$ and $R$ when 
$\epsilon_i\to 0$, $R\to\infty$.
Contributions of the monodromy-free punctures\ \eqref{sopossaposa} 
to the 
functional \eqref{tystsaooiausais} are finite\footnote{Unless the some
  of the holomorphic and antiholomorphic punctures coincide.}
and therefore there is 
no need to include additional
regularization terms to the action.
It is now easy to show that
\bea\label{sospasapo}
\eta^{\rm (reg)}_i=\frac{1}{2}\ \frac{\partial {\cal A}_{\rm Liouv}^*}{\partial p_i}\ ,
\eea
where ${\cal A}^*_{\rm Liouv}$ stands for the
stationary value of the functional \eqref{tystsaooiausais}
calculated on the field configuration
$\eta$ defined by  Eq.\eqref{siisaoiso}. 
More generally, the stationary value of the Liouville functional
depends on $6+L+{\bar L}$ variables,
\bea\label{iisoosaiosa}
{\cal A}^*_{\rm Liouv}=
{\cal A}^*_{\rm Liouv}\big(\,p_1,p_2,p_3\,|\,z_1,z_2,z_3;x_1,\ldots x_L;
{\bar y}_1,\ldots {\bar y}_{\bar L}\,\big)\ ,
\eea 
and 
its total deferential is given by \cite{zt,Zamolodchikov:1995aa}
\bea\label{aopsasao}
\rd {\cal A}^*_{\rm Liouv}=2\sum_{i=1}^3\eta_i\,\rd p_i-
\sum_{i=1}^3\big(c_i\,\rd z_i+{\bar c}_i\, \rd {\bar z}_i\big)-
\sum_{a=1}^LC_a\,\rd x_a-\sum_{b=1}^{\bar L}{\bar C}_b\,\rd {\bar y}_b\ .
\eea
As it follows from \eqref{aopsaps},
${\cal A}^*_{\rm Liouv}$ can be expressed in terms
of $F$  and
its antiholomorphic counterpart ${\bar F}$:
\bea\label{sopaapsoapspo}
{\cal A}^*_{\rm Liouv}=-F-{\bar F}\ .
\eea
The coordinate-independent constant $F_0$ in Eq.\eqref{apossosp} 
has not yet been fixed. 
Therefore there is no need to add a  $p_i$-dependent constant in 
\eqref{sopaapsoapspo}, as it can always be absorbed by $F_0$
and ${\bar F}_0$.
The number and positions of the holomorphic and antiholomorphic monodromy-free
punctures are fully independent. For instance, we can 
consider the general holomorphic GHO, whereas
the antiholomorphic differential operator \eqref{usyooapossap} is  reduced
to the pure hypergeometric oper, i.e. ${\bar L}=0$.
Then,  Eqs.\eqref{osopposopas},\eqref{sospasapo} and \eqref{sopaapsoapspo},
imply that the additive normalization of $F$
can be chosen to satisfy the relation
\eqref{sopasapsissu}. Of course, a similar relation holds for ${\bar F}$
and $\re^{{\bar \omega}_i}$.

\section{\label{sectwo}Perturbed Generalized Hypergeometric Oper}
 
\subsection{Definition of PGHO}

Consider the universal cover of the Riemann sphere  
with three marked points $z_1,z_2,z_3$ and 
${\cal P}(z)\,(\rd z)^2$, where ${\cal P}(z)$ is given by \eqref{oasioaq} 
with   positive  parameters $a_i$. If  $a_i$ satisfy the constraint
\eqref{iasoaisa}, the quantity 
${\cal P}(z)\,(\rd z)^2$
transforms as a
quadratic differential under $\mathbb{PSL}(2,\mathbb{C})$
transformations and the punctures  $z_1,z_2,z_3$ on the Riemann sphere
can still  be sent to any desirable positions.

Suppose  we are also  given a GHO,\ 
${\cal D}=-\partial_z^2+T_L(z)$ which 
has its first three punctures at the
branching points of ${\cal P}(z)$ plus $L$ monodromy-free punctures 
at $z=x_a$ ($a=1,\ldots L$). Remind, that previously we have required that 
the parameters $p_i$ obey the constraints 
\eqref{aopsaospap}. In what follows we will impose 
 somewhat stronger constraints on these parameters. Namely
we replace \eqref{aopsaospap} by
\bea\label{posposopa}
0<p_i<\frac{a_i}{4}\ \ \ \ \ \ \ \ \ \ \ \ \ (i=1,2,3)\ .
\eea
The r${\hat {\rm o}}$le of this constraint will be explained 
in Section\,\ref{sewd} bellow.
An immediate object of our interest is an ODE of the form
\bea\label{opspspsao}
{\cal D}(\lambda)\,\psi=0\ ,\ \ \ \ \ \ \ \ \ \  \ \ \ \ \ \ \ \ \ 
{\cal D}(\lambda)=
 -\partial^2_z+T_L(z)+\lambda^2\,{\cal P}(z)\ ,
\eea
where $\lambda$ stands for an arbitrary complex parameter.
The properties of the
differential equation \eqref{opspspsao} are essentially affected  by the  presence of the
$\lambda$-dependent term.
Nevertheless, one can still find particular values of its parameters 
to make the marked points $z=x_a \ (a=1,\ldots L)$  
to be monodromy-free
punctures for arbitrary values of $\lambda$.
Indeed, the conditions \eqref{opsoa} can be easily generalized 
for $\lambda\not=0$. In this case the system of algebraic system
\eqref{hsgspossaopa}-\eqref{osaospa}
is extended by additional $L$ equations
\bea\label{saopsoposa}
C_a=-\partial_z\log{\cal P}(z)\big|_{z=x_a}=\sum_{i=1}^3\frac{2-a_i}{x_a-z_i}\ \ \ \ \ \ 
\ \ \ \ \ (a=1,\ldots L)\  ,
\eea 
which determine the values of $x_a$ $(a=1,\ldots L)$. These algebraic
equations have a finite discrete set of solutions. Therefore,
for any given $L$, there only a finite number ${\cal N}_L$ of 
sets of monodromy-free punctures
\bea\label{sisoaisiosaoi}
{\cal A}^{(L)}_{\bf p}=\big\{\,\big(\,x^{(\alpha)}_{1},
\ldots x^{(\alpha)}_{L}\,\big)\,\big\}_{\alpha=1}^{{\cal N}_L}\ \ \ \ \ \ \ \ \ 
({\cal N}_L<
\infty) \, .
\eea
Notice  that
Eqs.\eqref{hsgspossaopa}-\eqref{osaospa},\,\eqref{saopsoposa} are 
symmetric upon permutations of $(x_1,\ldots x_L)$, therefore 
we will not distinguish sets, which differ only by a permutation 
of the positions of the punctures.

Let us illustrate the situation on the simplest  $L=1$ example.
As in Appendix\,\ref{app1}, we  set
$(z_1,z_2,z_2)= (0,1,\infty)$, so that
together with \eqref{soiasisisa}
one has an additional relation
\bea\label{sisiaiso}
y=a_1-1+a_3\, x\ .
\eea
This leads to a cubic equation for the position $x$ of the 
monodromy-free puncture:
\bea\label{sopspspa}
a_3\,s_3\, x^2\,(x-1)+s\, x\,(x-1)-
a_1\, s_1\, (x-1)-a_2\, s_2\, x=0\ ,
\eea
where
$s_i=-a_i(a_i-2)-4\,\delta_i$ and  $s=(s_2-s_1)(a_3-1)+s_3(a_1-1)$.
Thus, there are only three different positions $\{x^{(\alpha)}\}_{\alpha=1}^3$
for a monodromy-free puncture, determined by the
roots of \eqref{sopspspa}.

For $L\leq 3$ one can numerically check that
for generic  values of
$a_i$ and $p_i$\ \eqref{iasoaisa},\,\eqref{posposopa}, 
the number    of
solutions   of the algebraic 
system \eqref{hsgspossaopa}-\eqref{osaospa} and \eqref{saopsoposa}
({\em modulo} permutations)
is given by
\bea\label{saoposapo}
{\cal N}_1=3\ ,\ \ \ \ \ \ \ \ {\cal N}_2=9\ ,\ \ \ \ \ \ \ \ \ {\cal N}_3=22\ .
\eea
In many extents
these equations
are similar to the Bethe Ansatz equations.
In particular, using  Eq.\eqref{aopsaps},  they  
can be written in a compact Yang-Yang form
\bea\label{aoossaopaspo}
\frac{\partial Y }{\partial x_a}=0\ \ \ \ \ \ \
\ \ \ \ \  \ (a=1,\ldots L)\ ,
\eea
where
\bea\label{saoposa}
Y=-\sum_{a=1}^L\log {\cal P}(x_a)-F\ .
\eea

Once 
the algebraic  system \eqref{aoossaopaspo}
is solved, the function $T_L(z)$ in
\eqref{opspspsao} can be written in the form
\bea\label{oaposaosa}
T_{L}(z)=T_0(z)+L\,\sum_{i=1}^3\frac{a_i}{(z-z_j)(z-z_k)}-\sqrt{{\cal P}(z)}\
\frac{\partial }{\partial z}\sum_{a=1}^L\frac{2}{(z-x_a)\sqrt{{\cal P}(z)}}\ ,
\eea
where
\bea\label{soysypos}
T_0(z)=
-\sum_{i=1}^3\Big(\,\frac{\delta_i}{(z-z_i)^2}+\frac{c^{(0)}_i}{z-z_i}\,
\Big)\ ,
\eea
and
\bea\label{sopspsosa}
c^{(0)}_i=
\frac{\delta_i+\delta_j-\delta_k}{z_j-z_i}+\frac{\delta_i+\delta_k-\delta_j}{z_k-z_i}\ ,\ \  \ \ \ (i,j,k)={\tt perm}(1,2,3)\ .
\eea
We will refer to the differential operator ${\cal D}(\lambda)$ of
the form  \eqref{opspspsao} with ${\cal P}(z)$ and $T_{L}(z)$ are
given by \eqref{oasioaq} and  \eqref{oaposaosa}, respectively, as
Perturbed Generalized Hypergeometric Oper (PGHO). 
The finite set ${\cal A}^{(L)}_{\bf p}$ \eqref{sisoaisiosaoi} can be regarded as a moduli space of the PGHO's.
It is a finite discrete subset in the  moduli space  of GHO's.
(Notice that 
we   slightly modify the notation used in the introduction  by including the  subscript
${\bf p}=(p_1,p_2,p_3)$.)

\subsection{Wilson loop for  PGHO}

As we have just explained, the position of the punctures 
$x_a$ ($a=1,\dots L$) can be specially chosen so that solutions
of  ODE \eqref{opspspsao} still remain
single-valued in the vicinity of these points.
However, contrary to $T_L(z)$, the term $\lambda^2\, {\cal P}(z)$
is not single-valued on the punctured sphere.
Thus, even with  the special choice of $x_a$,
the monodromy group of the  differential operator ${\cal D}(\lambda)$
turns out  to be essentially different
from that in the case   $\lambda=0$.
Here we begin to explore the monodromy properties of PGHO.

\subsubsection{\label{defwil}Definition of the Wilson loop}

Let us consider the contour $\gamma_P$
depicted in   Fig.\,\ref{fig1av}.
It is usually called the Pochhammer contour (loop).
As an element of the fundamental group $\pi_1(\mathbb{CP}^1\backslash\{z_1,z_2,z_3\})$,
it
can be expressed in terms of the elementary
loops $\gamma_i$, $\gamma_j$ and $\gamma_k$ 
which wind around the punctures  $z_i$,  $z_j$ and $z_k$, respectively:
\bea\label{oaposopasa}
\gamma_P=\gamma_k\circ\gamma_i\circ\gamma_j \ \ \ \ \ \ \ \ \ \
\ \big(\,
\gamma_i\circ\gamma_k\circ\gamma_j=1, \ \ \ (i,j,k)={\tt circle\  perm}(1,2,3)\,\big)\ .
\eea 
Since the Pochhammer loop   winds 
around each of the three 
punctures and the relation \eqref{oasioaq} is imposed, 
the value of the function ${\cal P}(z)$
does not change upon the analytic  continuation along the contour $\gamma_P$.
Therefore the coefficients of PGHO
return to their original
values 
and it makes sense to introduce the quantity
\bea\label{osposoappsa}
{\cal W}(\lambda)={\rm Tr}\big[{\boldsymbol M}(\gamma_P|\lambda)\big]\ ,
\eea
where ${\boldsymbol M}(\gamma_P|\lambda)$ is the monodromy matrix
for ${\cal D}(\lambda)$   corresponding to the 
Pochhammer loop. A significant advantage of 
${\cal W}(\lambda)$ is 
that it does not depend on the precise
shape of the integration contour. In particular, it
is not sensitive to deformations of $\gamma_P$
which sweeps through
the monodromy-free punctures.
In what follows we will refer to \eqref{osposoappsa} as the {\em
  Wilson loop} (corresponding to PGHO ${\cal D}{(\lambda)}$).

The second order differential operator ${\cal D}{(\lambda)}$
depends analytically   on $\lambda^2$ and hence
${\cal W}(\lambda)$ is an entire function of $\lambda^2$, i.e.,
the series expansion
\bea\label{opososapo}
{\cal W}(\lambda)={\cal W}_0+\sum_{n=1}^\infty {\cal W}_n\ \lambda^{2n}
\eea
converges for any complex $\lambda$.
Its value at  $\lambda=0$  can
be  found using 
Eqs.\eqref{raospposa},\,\eqref{skisosai} from Section\,\ref{sec13}:
\bea\label{opasospaos}
&&{\cal W}_0={\rm Tr}\big[\,({\boldsymbol M}^{(i)})^{-1}
({\boldsymbol M}^{(j)})^{-1}\,
{\boldsymbol M}^{(i)}\,{\boldsymbol M}^{(j)}\,\big]=2\, \big(2+c(4p_1)+c(4p_2)+c(4p_3)+\\
&& c(2p_1+2p_2+2p_3)
+
c(2p_1+2p_2-2p_3)+c(2p_1-2p_2+2p_3)+c(-2p_1+2p_2+2p_3)\,\big)\, .
\nonumber
\eea
Note, that this expression does not depend on the number of the
monodromy-free punctures $L$. 
Higher expansion coefficients in the series \eqref{opososapo}
can, in principle, be calculated using the standard perturbation
theory. 

\subsubsection{\label{usaystsa}Large-$\lambda$ asymptotic expansion}

\begin{figure}
\centering
\includegraphics[width=6.  cm]{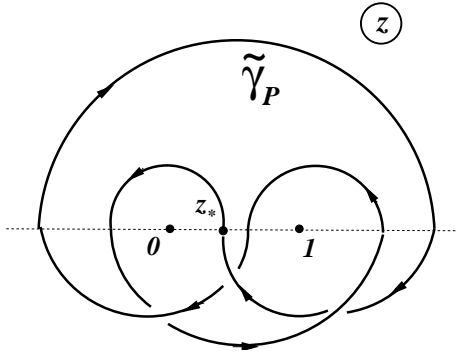}
\caption{The Pochhammer  loop on the complex plane with the punctures at
$z_1=0,\ z_2=1$ and $z_3=\infty$.}
\label{fig1avr}
\end{figure}

The leading large-$\lambda$ asymptotic of the Wilson loop  can be obtained
within  the
WKB approach. It is easy to see that
\bea\label{ospsopsasap}
{\cal W}(\lambda)\asymp 2\, \cosh\Big( \lambda\, \oint_{\gamma_P}\rd z\ \sqrt{{\cal P}(z)}
+o(1)\,\Big)\ 
\ \ \ \ \ {\rm as}\ \ \ \ |\lambda|\to\infty\ .
\eea
Here the r.h.s.  is written  as a sum of two WKB exponents.
Of course, for different values of  $\arg(\lambda^2)$
only one   term dominates whereas another exponent  should be neglected.
The quantity $\sqrt{{\cal P}(z)}$ is a multivalued 
function on $\mathbb{CP}^1\backslash\{z_1,z_2,z_3\}$ whose  phase has
not been yet uniquely specified.
To resolve this phase ambiguity,  we consider the
M\"obius transformation which sends $(z_1,z_2,z_2)$ to $(0,1,\infty)$.
With this change of the  integration variable,  the integral in \eqref{ospsopsasap}
transform to the form
\bea\label{sospsaopsa}
\oint_{\gamma_P}\rd z\ \sqrt{{\cal P}(z)}
=\re^{-\frac{\ri\pi}{2} (a_1+a_2)}\
\int_{{\tilde \gamma}_P}\rd z\ z^{\frac{a_1}{2}-1}(1-z)^{\frac{a_2}{2}-1}\ .
\eea
The Pochhammer contour now looks as in Fig.\,\ref{fig1avr}.
Let us choose the base point $z_*\in {{\tilde \gamma}_P}$
within the real segment $[0,1]$ and assume that
$z_*^{\frac{a_1}{2}-1}(1-z_*)^{\frac{a_2}{2}-1}>0$.
Then the phase of the integrand in \eqref{sospsaopsa} is determined
unambiguously through the analytic continuation along the integration contour.
This convention removes the phase ambiguity  of ${\cal P}(z)$  for
$z\in \gamma_P$. 
The integral 
which appears in the r.h.s. of
\eqref{sospsaopsa} is well known in the theory of the hypergeometric
equation: 
\bea\label{opopsapospa}
\oint_{{\tilde \gamma}_P}\rd z\ z^{\alpha-1}(1-z)^{\beta-1}=
\big(1-\re^{2\pi\ri\alpha}\big)\, \big(1-\re^{2\pi\ri\beta}\big)\ 
\frac{\Gamma(\alpha)\Gamma(\beta)}{\Gamma(\alpha+\beta)}\ .
\eea
We can now rewrite Eq.\eqref{ospsopsasap} in the form
\bea\label{usyssoaoopsa}
\log 
{\cal W}(\lambda)\asymp  -q_0\ \lambda+o(1)
\ \ \ \ \ {\rm as}\ \ \ \ |\arg(\lambda)|<\frac{\pi}{2}\ ,\ \ \ 
|\lambda|\to\infty\ ,
\eea
with
\bea\label{soapapos}
q_0=-\frac{4\pi^2}{\prod_{i=1}^3\Gamma(1-\frac{a_i}{2})}\ .
\eea

It is not difficult to extend the above leading  asymptotics to
a complete asymptotic expansion for large values of $\lambda$.
For this purposes, we  perform the change of variables 
in  ODE \eqref{opspspsao}
\bea\label{siisaoi}
w=\re^{\frac{\ri\pi}{2}\, (a_1+a_2)}\ 
\int \rd z\ \sqrt{{\cal P}(z)}\ ,\ \ \ \  \ \ \ \ \ \ \ 
\psi(z)\ ({\rd z})^{-\frac{1}{2}}={\hat \psi}(w)\ ({\rd w})^{-\frac{1}{2}}\ .
\eea
This transformation
brings Eq.\eqref{opspspsao} to the form of an 
ordinary Schr${\ddot {\rm o}}$dinger 
equation
\bea\label{kaajkasj}
\big(-\partial_w^2+{\hat T}_{L}(w)+\lambda^2\, \big)\ {\hat \psi}=0\ ,
\eea
with the potential 
\bea\label{sopspo}
{\hat T}_{L}={\cal P}^{-1}\ \bigg(\,T_{L}+
\frac{4\,{\cal P}\partial^2_z{\cal P}-5\, (\partial_z {\cal P})^2}{16\, 
{\cal P}^2}\, \bigg)\ .
\eea
It is well known how to develop  the large-$\lambda$ 
asymptotic expansion of
monodromy coefficients of Eq.\eqref{kaajkasj}.
The procedure leads to the following asymptotic series
\bea\label{aopsaosapoas}
\log{\cal W}(\lambda)\asymp  -q_0\ \lambda+\sum_{n=1}^\infty c_n\ q^{(L)}_{2n-1}\ \lambda^{1-2n}+
O(\lambda^{-\infty}) 
\ \ \ \ \ {\rm as}\ \ \ \ |\arg(\lambda)|<\frac{\pi}{2}\ ,\ \ \
|\lambda|\to\infty\ ,
\eea
where 
\bea\label{aoopsaap}
c_n=\frac{(-1)^n}{2n!}\ \frac{\Gamma(n-\frac{1}{2})}{\sqrt{\pi}}
\eea
and
\bea\label{opsoposaaps}
q^{(L)}_{2n-1}=\re^{\ri\pi (n-\frac{1}{2})\, (a_1+a_2)}\ 
\oint_{{\hat \gamma}_P}\rd w\ U_{n}\big[{\hat  T}_{L}\,\big]\ .
\eea
In the last formula  $U_{n}[\,  {\hat u}\, ]$ are
homogeneous $({\rm grade}({\hat u})=2, \ {\rm grade}(\partial )=1,\
{\rm grade}(U_n)=2n )$
differential polynomials  in ${\hat u}$ of the degree
$n$  (known as the Gel'fand-Dikii polynomials \cite{Gelfand}),
\bea\label{iusksksak}
U_{n}[\,{\hat u}\,]=
\frac{(-1)^n}{(2n-1)\, c_n}\ {\hat \Lambda}^n\cdot 1\ .
\eea
Here
\bea\label{reqssksa}
{\hat \Lambda}=-{\textstyle \frac{1}{ 4}}\
\partial^2+ {\hat u}-{\textstyle \frac{1}{ 2}}\  {\partial}^{-1}\  {\hat u}'  \, ,
\eea
and prime stands for the derivative. Thus,
\bea\label{rwesksksa}
U_0[\, {\hat u}\,]&=& 1\ ,\nonumber\\[.2cm]
U_1[\, {\hat u} \,]&=&  {\hat u}\ ,\\[.2cm]
U_{2}[\,{\hat u} \,]&=& 
{\hat u}^2-{\textstyle \frac{1}{ 3}}\, {\hat u}''\ , \nonumber\\[.2cm]
U_{3}[\, {\hat u}\,]&=& 
{\hat u}^3-{\textstyle \frac{1}{2}}\, ( {\hat u} ')^2-
{\hat u}\, ''+
{\textstyle \frac{1}{ 5}}\ {\hat u}''''\ , \nonumber\\[.2cm]
U_{n}[\,{\hat u}\,]&=&
{\hat u}^n+\ldots\ ,\nonumber
\eea
where the last line shows the overall normalization of the polynomials.

There is no need to do describe the contour ${\hat \gamma}_P$ in
\eqref{opsoposaaps} explictly, since 
we now change the integration variable 
in Eq.\eqref{opsoposaaps} back to the original coordinate $z$.
In this way one obtains
\bea\label{ysstopsaaspaps}
q^{(L)}_1&=&\oint_{{ \gamma}_P}\frac{\rd z}{{\cal P}^{\frac{1}{2}}}\ 
\ \bigg(\,T_{L}+
\frac{4\,{\cal P}\partial^2_z{\cal P}-5\, (\partial_z {\cal P})^2}{16\,
{\cal P}^2}\, \bigg)\nonumber\\
q^{(L)}_3&=&\oint_{ \gamma_P}\frac{\rd z}{{\cal P}^{\frac{3}{2}}}\ 
\ \bigg(\,T_{L}+
\frac{4\,{\cal P}\partial^2_z{\cal P}-5\, (\partial_z {\cal P})^2}{16\,
{\cal P}^2}\, \bigg)^2\  .
\eea
(Notice that in the  derivation 
 of  the second formula we  dropped  the  term  $-\frac{1}{3}\ \partial_w^2 
{\hat T}_{L}$ in  Eq.\eqref{opsoposaaps} with $n=2$,   which
do not contribute  to the integral.)
Of course, it is straightforward to perform the change of variables
in Eq.\eqref{opsoposaaps} for any given $n$. We do not present
explicit formulae for $n>2$, but note that
\bea\label{sosaposa}
q^{(L)}_{2 n-1}=\oint_{ \gamma_P}\frac{\rd z}{{\cal P}^{n-\frac{1}{2}}}\
\ \big(\,(\,T_{L}\,)^n+\ldots\,\big)\ ,
\eea
where the omitted terms contain derivatives and the
lower powers of $T_{L}$.

\subsubsection{\label{expan}Expansion coefficients $q_{2n-1}^{(0)}$}

Using the formulae \eqref{ysstopsaaspaps}, \eqref{sosaposa}, one can
perform some explicit calculations of the coefficients in the asymptotic series
\eqref{aopsaosapoas}.
Let us first consider  of  the perturbed
hypergeometric oper, i.e., PGHO without monodromy-free punctures.

Using Eqs.\eqref{oaposaosa},\,\eqref{ysstopsaaspaps}
 (specialized for  the case $L=0$) 
and the integral \eqref{opopsapospa}, one can show that
\bea\label{sopsopsaspaa}
q_1^{(0)}=\frac{8\pi^2}{\prod_{i=1}^3\Gamma(\frac{a_i}{2})}\ 
\bigg(\,\sum_{i=1}^3\frac{P_i^2}{4}-\frac{1}{8}\, \bigg)\ ,
\eea
and
\bea\label{sjusyu}
q^{(0)}_{3}&=&-\frac{2\pi^2}{3\,\prod_{i=1}^3\Gamma(\frac{3a_i}{2})}\ \bigg[\,
 \sum_{i=1}^3E_i\, \Big(\frac{P_i^4}{16}-\frac{P^2_i}{ 16}+\frac{1}{ 192}\Big)\\
&+&
\sum_{
i\not= j}E_{ij}\, \Big(\frac{P^2_i}{4}-\frac{1}{ 24}\Big)
\Big(\frac{P^2_j}{4}-\frac{1}{ 24}\Big)+
\frac{1}{ 240}\
\sum_{i=1}^3H_{i}\ \,\bigg]\ ,
\nonumber
\eea
where
\bea\label{osososap}
P_i=\frac{2p_i}{\sqrt{a_i}}\ ,
\eea
and
where the numerical coefficients $E_i,\ E_{ij}$ and $H_i$ are given by
\bea\label{kasuaiu}
&&E_{i}=a_i\, (3a_j-2)\,(3a_k-2)
\nonumber\\
&&E_{ij}=3\, a_i\, a_j\, (3a_k-2)
\\
&&H_i=8-a_i^2-9\ (a_1a_2+a_2a_3+
a_3a_1)+
15\ a_1a_2a_3
\ .\nonumber
\eea
The indices $(i,\,j,\,k)$ represents any permutation of the numbers
$(1,2,3)$.
For $n>2$ the calculation of $q_{2n-1}^{(0)}$ is straightforward, but rather long.
It is much easy to   establish the following general structure:
\bea\label{aopspaaspo}
q^{(0)}_{2n-1}=R_{n}(P_1^2,P_2^2,P_3^2)
\ , 
\eea
where  
$R_{n}$ stands   for  $n$-the degree polynomials in  
the variables  $P_i^2$  
\bea\label{spospospa}
R_{n}(P_1^2,P_2^2,P_3^2)=
\sum_{i+j+k=n} R^{(n)}_{ijk}\ \ P^{2i}_1\,P^{2j}_2\,P^{2k}_3
+\ldots\ 
\eea
(the  dots   represent the sum of monomials of degrees lower than $n$).
One can show that
\bea\label{usisusspospospa}
&&R^{(n)}_{ijk}=
\frac{(-1)^{n-1}\ 2^{5-2n}  \pi^2}{\prod_{i=1}^3\Gamma((n-\frac{1}{2})\,a_i)}\ \ 
\frac{ n!\, \big(a_1(\frac{1}{2}- n)\big)_{n-i} 
\big(a_2\,(\frac{1}{2}- n)\big)_{n-j}\,\big(a_3\,(\frac{1}{2}- n)\big)_{n-k}}
{ i!\,j!\,k!\ \, (2n-1)^3\  a^{1-i}_1a^{1-j}_2a^{1-k}_3}\ .
\eea

\subsubsection{\label{subsbubsub}Expansion coefficients $q_{1}^{(L)}$ and $q_{3}^{(L)}$ for $L\geq 1$}

It is not difficult to calculate $q_{1}^{(L)}$ for arbitrary $L$. 
Indeed, the third term in \eqref{oaposaosa} do not contribute to the 
integral  \eqref{ysstopsaaspaps} for $q_{1}^{(L)}$.
The contribution of the first 
term in \eqref{oaposaosa} is given by \eqref{sopsopsaspaa}.
The second term in \eqref{oaposaosa} 
gives a contribution proportional to $L$.  The final result reads as
\bea\label{asaoisaoisaoi}
q_{1}^{(L)}=q_{1}^{(0)}+\frac{8\pi^2}{\prod_{i=1}^3\Gamma(\frac{a_i}{2})}\ \ L\ .
\eea

The calculation of $q_{3}^{(L)}$ is very cumbersome
and we do not describe it
here. 
Bellow we quote the result
which is  expressed in terms of
the parameters $\delta_i,\ a_i\ (i=1,2,3)$ and  the coordinates
$x_a \ (a=1,\ldots L)$ 
of the monodromy-punctures. Also, it is assumed that  $(z_1, z_2,z_3)=(0,1,\infty)$;
\bea\label{usyisussaospsao}
&&q^{(L)}_3=q_3^{(0)}-\frac{2\pi^2}{3\,\prod_{i=1}^3\Gamma(\frac{3a_i}{2})}\ \bigg(\, 
\sum_{j>k}Q^{(L)}_{jk}+
 Q_0^{(L)}+Q_1^{(L)}\
\sum_{j=1}^L x_j
+Q_2^{(L)}  
\ 
  \sum_{j=1}^L x_j^2\, \bigg)\  .
\eea
Here
\bea\label{osisasaisoaiso}
Q^{(L)}_{jk}=\frac{3}{2}\, (2-3\,a_3)
\bigg( (3\,a_3-4)\, \frac{(x_j+x_k)^2\, (2-x_j-x_k)^2}{(x_j-x_k)^2}
+
(4\,a_3^2-3\,a_3-4) (x_j-x_k)^2\,\bigg) 
\eea
and
\bea\label{osaospapopoaspa}
Q_0^{(L)}&=& \bigg(\, \frac{1}{4}\ 
\big(\,36\, a_1^2\, a_3^2- 57\, a_1^2\, a_3 + 15\, a_1\, a_3^2+ 6\, a_1^2-126\, a_3^2 - 48\, 
a_1\, a_3
\nonumber\\
&+& 36\, a_1
 +  252\, a_3-112\,\big)+6\, (2 - 3\, a_3) (4 - a_1 - 3\, a_3)\, \delta_1+
6\, a_1\, (2 - 3\, a_3)\, \delta_2\nonumber\\
& +&
6\, a_1\, (2 - 3\, a_2)\, \delta_3\, \bigg)\ L + 3\, (2 - 3\, a_3) (4 - 4\, a_1 + a_1^2 - 3\, 
a_3 + 3\, a_1
\, a_3)\  L^2
\eea
and
\bea\label{oiosipaapospasop}
Q_1^{(L)}&=&6\ 
\big(\, 3\, a_1\, a_3^3  - 12\, a_1\, a_3^2- 2\, a_3^3+ 14\, a_1\, a_3 + 15\, a_3
^2 - 4\, a_1 
- 22\, a_3+8\,\big)
\nonumber\\
&+&12\, (1 - a_3) (2 - 3\, a_3) (\delta_2-\delta_1)+12\, (2 - 2\, a_1 + a_3 - 3\, a_3^2\,)\, 
\delta_3
\nonumber\\
&+&6\, (3\, a_3-2) (4 - 2\, a_1 - 3\, a_3 + 2\, a_1\, a_3\,)\  L\ ,\\
Q_2^{(L)}&=& 3 \ (3\,a_3-2)\,  \big(\,  
2 \, ( a_3^2-2)\, (L-1)+ 4\, a_3\, \delta_3+ (a_3-2)\,  a_3^2\,
\big)\ .\nonumber
\eea

\section{\label{jassysa}Hidden algebraic  structures behind   PGHO}
 
We have already mentioned a remarkable property of the
algebraic system \eqref{hsgspossaopa}-\eqref{osaospa},\,\eqref{aoossaopaspo}. Our numerical work shows that for given $L$
the  number  ${\cal N}_L$ of its solutions (i.e., the cardinality of the set 
${\cal A}^{(L)}_{\bf p}$ \eqref{sisoaisiosaoi})
does not depend on parameters, at least for
generic values of   $a_i$ and $p_i$ \eqref{iasoaisa},\,\eqref{posposopa}. For $L\leq 3$, the integers ${\cal N}_L$
are quoted in \eqref{saoposapo}. 
In this pattern
one can recognize 
the first values for the number of partitions of the integer $L$
into integer parts of three kinds, which we denote as ${\mathsf
  p}_3(L)$. This sequence is generated by the series 
\bea\label{aoposapos}
\sum_{L=0}^{\infty}
\, {\mathsf p}_3(L)\, q^L=\displaystyle
\prod_{k=1}^\infty\frac{1}{(1-q^k)^3}=1 + 3\, q + 9\, q^2 + 22\, q^3 + 
51\, q^4 + 108\, q^5 +\ldots\ .
\eea
We now interrupt our  formal study  to discuss
remarkable algebraic structures behind PGHO.

Introduce the 
three-component chiral Bose field $ {\boldsymbol \phi}=(\phi_{1}, \phi_{2}, \phi_{3})$, 
i.e. the operator valued
function
\bea\label{sakshs}
\phi_i(u)=  \frac{1}{2}\ ( \mathbb{Q}_i+ \mathbb{P}_i\, u)-\ri\ 
\sum_{n\not=0} \frac{a_i(-n)}{n}\
\re^{\ri  n u}\, \ \ \ \ \ \ \ \ \ \ \ (i=1,2,3)\ ,
\eea
where $\mathbb{Q}_i$, $\mathbb{P}_i$ and $a_i(n)$ 
are operators  satisfying the commutation relations of the
Heisenberg algebra
\bea\label{alksu}
[\, \mathbb{Q}_i,\mathbb{P}_j\,]=2\ri\ \delta_{ij}\ ,\ \ \ \ \ 
[\, a_i(n)\, , a_j(m)\, ]={\frac{n}{ 2}}\ \delta_{ij}\ \delta_{n+m,0}\ .
\eea

Let $P_{s+1}(\partial {\boldsymbol \phi},\
\partial^2 {\boldsymbol \phi},\ldots)$ be a local
field of spin $s+1$, which is 
a local polynomial of
$\partial {\boldsymbol \phi}$
and its higher derivatives ( $\partial$ stands for $\frac{\partial}{\partial u}$ here).
All  such fields are periodic in $u$, therefore
one can introduce the integral,
\bea\label{ospssapo}
\mathbb{I}[P_{s+1}]=\int_0^{2\pi}\frac{\rd u}{2\pi}\ P_{s+1}(\partial {\boldsymbol \phi},\
\partial^2 {\boldsymbol \phi},\ldots)\ .
\eea
Bellow 
the shortcut notation $\mathbb{I}_{s}$ for
$\mathbb{I}[P_{s+1}]$ is used.
Suppose we are given a special infinite  sequence of operators 
$\mathbb{I}_s$
(corresponding to special infinite  sequence of the polynomials $P_{s+1}$)
which are mutually commutative operators,
\bea\label{sasaisaisao}
[\,\mathbb{I}_s,\, \mathbb{I}_{s'}]=0\ .
\eea
We will refer to the operators  $\{\mathbb{I}_{s}\}$ as the  (chiral) local Integral
of Motions (IM).

A complete algebraic classification of all possible 
infinite sets of  local IM seems to be a hopeless task.
However some  non trivial examples   are  available. Among them there is
a two-parameter family  discovered by Fateev in
\cite{Fateev:1996ea}. The first two representatives from this set are 
given by
\bea\label{isusoposap}
{\mathbb I}_1 = \int_0^{2\pi} \frac{\rd u}{  2\pi}\,
\sum_{i=1}^3 (\partial { \phi_i })^2 \,,
\eea
and
\bea\label{susyoposap}
\mathbb{I}_3
&=&\frac{1}{3}\ \int_0^{2\pi} \frac{\rd u}{  2\pi}\, 
\bigg[\, \sum_{j=1}^3E_j\ \big(\partial \phi_j\big)^4+
\sum_{
m\not=j}E_{mj}\ \big(\partial \phi_m\big)^2\big(\partial \phi_j\big)^2
\nonumber \\ &+&
 \sum_{j\not=k\not=m} K_j\
\partial^2 \phi_j\, \partial \phi_k\, \partial \phi_m+
\sum_{j=1}^3H_{j}\ \big(\partial^2 \phi_j\big)^2\, \bigg]\ .
\eea
Numerical coefficients in the last formula depends on 
three parameters $\alpha_1,\ \alpha_2$ and $\alpha_3$ obeying the quadratic
constraint  $\alpha^2_1+\alpha^2_2+\alpha^2_3=\frac{1}{2}$.
It turns out that  $E_j$, $E_{mj}$ and $H_{j}$ are
given by Eqs.\eqref{kasuaiu}, provided the parameters 
are identified  as
\bea\label{paaossasai}
\alpha_i={\textstyle\frac{1}{2}}\ \sqrt{a_i}\ ,
\eea
whereas
\bea\label{oposaoosp}
K_j=32\ri\ \alpha_1\, \alpha_2\, \alpha_3\ \big(1-6 \alpha^2_j\big)\ .
\eea
An explicit form for  the  higher spin  representatives  is not available.
However it is known
that \cite{Lukyanov:2012wq}
\bea\label{aaosapospa}
\mathbb{I}_{2n-1}=2^{2n}\ 
\int_0^{2\pi} \frac{\rd u}{  2\pi}\,
\bigg[\, \sum_{i+j+k=n}^3 C^{(n)}_{ijk}\ 
\big(\partial \phi_1\big)^{2i}\big(\partial \phi_2\big)^{2j}
\big(\partial \phi_3\big)^{2k}+\ldots\, \bigg]\ ,
\eea 
where the dots stand for the terms, involving higher derivatives of
$\phi_i$ and  the constant $C^{(n)}_{ijk}$ is given by \eqref{saosopsaosa}.
There are good reasons to expect (see Ref.\cite{Lukyanov:2012wq} and Section\,\ref{secnn} bellow) that
$\mathbb{I}_1$ and $\mathbb{I}_3$ are just the first two 
representatives of an 
infinite two-parameter family of mutually commuting IM,
$\{\mathbb{I}_{2n-1}\}_{n=1}^\infty$.

Let ${\cal F}_{\bf P}$\  with ${\bf P}=(P_1,P_2,P_3)$ be the Fock space,
i.e., the space generated by the action of $a_i(n)$ with $n<0$ on the vacuum
state $|\,  {\bf P}\, \rangle$ which satisfies the equations
\bea\label{shahksa}
{\mathbb P}_i\ |\,  {\bf P}\, \rangle=P_i\ |\,  {\bf P}\, \rangle\ ,\ \ \ \ \ \ \ \ 
 a_i(n)\ |\,  {\bf P}\, \rangle=0\, ,\ \ \ \ \ n=1, 2, 3\ldots
\ .
\eea
The space ${\cal F}_{\bf P}$ naturally splits into the
sum of finite dimensional ``level subspaces''
\bea\label{lksasl}
{\cal F}_{\bf P}=\oplus_{L=0}^{\infty}\,  {\cal F}^{(L)}_{\bf P}\ ;\ \ \ \ \ \ \ \ \ 
{\mathbb L}\, {\cal F}^{(L)}_{\bf P}=L\ {\cal F}^{(L)}_{\bf P}\ ,
\eea
where
\bea\label{ospospasopsa}
{\mathbb L}=2\ \sum_{i=1}^3\sum_{n=1}^\infty a_i(-n)\,a_i(n)\ .
\eea
The dimensions of the level subspaces do not depends on  ${\bf
  P}$. Obviously, it coincides with the number of integer partitions
of $L$ \ into parts of three kinds, defined in \eqref{aoposapos}, 

\bea\label{oapopossap}
\dim\big[\,{\cal F}^{(L)}_{\bf P}\,\big]={\mathsf p}_3(L)\,.
\eea

The grading operator ${\mathbb L}$ essentially
coincides with ${\mathbb I}_1$ \eqref{isusoposap}:
\bea\label{soosapaspas}
{\mathbb I}_1={\mathbb L}+\sum_{i=1}^3\Big(\frac{P^2_i}{4}-\frac{1}{24}\,\Big)\  .
\eea
Therefore all local IM from the Fateev family act invariantly in the
level subspaces ${\cal F}^{(L)}_{\bf P}$. The diagonalization
of ${\mathbb I}_{2n-1}$ in a given level subspaces reduces
to a finite-dimensional matrix problem
which however rapidly becomes very complex for
higher levels.

Of course, the highest weight vector of 
the Fock space (the ``vacuum'' vector) 
is an  eigenvector for all integrals of motion
${\mathbb I}_{2n-1}$. Let $I_{2n-1}^{(0)}$ be
the corresponding eigenvalues. The results from Section\,\ref{expan} 
and Eqs.\eqref{isusoposap},\,\eqref{susyoposap}
imply  that for $n=1$ and $n=2$
the following relation  holds
\bea\label{sospsaps}
q^{(0)}_{2n-1}=
\frac{(-1)^{n-1}\ 2^{5-2n}\pi^2}{\prod_{i=1}^3\Gamma\big(2\,(2n-1)\,\alpha^2_i\big)}\ \  I^{(0)}_{2n-1}\ ,
\eea
where the parameters $a_i$ and $p_i$ of $q^{(0)}_{2n-1}$ 
are related to $\alpha_i$ and  the zero 
mode momentum $P_i$ as in Eqs.\eqref{paaossasai} and \eqref{osososap}, respectively. 
Moreover, for any value of $n$ 
both sides of \eqref{sospsaps}
are polynomials in the variables $(P_1)^2,\, (P_2)^2,\,(P_3)^2$ of the 
degree $n$. Comparing  
Eqs.\eqref{aopspaaspo},\,\eqref{spospospa} with
\eqref{aaosapospa},\,\eqref{saosopsaosa}, it is easy to check that  
all leading $n$-th degree monomials are exactly the same in the  both sides.
Thus one can reasonably expect that \eqref{sospsaps}, involving the
vacuum eigenvalues of the integral of motion and the expansion
coefficients of the Wilson loop for the PGHO with $L=0$ 
holds exactly for any value of $n\geq 1$.

Actually, we expect that \eqref{sospsaps} can be extended to the
relation between the whole spectrum of ${\mathbb I}_{2n-1}$ in any
level subspace ${\cal F}^{(L)}_{\bf P}$  and
admissible values of $q^{(L)}_{2n-1}$ associated with the different
PGHO's with $L$ monodromy-free punctures.  
Indeed, 
for $a_i$ and $p_i$  restricted as in
\eqref{iasoaisa},\,\eqref{posposopa}, 
the number of solutions of
the algebraic system \eqref{aoossaopaspo}, ${\cal N}_L$, is expected to 
coincide with $\dim\big[\,{\cal F}^{(L)}_{\bf P}\,\big]$. 
As before, let ${\cal A}_{\bf p}=\big\{
\big(\,x_1^{(\alpha)},\ldots x_L^{(\alpha)}\,\big)\big\}_{\alpha=1}^{{\cal N}_L}$
be the whole  set of such solutions. 
With a chosen  representative  $\big(\,x_1^{(\alpha)},\ldots x_L^{(\alpha)}\,\big)\in {\cal A}_{\bf p}$,
one can associate
an  infinite sequence of the expansion coefficients $q_{2n-1}^{(L,\alpha)}$.
In the case $n=1$ and $n=2$  explicit formulae 
are presented  in Section\,\ref{subsbubsub}. 
From the other side,
let $\big\{\,I_{2n-1}^{(L,\beta)}\,\}_{\beta=1}^{{\cal N}_L}$ be a sets of eigenvalues of the 
${\cal N}_L\times {\cal N}_L$-matrix  of ${\mathbb I}_{2n-1}$ acting in the level subspace ${{\cal F}^{(L)}_{\bf P}}$.
We expect that, up to the overall normalization factor, 
the set $\big\{\,q_{2n-1}^{(L,\alpha)}\,\big\}_{\alpha=1}^{{\cal N}_L}$
coincides with
$\big\{\,I_{2m-1}^{(L,\beta)}\,\big\}_{\beta=1}^{{\cal N}_L}$ 
for any  fixed $n$ and $L$. Thus the subscripts $\alpha$ and
$\beta$ can be identified and Eq.\eqref{sospsaps} is generalized as follows:
\bea\label{isussospsaps}
q^{(L,\alpha)}_{2n-1}=
\frac{(-1)^{n-1}\ 2^{5-2n}\pi^2}{\prod_{i=1}^3
\Gamma\big(2\,(2n-1)\,\alpha^2_i\big)}\ \  I^{(L,\alpha)}_{2n-1}\ .
\eea
For $m=1$,  ${\mathbb I}_{1}|_{{{\cal F}^{(L)}_{\bf P}}}\propto {\boldsymbol  1}_{{\cal N}_L\times {\cal N}_L}$
and \eqref{isussospsaps} follows from \eqref{asaoisaoisaoi}.
Unfortunately we do not know how to prove  this  
remarkable relation for $n>1$.
However,  an  explicit form of  ${\cal N}_L\times {\cal N}_L$-matrices
${\mathbb I}_{3}|_{{{\cal F}^{(L)}_{\bf P}}}$ is available
and the conjectured relation  has been  tested numerically  for $L\leq 3$ and  a wide range of parameters 
$a_i$ and $p_i$ from  the domain \eqref{iasoaisa},\,\eqref{posposopa}.
The  numerical work also suggests that, for generic  values of the parameters, the eigenvalues of
the matrices ${\mathbb I}_{3}|_{{{\cal F}^{(L)}_{\bf P}}}$
are not degenerate.
With  this observation, one may expect that the joint  eigenvectors
of the commuting family of  IM,
\bea\label{sopssaop}
|\,L, \alpha\,\rangle\in {\cal F}^{(L)}_{\bf P}\ :
\ \ \  \ \  {\mathbb I}_{2n-1}\ |\,L, \alpha\,\rangle= { I}^{(L,\alpha)}_{2n-1}\ |\,L, \alpha\,\rangle\ ,
\eea
form  a non-degenerate   basis in each level subspace  ${\cal F}^{(L)}_{\bf P}$.
Therefore there exists a  bijection between the moduli space ${\cal A}_{\bf p}^{(L)}$
of  PGHO's with $L$ monodromy-free punctures  and the level-$L$ joint  eigenbasis
$\big\{|\,L, \alpha\,\rangle\}_{\alpha=1}^{{\cal N}_L}$.

\section{\label{seccon}Connection matrices  for  PGHO}

In the previous sections we have discovered  interesting properties of the Wilson loop
(defined in
\eqref{osposoappsa}) 
by studying its asymptotic expansions at
large values of $\lambda$, using the WKB approximation. Even though this
asymptotic analysis has led to remarkable insights into the algebraic structure
of the problem, considered in Section\,\ref{jassysa}, it does not solve the  
mathematical problem of an exact calculation of the Wilson loop
as entire functions of the
variable $\lambda^2$. 
In this section we address this problem. 
Actually, here we solve a more general problem of an exact calculation 
of all connection matrices for the PGHO. By doing this we employ and
extend ideas and methods 
previously developed in \cite{Bazhanov:1994ft,Bazhanov:1996dr,Bazhanov:1998dq,Bazhanov:1998wj}.
The matrix elements of the connection 
matrices are entire
functions of $\lambda^2$. 
Additional information about their analytic properties, 
namely, asymptotic distributions of their zeroes,  
is deduced from the standard WKB analysis. 
We use various symmetries of the differential operator \eqref{opspspsao} 
and derive a system of functional relations, which allows one to completely
determine all the connection matrices. 
Interestingly, these functional relations have only a
discrete (albeit infinite) set of solutions, 
which possess the required analytic
properties. We conjecture that these solutions precisely 
correspond to PGHO's with an arbitrary number of monodromy-free punctures.
The results are supported by several  analytical and numerical
checks for PGHO with  $L=0$.

\subsection{\label{sewd}Functional relations for the connection matrices}
\begin{figure}
\psfrag{a}{ $ z_i$}
\psfrag{c}{ $ z_{j}$}
\psfrag{e}{ $ z_{k}$}
\psfrag{b}{ $ {\boldsymbol S}^{(j,i)}(\lambda)$}
\psfrag{d}{ $ {\boldsymbol S}^{(i+2,i+1)}(\lambda)$}
\psfrag{f}{ $ {\boldsymbol S}^{(i,i+2)}(\lambda)$}
\centering
\includegraphics[width=6.  cm]{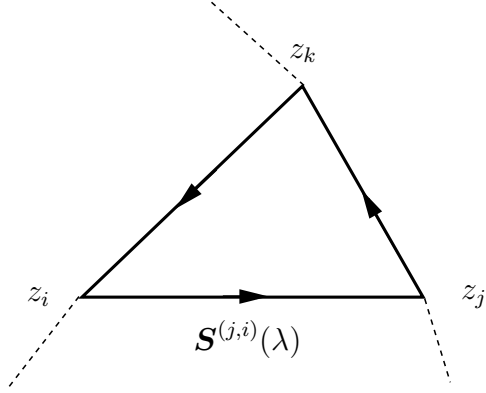}
\caption{The Riemann sphere with cuts. The dashed lines represent    cuts which extended from the branching 
points of ${\cal P}(z)$ to $z=\infty$. The connection matrixes are associated with oriented links.}
\label{fig5avr}
\end{figure}

A proper definition of the bases of
solutions \eqref{opaosap} for $\lambda\not=0$
requires some additional considerations. 
First of all, one needs to take into account that 
(unlike the $\lambda=0$ case) analytic continuations along infinitesimal
loops around the singular points $z_1$, $z_2$ and $z_3$ 
affects the PGHO itself. 
Therefore, in order to define solutions 
by asymptotic conditions at these points one needs to   
make suitable brunch cuts. 
Let us chose an extra  point,   say $z=\infty$, 
and cut the Riemann sphere along the lines, connecting  
this point with the branching points of ${\cal P}(z)$.
In  Fig.\,\ref{fig5avr} these cuts are shown by the dashed lines. 
Next, the asymptotic conditions \eqref{opaosap}
must be slightly modified
\bea\label{chi-sol2}
\chi_\sigma^{(i)}\to\frac{1}{\sqrt{2p_i}}\  (z-z_i)^{\frac{1}{2}+\sigma
  p_i}\,\Big(1+O\big((z-z_i)^{\frac{a_i}{2}}\big)\Big)
\qquad {\rm as}\ \ \ \ \ \ \ \ \ z\to z_i\ ,
\eea
since the order of the correction term is changed with respect to that in the 
$\lambda=0$ case. The above conditions uniquely define
the solutions $\chi^{(i)}_\sigma(z)$ provided that the parameters
$p_i$ satisfy an additional constraints $0<{p}_i<\frac{a_i}{4}$, which
were already enforced in Eq.\eqref{posposopa} above.

The connection matrices ${\boldsymbol S}^{(j,i)}(\lambda)$ for
$\lambda\not=0$ can  be  defined in the same way \eqref{sksssopsosp}
as in the case of unperturbed GHO: 
\bea\label{S-def}
{\boldsymbol \chi}^{(i)}={\boldsymbol \chi}^{(j)}
\ {\boldsymbol S}^{(j,i)}(\lambda)\ .
\eea
They satisfy the same relations  
\eqref{saopsapo} as for $\lambda=0$:
\beq\label{Slambda}
\det\big({\boldsymbol S}^{(j,i)}(\lambda)\big)=1\,,\qquad\ \ 
{\boldsymbol S}^{(i,j)}(\lambda)\,{\boldsymbol S}^{(j,i)}(\lambda)={\boldsymbol I}\,,\qquad\ \ 
{\boldsymbol S}^{(i,k)}(\lambda)\,{\boldsymbol S}^{(k,j)}(\lambda)\,{\boldsymbol
  S}^{(j,i)}(\lambda)={\boldsymbol I}\  .
\eeq
Throughout this section we assume that $(i,j,k)$ is a cyclic
permutation of $(1,2,3)$.
In Fig.\,\ref{fig5avr} the 
matrices ${\boldsymbol
  S}^{(j,i)}(\lambda)$\   are associated with
the oriented lines connecting the points $z_{i}$ and $z_{j}$. 

The main r${\hat{\rm o}}$le in the following analysis belongs to
symmetry transformations which essentially allows one to connect
solutions ${\boldsymbol \chi}^{(i)}$ on different sheets of the
Riemann surface of the PGHO. 
Let 
\bea\label{O-def}
\wh\Omega_i:\qquad z\mapsto \gamma_i\circ z\,,\qquad \lambda\mapsto q_i^{-1}\,
\lambda\, \qquad (i=1,2,3)
\eea
be a transformation, involving a translation of the independent 
variable $z$ along the contour $\gamma_i$, accompanied by the
substitution $\lambda\mapsto q_i^{-1}\, \lambda$, where 
\bea\label{qi-def}
q_i=\re^{\ri\pi a_i}\,, \qquad\ \  q_1\,q_2\,q_3=1\ .
\eea
It is easy to see, that the  substitutions \eqref{O-def}  leave 
PGHO  unchanged.  Therefore
they act as linear transformations in the space of solutions. 
Namely, in the basis ${\boldsymbol\chi}^{(i)}$ they read 
\begin{subequations}\label{N123}
\bea\label{N123a}
\wh\Omega_i\big({\boldsymbol\chi}^{(i)}\big)&=&-{\boldsymbol
  \chi}^{(i)}\,\re^{-2\pi \ri p_i\sigma_3}\\[.4cm] 
\wh\Omega_j\big({\boldsymbol\chi}^{(i)}\big)
&=&-{\boldsymbol\chi}^{(i)}\,{\boldsymbol
  S}^{(i,j)}(\lambda)\, 
\re^{-2\pi \ri p_j \sigma_3}\, {\boldsymbol S}^{(j,i)}(\lambda\,
q_j^{-1}) \label{N123b}\\[.4cm] 
\wh\Omega_k\big({\boldsymbol\chi}^{(i)}\big)
&=&-{\boldsymbol\chi}^{(i)}\,{\boldsymbol
  S}^{(i,k)}(\lambda)\,  
\re^{-2\pi \ri p_k \sigma_3} \,{\boldsymbol S}^{(k,i)}(\lambda\,
q_k^{-1})\ .\label{N123c} 
\eea
\end{subequations}

The most fundamental property of the differential operator
\eqref{opspspsao} is that   a combined transformation 
$\wh\Omega_k\circ\wh\Omega_j\circ\wh\Omega_i$,
where $(i,j,k)$ is a cyclic permutation of $(1,2,3)$, is equivalent to
the identity transformation in the space of solutions of \eqref{opspspsao},
\beq\label{prop1}
\wh\Omega_k\circ\wh\Omega_j\circ\wh\Omega_i\big({\boldsymbol\chi}^{(i)}\big)={\boldsymbol\chi}^{(i)}\ .
\eeq
The proof follows from the relation \eqref{qi-def} and the fact that 
 $\gamma_k\circ\gamma_j\circ\gamma_i$ is a contractible contour, which
loops around a regular point ($z=\infty$) of the PGHO 
(see Fig.\,\ref{fig1ay}). 
Combining \eqref{N123} and \eqref{prop1} with the definition
\eqref{S-def} one easily obtains
\beq\label{S-rel2}
{\boldsymbol S}^{(i,k)}(\lambda)\ \re^{-2\pi \ri p_k \sigma_3}\ 
{\boldsymbol S}^{(k,j)}(\lambda\,q_k^{-1})\ \re^{-2\pi \ri p_j \sigma_3}\ 
{\boldsymbol S}^{(j,i)}(\lambda\,q_i)\ \re^{-2\pi \ri p_i \sigma_3}=
-{\boldsymbol I}\ .
\eeq
Consider now the transformation
$\wh\Omega_k\circ\wh\Omega_i\circ\wh\Omega_j$, where the indices $i$
and $j$ are interchanged with respect to \eqref{prop1}. 
Repeating the above arguments (again with an account of \eqref{qi-def})
one can show
that this transformation is equivalent to a linear transformation of
solutions  
\beq\label{prop4}
\wh\Omega_k\circ\wh\Omega_i\circ\wh\Omega_j\big({\boldsymbol\chi}^{(i)}\big)
={\boldsymbol\chi}^{(i)}\,\boldsymbol{M}(\gamma_P|\lambda)\  ,
\eeq
where $\boldsymbol{M}(\gamma_P|\lambda)$ can be interpreted as  a monodromy matrix
of the Pochhammer loop  depicted in Fig.\,\ref{fig1av}.
Then
using \eqref{N123} one obtains,
\beq\label{S-rel3}
{\cal W}(\lambda)=-{\rm Tr}\Big[\re^{-2 \pi \ri p_i \sigma_3}\, 
\boldsymbol{S}^{(i,j)}(\lambda\,q_j)\,
\re^{-2 \pi i p_j \sigma_3}\, 
\boldsymbol{S}^{(j,k)}(\lambda)\,
\re^{-2 \pi \ri p_k \sigma_3}\, 
\boldsymbol{S}^{(k,i)}(\lambda\, q_k^{-1})\,\Big]\ .
\eeq
We would like to stress that the above considerations apply to all
PGHO's with an arbitrary number of the monodromy-free punctures. 
This means that the connection matrices will always satisfy the same
relations \eqref{Slambda}, \eqref{N123}, \eqref{S-rel2} and
\eqref{S-rel3}, even though these matrices depend
on a set of 
the monodromy-free punctures. Note, in particular, 
Eqs.\eqref{Slambda} and \eqref{S-rel2} forms a system of functional
relations for the coefficients of the connection
matrices.  A simple inspection shows that there are only nine
independent relations among \eqref{Slambda} and \eqref{S-rel2}  
for twelve different coefficients. 
Nevertheless, as we shell see
below, these functional relations together with appropriate 
analyticity assumptions completely determine all these 
coefficients. More precisely, the relations have an
infinite discrete set of solutions, corresponding the PGHO's with
arbitrary number of the monodromy-free punctures.

For further references note, that the elements of 
the connection matrices, 
are simply related to the
Wronskians of the basic solutions, 
\bea\label{aoasposa}
{\tt W}[\chi_{\sigma'}^{(j)},\chi_{\sigma}^{(i)}]=
-\sigma'\,{ S}^{(j,i)}_{-\sigma'\sigma}(\lambda)\ .
\eea
In what follows
the set of functions
${ W}_{\sigma'\sigma}^{(k)} (\lambda)$ defined
through the relation
\bea\label{ystsosapsassasss}
{\tt W}[\chi_{\sigma'}^{(j)},\chi_{\sigma}^{(i)}]
=-\ri\ 
\re^{\ri\pi\sigma' p_j}\
 \ { W}_{\sigma'\sigma}^{(k)} (\lambda)\ \re^{-\omega_j\sigma'-\omega_i\sigma}\ ,
\eea
will be referred as {\it connection coefficients}.
For $\lambda=0$ this definition   coincides with   Eq.\eqref{osapsassasss} and therefore
$W_{\sigma'\sigma}^{(k)}(0)$ coincides with $A^{(k)}_{\sigma'\sigma}$  from Eq.\eqref{ospopsassss}. 
As well as the Wilson loop,
the connection coefficients
are entire functions of the variable $\lambda^2$,
i.e.
they  can be
represented by 
power series in $\lambda^2$
with infinite radius of convergence.
Our next goal is to describe  their characteristic  properties.

\subsection{Large-$\lambda$ asymptotic}

Consider the large $\lambda$ behavior of the connection coefficients.
In the leading order one has
\bea\label{oisiossaoasioa}
 W_{\sigma'\sigma}^{(k)}(\lambda)
\sim\exp\Big(\,\lambda\int_{z_i}^{z_{j}}\rd z\ \sqrt{{\cal P}(z)}\,\Big)\ \  \ \ \ \ \ \ \ 
(i=1,2,3\, ,\ \ \ z_i\sim z_{i+3})\ ,
\eea
where the integrals taken along the oriented links depicted  in Fig.\,\ref{fig5avr}. Introduce the
constants $r_k>0$ and $b_k$:
\bea\label{oaisoasisiao}
\int_{z_i}^{z_{j}}\rd z\ \sqrt{{\cal P}(z)}=r_{k}\ \re^{\ri\pi b_{k}}\ .
\eea
Then Eq.\eqref{oisiossaoasioa} can be equivalently written in the form
\bea\label{sopsspsaop}
 W_{\sigma'\sigma}^{(k)}\big( \re^{-\ri\pi b_{k}}
\lambda\big)\asymp\exp\big(r_{k}\lambda+O(\log\lambda)\,\big)\ \ 
\ \ \ \ \ \ \ \ \ \ \ \ \ 
\big(\,\lambda\to\infty\, ,\ \ \ \ |\arg(\lambda^2)|<\pi\,\big)\ .
\eea
Note that, as it follows from the definition \eqref{oaisoasisiao}, the positive constant $r_k$ is given by
\bea\label{ospspsa}
r_{k}={\textstyle \frac{1}{\pi}}\
\sin\big({\textstyle \frac{\pi}{2}}\,  a_{k}\big)\ \prod_{n=1}^3\Gamma\big({\textstyle \frac{a_n}{2}}\big)
\eea
(here $k=1,2,3$ and  $a_0\sim a_3$),
whereas $b_k$ satisfy the relations
\bea\label{ospspsausys}
\re^{\ri\pi(b_{i}-b_{k})}=-\re^{-\frac{\ri \pi}{2}\ a_{j}}\ .
\eea
To assign precise meaning  to an individual phase factor  $\re^{-\ri\pi b_k}$ 
in Eq.\eqref{sopsspsaop}, one needs to resolve the overall phase ambiguity of
$\sqrt{{\cal P}(z)}$. 
Following the  procedure from  Section\,\ref{usaystsa}
we send $(z_1,z_2,z_3)$ to $(0,1,\infty)$, then
\bea\label{sossisai}
\int_{z_1}^{z_2}\rd z\ \sqrt{{\cal P}(z)}=\re^{-\frac{\ri\pi}{2} \,
(a_1+a_2)}\ \int_{0}^1\rd z\ z^{\frac{a_1}{2}-1}\ (1-z)^{\frac{a_2}{2}-1}\ .
\eea
Assuming that the integrand in the l.h.s.  of this equation is positive    for $0<z<1$, one finds
$\re^{\ri\pi b_3}=\re^{-\frac{\ri\pi}{2} (a_1+a_2)}$. Together with \eqref{ospspsausys}, this implies
\bea\label{usystsaospsopsao}
\re^{\ri\pi b_1}=
\re^{ \frac{\ri\pi}{2} (a_3-a_2)}   \ ,\ \ \ \ \ \ \ \ 
\re^{\ri\pi b_2}=-\re^{-\frac{\ri\pi a_2}{2}}\ ,\ \ \ \ \ \ \ \
\re^{\ri\pi b_3}=-\re^{ \frac{\ri\pi a_3}{2}}\ .
\eea

In fact,
it is not difficult to calculate explicitly
the subleading  term in the asymptotic formula \eqref{sopsspsaop}.
In order to simplify formulae bellow  we make use the notation
\bea\label{sospsosapo}
A^{(k)}_{\sigma'\sigma}(\lambda)\equiv
W_{\sigma'\sigma}^{(k)}(\ri\,  \re^{-\ri\pi b_k}\ \lambda)\ ,\ \ \ \ \ \ \ \ 
A^{(k)}_{\sigma'\sigma}(0)\equiv A^{(k)}_{\sigma'\sigma}\ ,
\eea
where $A^{(k)}_{\sigma'\sigma}$ is given by Eq.\eqref{ospopsassss}.
Then
\bea\label{aspoaskasus}
A_{\sigma'\sigma}^{(k)}(\ri\lambda)\asymp
\frac{\big( \Lambda_{j}(\lambda)\,\big)^{\sigma'}\, \big( \Lambda_i(\lambda)\,\big)^{\sigma}}
{\sqrt{ 4\,s(\frac{2p_i}{a_i}) s(\frac{2p_{j}}{a_{j}})}}\ \ \re^{r_{k}\, \lambda}\ \ 
 \big(1+O(\lambda^{-1})\big)\ ,
\eea
where
\bea\label{ossaaosp}
 \Lambda_i(\lambda)=
 \Big(\frac{\lambda}{a_i}\Big)^{-\frac{2 p_i}{a_i}}\ \
\sqrt{\frac{\Gamma(1+\frac{2p_i}{a_i})}{\Gamma(1-\frac{2p_i}{a_i})}}\ \
\re^{\omega_i}\ \ \Big(\frac{z_{jk}}{z_{ji}z_{ik}}\Big)^{- p_i}\ .
\eea
The above formula can be applied for large $\lambda^2$  such that   $|\arg(\lambda^2)|<\pi$.
In the case  of real   $\lambda^2<0$, i.e. when 
$\lambda=\ri\, |\lambda|$, the asymptotic is  given by
\bea\label{ysstaspoaskasus}
A_{\sigma'\sigma}^{(k)}\big( |\lambda|\big)\asymp
\frac{\big( \Lambda_{j}(|\lambda|)\,\big)^{\sigma'}\,
\big( \Lambda_i(|\lambda|)\,\big)^{\sigma}}{\sqrt{ s(\frac{2p_i}{a_i}) s(\frac{2p_{j}}{a_{j}})}}\ \,
\cos\big(\,r_k\,|\lambda|-
{\textstyle \frac{\sigma\pi p_i}{a_i}}-{\textstyle\frac{\sigma'\pi  p_{j}}{a_{j}}}+O(\lambda^{-1})\big)\, .
\eea

As it was discussed at the end of Section\,\ref{sec13}, 
the combinations \eqref{asospasoaso}, which appear  in the formula \eqref{ossaaosp}, are
functions of the $L$ projective invariants. 
In the case $L=0$, they are given by equation  \eqref{saopopsa} from Appendix\,\ref{app2}.
For this reason it is convenient to write
the   subleading terms in the asymptotic formulae \eqref{aspoaskasus} and  \eqref{ysstaspoaskasus} as
\bea\label{sisiossaio}
\big(\,\Lambda_j(\lambda)\,\big)^{\sigma'} \big(\,\Lambda_i(\lambda)\,\big)^\sigma&=&
\big(\,S( \sigma' p_j |p_k+p_i)\,S( \sigma' p_j|p_k-p_i)\, S( \sigma p_i |p_j+p_k)\,S( \sigma p_i|p_j-p_k)\,
\big)^{\frac{1}{4}}\
\nonumber\\
&\times&  \big(g^{(L,\alpha)}_j\big)^{\sigma'}\
 \big(g^{(L,\alpha)}_i\big)^{\sigma}\ ,
\eea
where
\bea\label{asosaosapao}
S(p_i|q)=
\Big(\frac{\lambda}{a_i}\Big)^{-\frac{4 p_i}{a_i}}\
\frac{\Gamma(\frac{1}{2}+p_i-q)\, \Gamma(\frac{1}{2}+p_i+q)}
{ \Gamma(\frac{1}{2}-p_i-q)\, \Gamma(\frac{1}{2}-p_i+q)}\
\frac{\Gamma(1-2 p_i)\, \Gamma(1+\frac{2p_i}{a_i})}{\Gamma(1+2p_i)\, \Gamma(1-\frac{2p_i}{a_i})}\ ,
\eea
and $g^{(L,\alpha)}_i$  $(g^{(0)}_i=1)$ stand for  $\lambda$-independent  constants corresponding to  a  given set of
monodromy\,-\,free punctures
\eqref{sisoaisiosaoi}.\footnote{\label{footn8}In the case $L=1$, the formula \eqref{soaoposap} from Appendix\,\ref{app2} leads to
$$g^{(1,\alpha)}_i=
\ri\
\frac{\vartheta_2(u^{(\alpha)}-u_i,q)}{\vartheta_1(u^{(\alpha)}-u_i,q)}\
\ \ \frac{\vartheta_3(u_j-u_k,q)}{\vartheta_4(u_j-u_k,q)}\ \ \ \ \ \ \ \ \ (\alpha=1,2,3)\ ,
$$
where $u^{(\alpha)}$ are the values of uniformizing parameter $u$ \eqref{soosapsao} corresponding to
the roots $\big(x^{(\alpha)},y^{(\alpha)}\big)$ of the
system of  two equations \eqref{soiasisisa} and \eqref{sisiaiso}
 (i.e. $\big(x^{(\alpha)},y^{(\alpha)}\big)=\big(x(u^{(\alpha)}),\,y(u^{(\alpha)})\big)$, where
functions $x=x(u)$ and $y=y(u)$ are given by \eqref{osospsa}).}

The asymptotic  formula \eqref{aspoaskasus} can be extended to the following systematic asymptotic series
\bea\label{ystaspoaskasususys}
A_{\sigma'\sigma}^{(k)}(\ri\lambda)\asymp
\frac{\big( \Lambda_{j}(\lambda)\,\big)^{\sigma'}\, \big( \Lambda_i(\lambda)\,\big)^{\sigma}}
{\sqrt{ 4\,s(\frac{2p_i}{a_i}) s(\frac{2p_{j}}{a_{j}})}}\ \ \re^{r_k\, \lambda}\ \ \  B^{(k)}(\lambda)\
X^{(j)}_{\sigma'}\Big(\re^{-\frac{\ri\pi a_j}{2}}\lambda\Big)\ X^{(i)}_\sigma(\lambda)\ .
\eea
Here the quantity $B^{(k)}(\lambda)$ 
is a  formal power series
\bea\label{saopsspas}
B^{(k)}(\lambda)=\exp\bigg(\, \sum_{n=1}^\infty\  \frac{c_n\    q^{(L)}_{2n-1}}{
4\, \sin\big(\, \pi\,( n-\frac{1}{2})\, a_i\,\big)
\,\sin\big(\, \pi\, ( n-\frac{1}{2})\, a_{j}\,\big)}\ \ \lambda^{1-2 n}\, \bigg)\ ,
\eea
where  $ q^{(L)}_{2n-1}$ stand for  the expansion coefficient for 
the Wilson loop \eqref{aopsaosapoas} and the numerical coefficients $c_n$
are defined by Eq.\eqref{aoopsaap}.
Similarly, the symbol  $X^{(i)}_\sigma (\lambda)$  in
\eqref{ystaspoaskasususys} denotes the formal 
power series expansion in fractional powers of $\lambda$, namely,\footnote{\label{footn9}An
explicit form of the expansion coefficients $ x^{(i)}_{\sigma,n}$
are not known. The sole exclusion is the  first coefficient in  the  case of  PGHO with 
$L=0$, which reads explicitly
$$x^{(i)}_{\sigma, 1}|_{
L=0}=\Big(\frac{2}{a_i}\Big)^{-\frac{2}{a_i}}\ \
\frac{\Gamma(\frac{1}{a_i}) \Gamma(\frac{1}{2}-\frac{1}{a_i})}{4\,\sqrt{\pi}}\ \
\frac{\Gamma(1+\frac{1}{a_i}+\frac{2\sigma p_i}{a_i})}
{\Gamma(-\frac{1}{a_i}+\frac{2\sigma p_i}{a_i})}\ \
\bigg(\,\frac{ p_{j}^2-p_{k}^2}{  p_i^2-\frac{1}{4}}+\frac{a_{j}-a_{k}}{2+a_i}
\,\bigg)\ .
$$
}
\bea\label{sospospsoa}
X^{(i)}_\sigma (\lambda)=\exp\bigg(\, 
\sum_{n=1}^\infty x^{(i)}_{\sigma,n}\  \lambda^{-\frac{2 n}{a_{i}}}\,\bigg)\ .
\eea

\subsection{Zeroes of $A_{\sigma'\sigma}^{(k)}(\lambda)$}

By definition \eqref{sospsosapo}, the functions  
$A_{\sigma'\sigma}^{(k)}(\lambda)$
are entire functions of $\lambda^2$. 
Let us discuss patterns their zeros
$\{\lambda_n^{(k)}\}_{n=1}^{\infty}$, so that
\beq
A_{\sigma'\sigma}^{(k)}(\lambda^{(k)}_n)=0\,,\qquad n=1,2,\ldots \ .
\eeq
Here we have omitted the indices $\sigma',\sigma$ in the notation of
zeroes. This dependence will be implicitly assumed.
We will also assume that the sign of $\lambda_n^{(k)}$ is fixed by the requirement $-\pi/2<\arg
\big(\lambda_n^{(k)}\big)\le \pi/2$.

Due to the cyclic symmetry, 
it is sufficient to consider one value of $k$,  say $k=3$.
Let us set  $(z_1,z_2,z_3)$ to $(0,1,\infty)$ and then make the
change  of  variables 
\eqref{siisaoi}. The transformation  $w=w(z)$ is the
Schwartz-Christoffel mapping which sends
$(0,1,\infty)$ to $(0, r_3,\, r_2\,\re^{\frac{\ri\pi}{2}\, a_1})$,
whereas the function $ {\hat \psi}$ satisfies 
the ordinary Schr\"odinger
equation \eqref{kaajkasj} with the potential ${\hat T}_L(w)$
 given  by \eqref{sopspo}.
Consider a zero $\lambda^{(3)}_n$
of the function $A^{(3)}_{\sigma'\sigma}(\lambda)$. It is
easy to see 
that if $\lambda^2=-(\lambda^{(3)}_n)^2$,  the  Schr\"odinger
equation \eqref{kaajkasj} has a solution ${\hat \psi}_n$ such that
\bea\label{soispossasaop}
{\hat \psi}_n(w)\sim
\begin{cases}
&w^{\frac{1}{2}+\frac{2p_1\sigma}{a_1}}\  \big(\,1+O(w)\,\big)\,,\qquad\qquad\qquad\qquad
  {\rm as}\ \ \ \  w\to 0\\[.3cm]
&(r_3-w)^{\frac{1}{2}+\frac{2p_2\sigma'}{a_2}}\ \big(\,1+O(w-r_3)\,\big)\,,\qquad
  \ \  \,
{\rm as}\ \ \ \  w\to r_3
\end{cases}\  .
\eea
If the parameters $p_i$ are restricted by the 
condition $0<\frac{2p_i}{a_i}<\frac{1}{2}$ (see Eq.\eqref{posposopa}),
the above asymptotic conditions lead to well-defined
spectral problems for all $\sigma,\sigma'=\pm1$. 
An immediate consequence of this fact is that all the  zeroes of
 $A^{(3)}_{\sigma'\sigma}(\lambda)$ are simple.

In the simplest case of the perturbed hypergeometric oper (i.e., for
$L=0$), the potential in the  Schr\"odinger 
equation \eqref{kaajkasj} 
is real and positive.
Therefore all the zeroes $\lambda^{(3)}_n$ are also real and positive.  
Then the large-$\lambda$ asymptotic formulae
\eqref{aspoaskasus}-\eqref{ysstaspoaskasus} imply that 
the zeroes  accumulate at the infinity along
the positive real axis and for large integer $n\gg
1$ one has
\bea\label{spssapsaissia}
\lambda^{(k)}_n\asymp\frac{\pi}{r_k}\
\Big(n+{ \frac{\sigma p_i}{a_i}}+{\frac{\sigma' p_{j}}{a_{j}}}
-{\frac{1}{2}}\,\Big)+O(n^{-1})\ .
\eea
(Because of cyclic symmetry, the last formula
is valid for any cyclic permutation $(i,j,k)$.)
For a general case of PGHO with $L>0$ the potential in the
equation \eqref{kaajkasj}, in general, becomes
complex-valued for $w\in[0,r_3]$, so that the zeros $\lambda^{(3)}_n$ 
also become complex.
However, they still remain simple and accumulate at infinity in the vicinity
of the positive real axis.  The asymptotic formula 
\eqref{spssapsaissia} continues to hold 
for $L>0$. Moreover, we would like to stress, 
that for large $n$ this formula gives  
the asymptotics of precisely the $n$-th zero $\lambda^{(k)}_n$ 
(in the sense that $n$ coincides with the number of zeroes, 
whose absolute value is less or equal than $|\lambda^{(k)}_n|$).
Similar considerations apply to all functions
$A^{(k)}_{\sigma'\sigma}(\lambda)$; they can be written in the form of
convergent products 
\bea\label{isusoispssaop}
A^{(k)}_{\sigma'\sigma}(\lambda)=A_{\sigma'\sigma}^{(k)}\ 
\prod_{n=1}^\infty\bigg(1-\frac{\lambda^2}
{\big(\lambda^{(k)}_n\big)^2}\bigg)\  ,
\eea
where $A_{\sigma'\sigma}^{(k)}$ is given by \eqref{ospopsassss}.
At this stage it is convenient to introduce spectral
$\zeta$-functions,
which capture all information
about the distribution
of zeroes $\lambda^{(k)}_n$,
\bea\label{zeta-def}
\zeta_k(\nu)=
\sum_{n=1}^\infty \big(\lambda^{(k)}_n\big)^{-\ri\nu}
\eea
(recall that we assume that $-\pi/2<{\rm arg}\, \big(\lambda_n^{(k)}\big)\leq \pi/2$).
As follows from the asymptotic formula \eqref{spssapsaissia}
the function 
$\zeta_k(\nu)$ is analytic in the lower half
plane $\Im m( \nu)\leq 0$, except the point $\nu=-\ri$, where it has a simple
pole with the residue $-\ri\, r_k/\pi$. Using these properties
the product formula \eqref{isusoispssaop} can be transformed into an integral representation
\bea\label{A-zeta}
\log\bigg(\frac{A^{(k)}_{\sigma'\sigma}(\ri\lambda)}
{A^{(k)}_{\sigma'\sigma}(0)}\bigg)=
r_k\, \lambda -\frac{1}{2}\int_{{\mathbb
    R}-\ri
  0}  \frac{\rd\nu}{\nu}\ \frac{\zeta_k(\nu)}
{\sinh\big(\frac{\pi\nu}{2}\big)}\ \ \lambda^{\ri\nu}\  .
\eea
Closing the integration contour in this formula in upper half plane
and comparing the result with the asymptotic expansion \eqref{ystaspoaskasususys} one
concludes that the function $\zeta_k(\nu)$ has  zeroes at $\nu=2\ri,\,4\ri,\ldots$ and
additional simple poles on the imaginary axis $\nu$ with the following residues
\bea\label{resu}
&&{\rm res}\big[ \zeta_k(\nu)\big]_{\nu=\ri
  (2n-1)}=\ri\ \ 
 \frac{  \Gamma(n+\frac{1}{2})\    q^{(L)}_{2n-1}}{
4\pi^{\frac{3}{2}}\, n!\, \sin\big(\, \pi\,( n-\frac{1}{2})\, a_i\,\big)
\,\sin\big(\, \pi\, ( n-\frac{1}{2})\, a_{j}\,\big)}\ 
\ ,\nonumber\\[.3cm]
&&{\rm res}\big[ \zeta_k(\nu)\big]_{\nu=\frac{2 \ri 
 n}{a_i}}=-\ri\ \frac{2\,n}{\pi a_i}\,\sin\Big(\frac{\pi
  n}{a_i}\Big)\,x^{(i)}_{\sigma,n} 
\ ,\label{residues}  \\[.3cm]
&&{\rm res}\big[ \zeta_k(\nu)\big]_{\nu=\frac{2\ri n
 }{a_j}}=-\ri\ \frac{2\,n}{\pi a_j}\,\sin\Big(\frac{\pi
  n}{a_j}\Big)\ (-1)^{n}\  x^{(j)}_{\sigma',n}\ ,
\nonumber
\eea
where $n=1,2,\ldots\ .$
Moreover it follows from \eqref{aspoaskasus} that 
\beq
\zeta_k(0)=-\frac{\sigma p_i}{a_i}-
\frac{\sigma' p_j}{a_j}
\eeq
and 
\bea\label{zetaprime}
\exp\big(-2\ri\,\zeta_k'(0)\big)=
\frac{\big( \Lambda_{j}(\lambda)\,\big)^{\sigma'}\, 
\big( \Lambda_i(\lambda)\,\big)^{\sigma}}
{A^{(k)}_{\sigma'\sigma}(0)\sqrt{ 4\,s(\frac{2p_i}{a_i}) s(\frac{2p_{j}}{a_{j}})}}\,\Bigg|_{\lambda=1}\ ,
\eea
where $\Lambda_i(\lambda)$ is defined in \eqref{ossaaosp}.

\subsection{\label{BAEQ}Bethe Ansatz equations}

The nine non-linear functional equations \eqref{Slambda} and \eqref{S-rel2}
involve too many unknown functions (twelve) and, in fact, appear
to be rather   
complicated for a direct analysis. Fortunately,
it is possible to reduce these equations to eight sets of rather 
compact equations of the Bethe Ansatz type,  
where each set involves only three unknown
functions. In principle, this could be done by direct
manipulations with the equations  \eqref{Slambda} and \eqref{S-rel2},
but here we prefer a more efficient approach involving
direct calculations of the Wronskians. It is based on the relation 
\eqref{prop1} and the following simple properties: 
\begin{enumerate}[{\bf (i)}]
\item The solutions ${\boldsymbol\chi}^{(i)}(z)$ are  simply
  transformed under the action of  $\wh{\Omega}_i$ with the same $i$
  (see  Eq.\eqref{N123a}). 
This follows from the fact that the asymptotic condition
\eqref{opaosap}, defining the 
solution $\chi^{(i)}_\sigma(z)$, does not involve the parameter $\lambda$.
\item For any two solutions $\psi_1(z)$ and $\psi_2(z)$ of \eqref{opspspsao} 
one has 
\beq\label{prop3}
{\wr}\big[\,\wh\Omega_i(\psi_1),\,\wh\Omega_i(\psi_2)\,\big]
=w(\lambda q_i^{-1})\ ,\qquad\ \ \ \ \ \ {\rm where} \ \ \ w(\lambda)=\wr\big[\psi_1,\psi_2\big]\ .
\eeq
The proof follows from the definition \eqref{O-def} and the fact
that the Wronskian in \eqref{prop3} does not depend 
on the
point $z$ where it is calculated. 
\end{enumerate}

Let $(i,j,k)$ be a cyclic permutation of $(1,2,3)$ and
$\sigma,\sigma',\sigma''=\pm1$. Together with functions
$A^{(k)}_{\sigma'\sigma}(\lambda)$ \eqref{sospsosapo} 
it is convenient to introduce additional notation $ T^{(i)}_\sigma$ 
\beq\label{T-def}
\wr\big[\,\wh\Omega_{j}(\chi^{(i)}_\sigma)\,,\,
\chi_\sigma^{(i)}\,\big]=\ri \  T^{(i)}_\sigma(-\ri\,\re^{\ri\pi b_i}\,\lambda)\ .
\eeq
{}From Eq.\eqref{N123b} it immediately follows that 
\beq\label{T-res1}
\ri\, T^{(k)}_\sigma(\lambda)=
-\re^{2 \pi \ri p_j}\, A^{(k)}_{+,\sigma}\big(\,\lambda q_j^{\frac{1}{2}}\,\big)\,
A^{(k)}_{-,\sigma}\big(\,\lambda q_j^{-\frac{1}{2}}\,\big)+
\re^{-2 \pi \ri p_j} A^{(k)}_{+,\sigma}\big(\,\lambda q_j^{-\frac{1}{2}}\,\big)\,
A^{(k)}_{-,\sigma}\big(\,\lambda q_j^{\frac{1}{2}}\,\big)\, .
\eeq
The same quantity can be calculated in a different way, using the
additional relations \eqref{prop1} and \eqref{prop3}. 
First, from \eqref{N123a} and \eqref{prop3} it follows that
\beq\label{wron1}
\wr\big[\wh\Omega_i\big(\chi^{(j)}_{\sigma'}\big)\,,\chi_\sigma^{(i)}\,\big]
= \sigma'\,
\re^{-2 \pi \ri \sigma  p_i }\,S^{(j,i)}_{-\sigma',\sigma}(\lambda q_i^{-1})\ .
\eeq
Similarly, using also the property \eqref{prop1}, one can easily show that 
\beq \label{wron2}
\wr\big[\,\wh\Omega_j\big(\chi^{(i)}_\sigma\big)\,,\,
  \chi^{(k)}_{\sigma''}\,\big]  
=\wr\big[\,\wh\Omega_k^{-1}\,\wh\Omega_i^{-1}\big(\chi^{(i)}_\sigma\big)\,,\,
  \chi^{(k)}_{\sigma''}\,\big]=
-\sigma \,
\re^{-2\pi \ri(\sigma p_i  -\sigma'p_k )} \, 
S^{(i,k)}_{-\sigma,\sigma''}(\lambda q_k) \ .
\eeq
Next, any three basic solutions $\chi^{(i)}_\sigma$, $\chi^{(j)}_{\sigma'}$
and $\chi^{(k)}_{\sigma''}$ are connected by the a linear relation
\beq\label{rel3}
\sigma' S^{(j,i)}_{-\sigma,\sigma'}(\lambda)\,\chi^{(k)}_{\sigma''}+
\sigma'' S^{(k,j)}_{-\sigma,\sigma''}(\lambda)\,\chi^{(i)}_{\sigma}+
\sigma S^{(i,k)}_{-\sigma'',\sigma}(\lambda)\,\chi^{(j)}_{\sigma'}=0\ .
\ee
Consider again the Wronskian in \eqref{T-def}. Expressing the second 
$\chi_\sigma^{(i)}$ therein from \eqref{rel3} and then using 
the previous relation \eqref{wron1}, \eqref{wron2}
one obtains 
\bea\label{TQ-eq}
T^{(i)}_\sigma(\lambda)\, A_{\sigma''\sigma'}^{(i)}(\lambda)&=&
\re^{-\ri\pi(\sigma p_i-\sigma' p_j -\sigma''p_k)}\,
A_{\sigma\sigma''}^{(j)}(\lambda q_k^{\frac{1}{2}})\,
A_{\sigma'\sigma}^{(k)}(\lambda q_j^{\frac{1}{2}})\nonumber\\
&+&\re^{\ri\pi(\sigma p_i-\sigma' p_j -\sigma''p_k)}\,
A_{\sigma\sigma''}^{(j)}(\lambda q_k^{-\frac{1}{2}})\,
A_{\sigma'\sigma}^{(k)}(\lambda q_j^{-\frac{1}{2}})\,.
\eea
Making simultaneous cyclic permutations 
of the indices $(i,j,k)$ and the values $(\sigma,\sigma',\sigma'')$
one obtains another two equations of the same type, which
contain the same three functions $A_{\sigma''\sigma'}^{(i)}(\lambda)$, \ \ 
$A_{\sigma\sigma''}^{(j)}(\lambda)$ and $A_{\sigma'\sigma}^{(k)}(\lambda)$ as in 
the equation \eqref{TQ-eq}. 
By definition, $T^{(i)}_\sigma(\lambda)$ is an entire function of
$\lambda^2$, therefore the l.h.s. of \eqref{TQ-eq} vanishes at all zeroes
of $A^{(i)}_{\sigma'',\sigma'}(\lambda)$. Proceeding in this way one
obtains a system of  
three coupled Bethe Ansatz type
equations for the position of zeroes
\beq
\label{BAE}
\left\{
\begin{array}{rcl}
\displaystyle
\re^{2\pi \ri(\sigma''p_k+\sigma' p_j-\sigma p_i)}\ 
\frac{A^{(k)}_{\sigma'\sigma}\big(\lambda_n^{(i)}\,q_j^{+\hf}\big)\,
A^{(j)}_{\sigma\sigma''}\big(\lambda_n^{(i)}\,q_k^{+\hf})}
{A^{(k)}_{\sigma'\sigma}\big(\lambda_n^{(i)}\,q_j^{-\hf})\,
A^{(j)}_{\sigma\sigma''}\big(\lambda_n^{(i)}\,q_k^{-\hf}\big)}
&=&-1 
\\[.9cm]
\displaystyle
\re^{2\pi \ri(\sigma'p_j+\sigma p_i-\sigma'' p_k)}\ 
\frac{A^{(j)}_{\sigma\sigma''}\big(\lambda_n^{(k)}\,q_i^{+\hf}\big)\,
A^{(i)}_{\sigma''\sigma'}\big(\lambda_n^{(k)}\,q_j^{+\hf})}
{A^{(j)}_{\sigma\sigma''}\big(\lambda_n^{(k)}
\,q_i^{-\hf})\,
A^{(i)}_{\sigma''\sigma'}\big(\lambda_n^{(k)}\,q_j^{-\hf}\big)}
&=&-1 
\\[.9cm]
\displaystyle
\re^{2\pi \ri(\sigma p_i+\sigma'' p_k-\sigma'p_j)}\ 
\frac{A^{(i)}_{\sigma''\sigma'}\big(\lambda_n^{(j)}\,q_k^{+\hf})
\,A^{(k)}_{\sigma'\sigma}\big(\lambda_n^{(j)}\,q_i^{+\hf}\big)}
{A^{(i)}_{\sigma''\sigma'}
\big(\lambda_n^{(j)}\,q_k^{-\hf}\big)\,
A^{(k)}_{\sigma'\sigma}\big(\lambda_n^{(j)}\,q_i^{-\hf})}
&=&-1
\\[.9cm]
\end{array}
\right.
\ ,
\eeq
where $n=1,2,\ldots$ and $\lambda^{(i)}_n$, $\lambda^{(j)}_n$ and 
$\lambda^{(k)}_n$ denote the zeroes of $A^{(i)}_{\sigma''\sigma'}(\lambda)$,\ 
$A^{(j)}_{\sigma\sigma''}(\lambda)$ and $A^{(k)}_{\sigma'\sigma}(\lambda)$,
respectively.     
Choosing $(\sigma,\sigma',\sigma'')=(\pm 1,\pm 1,\pm
1)$ 
one gets, eight different triples of the 
Bethe Ansatz type equation, where each set involves 
only three different functions. 
Any particular functions $A^{(i)}_{\sigma'\sigma}(\lambda)$
 enters into the two sets of these equations. 

As an immediate consequence 
one can derive an 
``asymptotically exact'' Bohr-Sommerfeld quantization condition for
the roots $\lambda^{(k)}_{n}$.  Substituting the asymptotic formula 
\eqref{ystaspoaskasususys} into \eqref{BAE} one obtains,
\bea\label{sospssaop}
\lambda=\lambda^{(k)}_n\ \ :\ \ \ \ \ \ 
r_k\, \lambda+\Phi^{(k)}_{\sigma \sigma'}(\lambda)\asymp \pi\
\Big(n+
{ \frac{\sigma p_i}{a_i}}+{\frac{\sigma' p_{j}}{a_{j}}}-{\frac{1}{2}}\,\Big)\ ,
\eea
where
\bea\label{saosospsao}
\Phi^{(k)}_{\sigma' \sigma}(\lambda)&=&
\sum_{n=1}^\infty\  \frac{(-1)^n\ c_n\    q^{(L)}_{2n-1}}{
4\, \sin\big(\pi ( n-\frac{1}{2})\, a_i\,\big)
\,\sin\big(\pi ( n-\frac{1}{2})\, a_{j}\,\big)}\ \lambda^{1-2n}\\
&-&
\sum_{n=1}^\infty \, x^{(i)}_{\sigma,n}\
\sin\big({\textstyle\frac{\pi n}{a_{i}}}\big)\
 \lambda^{-\frac{2 n}{a_{i}}}-
\sum_{n=1}^\infty (-1)^n\, x^{(j)}_{\sigma',n}\ \sin\big({\textstyle\frac{\pi
    n}{a_j}}\big)\ 
\lambda^{-\frac{2 n}{a_{j}}}\ .\nonumber
\eea

It is convenient to introduce a new function 
\beq\label{eps-def}
\epsilon_i(\lambda)=\ri \log\left(
\frac{A^{(j)}_{\sigma\sigma''}\big(\lambda\,q_k^{+\hf}\big)
\,A^{(k)}_{\sigma'\sigma}\big(\lambda\,q_j^{+\hf}\big)}
{A^{(j)}_{\sigma\sigma''}\big(\lambda\,q_k^{-\hf}\big)\,
A^{(k)}_{\sigma'\sigma}\big(\lambda\,q_j^{-\hf}\big)}\right)+2\pi\,\big(
\,\sigma p_i  -\sigma' p_j  -\sigma'' p_k\,  \big)
\eeq
and another two functions $\epsilon_j(\lambda)$ and $\epsilon_k(\lambda)$, which
are obtained from \eqref{eps-def} by simultaneous cyclic permutations
of the indices $(i,j,k)$ and $(\sigma,\sigma',\sigma'')$.  To simplify
the following equations we have omitted the indices 
$\sigma,\sigma',\sigma''$ in the notation of $\epsilon$-functions since
their arrangement is firmly connected to the indices $(i,j,k)$ and so
that they can always be restored.

We expect that  the Bethe Ansatz equations \eqref{BAE} combined with the
asymptotic formula \eqref{spssapsaissia} have an infinite 
number of solutions, corresponding to
PGHO's with different configurations of monodromy-free punctures.  
These solutions are distinguished by different phase 
assignments in the {\em logarithmic form} of the Bethe Ansatz 
equations \eqref{BAE},
\beq\label{BAE2}
\epsilon_i\big(\lambda^{(i)}_n\big)=\pi \,\big(\,2 m^{(i)}_n-1\,\big)\,, \qquad   
m^{(i)}_n\in {\mathbb Z}\ \qquad \ \ \ \ \ \ (i=1,2,3)\ ,
\eeq
which involve  
three sets of integers $\{m^{(i)}_n\}_{n=1}^{\infty}$,
\ $i=1,2,3$. These integers, of course, depend on the choice of branches of the
logarithm in the left hand side of \eqref{eps-def}. However, once
these branches are appropriately fixed,
every solution is characterized by a unique choice of 
$\{m^{(i)}_n\}$. In particular, 
for the PGHO without any monodromy free punctures
($L=0$ case) all
roots lie of the real axis and the integers $m^{(i)}_n$ exactly
coincide with $n$, 
\beq
m^{(i)}_n\equiv n\, , \qquad n=1,2,\ldots\, , \qquad \mbox{for} \qquad L=0\ .
\eeq 

Further, although we have previously assumed 
that the parameters $p_i,p_j,p_k$ obey the constraints \eqref{posposopa}, the resulting 
Bethe Ansatz equations \eqref{BAE} make sense for any
complex values of $p_i$.  Most importantly, their solutions continuously
depend on these parameters. Below we will use this fact to
enumerate all solutions of \eqref{BAE}, following the line of  Appendix A of Ref.\cite{Bazhanov:2003ni}.
Fix the values $\sigma,\sigma'$ and $\sigma''$ and assume that
\beq\label{pbig}
\sigma p_i\gg 1,\qquad \sigma' p_j\gg1, \qquad \sigma'' p_k \gg 1\,,
\eeq
and that $|p_i|$, $|p_j|$ and $|p_k|$ are of the same order of
magnitude.
Then the asymptotics \eqref{spssapsaissia} (as well as the numerical
analysis of \eqref{BAE2}) suggests that for sufficiently large values
of the parameters \eqref{pbig} {\em all roots} $\lambda^{(i)}_n$ \ will be 
ordered $|\lambda_{n+1}|-|\lambda_n|\sim O(1)$ and lie in a close vicinity of
the real axis. Then, if one uses the principal branch of the
logarithms in \eqref{eps-def}, all the  integers $m_n^{(i)}$ 
will be distinct and uniquely defined for every solution of \eqref{BAE2}.

Obviously, not every set of integers
$\{m^{(i)}_n\}$ corresponds to a solution of \eqref{BAE2}. Indeed, 
substituting \eqref{spssapsaissia} into \eqref{BAE2} one concludes
that the sequences of  integers $m_n^{(i)}$ stabilize at
large $n$, i.e.,
\beq
 m_n^{(i)}= n\, ,\qquad  \mbox{for sufficiently large\ }\ \  n\ .
\eeq
Thus, the infinite sets $\{m_n^{(i)}\}$ associated with different
solutions of \eqref{BAE2} only differ in
finitely many first entries.
Therefore the most general pattern for the set 
 $\{m_n^{(i)}\}$ can be obtained from the $L=0$ set 
($m_n^{(i)}\equiv n$, for all $n=1,2,\ldots,\infty$) by deleting 
a certain number of (positive) entries (we denote this number by ${M_i}$) 
and adding the same number
of distinct non-positive integer entries. 
It can be written as
\bea\label{mpk}
m^{(i)}_n=\left\{
\begin{matrix}
1-\tilde{\mu}^{(i)}_{{M_i}-n+1}\ , &{\rm for\ \ }
n=1,\, \ldots {M_i}\hfill\\[.4cm]
N_{n-M_i}(\mu^{(i)})\ ,&{\rm for\ \ } n\ge {M_i}+1\hfill
\end{matrix}
\right.\ .
\eea
Here $\mu^{(i)}=\{\mu^{(i)}_1,\mu^{(i)}_2,\ldots\, \mu^{(i)}_{M_i}\}$ 
and $\tilde{\mu}^{(i)}=\{\tilde{\mu}^{(i)}_1,\tilde{\mu}^{(i)}_2,
\ldots\,\tilde{\mu}^{(i)}_{{M_i}}\}$ 
denote two increasing
sequences  of positive integers
$1\le\mu^{(i)}_1<\mu^{(i)}_2<\ldots<
\mu^{(i)}_{M_i}$ and $1\le 
\tilde{\mu}^{(i)}_1<\tilde{\mu}^{(i)}_2<
\ldots<\tilde{\mu}^{(i)}_{{M_i}}$
with $M_i\ge0$;\   
and $N_\ell(\mu^{(i)})$,\  $\ell=1,2,\ldots\ $, \   denotes $\ell$-th
element of the increasing sequence of
consecutive positive
integers with deleted entries $\mu^{(i)}_n$, $n=1,\ldots M_i$:
\def\crossout#1{%
\hbox{\big/}\kern-.73em {#1}}
\bea\label{Ndef}
\big\{N(\mu^{(i)})\big\}=\big\{ 1,\, 2,\, \ldots\, \crossout{\mu^{(i)}_1},\, \ldots\, 
\crossout{\mu^{(i)}_2}, \ldots \big\}\ .
\eea

We conjecture that the solutions of the Bethe Ansatz equations
\eqref{BAE2}, associated with such set of integers $\{\mu^{(i)}\}$ 
and $\{\tilde{\mu}^{(i)}\}$ correspond to PGHO's with exactly 
\beq\label{set-eq}
L=\sum_{i=1}^3\, \sum_{\ell=1}^{M_i} \,
\big(\,\tilde{\mu}^{(i)}_\ell+{\mu}^{(i)}_\ell-1\,\big)
\eeq
monodromy-free punctures.
For a given value of $L$ the number of the integer sets 
$\big\{\tilde{\mu}^{(1)},
{\mu}^{(1)},
\tilde{\mu}^{(2)},
$
$
{\mu}^{(2)},
\tilde{\mu}^{(3)},
{\mu}^{(3)}
\big\}$, satisfying this equation is equal to ${\mathsf p}_3(L)$ (which is
the number of partitions of $L$ into integer parts of three kinds,
already defined in \eqref{aoposapos}). 

\subsection{Non-linear integral equations for $L=0$\label{nl-cft}}

The  entire function $A^{(k)}_{\sigma'\sigma}(\lambda)$ is completely
determined by its zeros $\lambda^{(k)}_n$ 
and the leading  asymptotic term in \eqref{aspoaskasus}. On the other hand,
the positions of the zeros  are restricted by the equation \eqref{BAE2}.
Mathematically, the   problem of reconstructing the
function $A^{(k)}_{\sigma'\sigma}(\lambda)$  from this data is 
similar to the one which  emerged long ago in the context
of the analytic Bethe Ansatz \cite{Baxter:1971, Baxter:1982, Sklyanin:1978, Reshetikhin:83}.
For the sine-Gordon model 
the problem was solved by Destri and
De Vega \cite{Destri:1992qk,Destri:1997yz}, who have reduced it to a single
complex non-linear integral equation.
Similar equation was earlier derived in 
the lattice $XXZ$-model in 
Ref.\cite{Klumper:1991}.
Here we consider the the non-linear  integral equations determining $A^{(k)}_{\sigma'\sigma}(\lambda)$
in 
the  simplest case of PGHO without monodromy-free
punctures, i.e, $L=0$.

Using \eqref{A-zeta} define spectral
$\zeta$-functions 
$\zeta_i(\nu)$, $\zeta_j(\nu)$ and $\zeta_k$, associated with 
 $A^{(i)}_{\sigma''\sigma'}(\lambda)$,
$A^{(j)}_{\sigma\sigma''}(\lambda)$ and  $A^{(k)}_{\sigma'\sigma}(\lambda)$,
respectively. It is convenient to introduce a new variable $\theta=\log(\lambda)$.
The Bethe Ansatz equations \eqref{BAE} allows one to derive a
non-trivial relation between $\epsilon$- and
$\zeta$-functions. For the case 
when all roots lie on the positive real axis, it reads (see
\cite{Destri:1992qk,Destri:1997yz} for  details of a similar derivation) 
\beq\label{zeta-a}
\zeta_i(\nu)
=\ri \nu\,\sum_{i=1}^3\,
\Phi_{il}(\nu)\,\int_{-\infty}^\infty
\frac{\rd \theta}{\pi}\  \re^{-\ri\nu\theta}\   
\Im m\Big[\log\big(1+\re^{-\ri\epsilon_l(\theta-\ri 0)}\big)\Big]\ \ \ \ \ \ \ (\Im m(\nu)>0)\ ,
\eeq
where
\bea\label{akasissau}
\Phi_{ii}(\nu)&=&
\frac{\sinh(\frac{\pi \nu}{2})\,\sinh(\frac{\pi\nu(a_j+a_k)}{2})}{2\cosh(\frac{\pi\nu}{2})
\sinh(\frac{\pi\nu a_j }{2})\sinh(\frac{\pi\nu a_k}{2})}\ ,\nonumber\\
\Phi_{ij}(\nu)&=&\Phi_{ji}(\nu)
=\frac{\sinh(\frac{\pi \nu}{2})}{2\cosh(\frac{\pi\nu}{2})
\sinh(\frac{\pi\nu a_k }{2})}\ \ \ \ \ \ (i\not= j)\ .
\eea
The integral
\eqref{zeta-a} converges in the half plane $\Im m(\nu)>0$, but it can be
analytically 
continued to the whole complex plane of $\nu$. In fact, as it was remarked before,  the function $\zeta_i(\nu)$
is analytic in the 
lower plane $\Im m(\nu)\leq 0$ except a simple pole at $\nu=-\ri$.
Combining the relations \eqref{A-zeta}, \eqref{eps-def} and
\eqref{zeta-a} 
it is easy to show that 
\beq\label{ddv-eq}
\epsilon_i(\theta)=2 r_i\, \re^\theta -\pi\, \big(
\,\sigma'\, {\textstyle \frac{2 p_j}{a_j}}+\sigma''\, {\textstyle \frac{ 2p_k}{a_k}}\,\big)+
 \sum_{l=1}^3
\int_{-\infty}^\infty\frac{\rd\theta'}{\pi}\  G_{il}(\theta-\theta') \,{\Im m}
\Big[\log\big(1+\re^{-\ri \epsilon_l(\theta'-\ri 0)}\,\big)\Big]\ ,
\eeq
where 
\bea\label{uasysa}
G_{il}(\theta)=
\int_{-\infty}^{\infty}\rd\nu\ \big(\,\Phi_{il}(\nu)-\delta_{il}\,\big)\ \re^{\ri\nu\theta}\ .
\eea

Notice that Eqs.\eqref{resu} and \eqref{zeta-a} imply the following relations
\bea
&& q^{(0)}_{2n -1} =\frac{8\,
  n!\,\sqrt{\pi}}{\Gamma(n-\frac{1}{2})}\  \sum_{i=1}^3 \sin\big(\,\pi
(n-\sh)\,a_i\,\big) 
\,f_i\big(\,\ri\, (2n-1)\,\big)\,,
\label{apspaospa}\\
&&x^{(i)}_{\sigma,n}|_{L=0}=\frac{1}{ a_i \cos\big(\frac{\pi n}{a_i}\big)}\ \Big(\,
(-1)^n\ f_j\big(\,\ri\,  {\textstyle \frac{2 n}{a_i}}\big)+
f_k\big(\,\ri\, {\textstyle\frac{ 2 n}{a_i}}\big)\, \Big) \ ,\label{spspsao}
\eea
where we use function
\beq
f_i(\nu)=\int_{-\infty}^{\infty} \frac{\rd\theta}{\pi} \,\re^{-\ri\nu
  \theta}\,
\Im m\Big[\log\big(\,1+\re^{-\ri\epsilon_i(\theta-\ri 0)}\,\big)\Big]\ ,
\eeq
which is analytic in the upper half-plane, $\Im m(\nu)>0$. 
The function $f_i(\nu)$ has a simple pole at $\nu=0$,
\beq
f_i(\nu)=-\frac{ \ri}{\nu}\ \big(
\sigma p_i-\sigma' p_j -\sigma'' p_k\big)
+f^{(0)}_i+O(\nu)\ .
\eeq
In a view of Eqs.\eqref{zetaprime} and \eqref{zeta-a}, 
it is easy to  see that

\bea\label{opssasap}
\big( \Lambda_{j}(\lambda)\,\big)^{\sigma'}\,
\big( \Lambda_i(\lambda)\,\big)^{\sigma}
\Big|_{\lambda=1\atop L=0}=\exp\Big(\,
f^{ i\,,\,j\,,\,k}_{\sigma,\sigma',\sigma''}-f^{ i\,,\, j\,,\, k}_{-\sigma,-\sigma',-\sigma''}\,\Big)\ ,
\eea
where
$$f^{ i\,,\,j\,,\,k}_{\sigma,\sigma',\sigma''}
={ \frac{1}{ 2a_i}}\ \big(
f_j^{(0)}+f_{k}^{(0)}\big)+
{ \frac{1}{ 2a_j}}\ \big(f_i^{(0)}+f_{k}^{(0)}\big)
\ .
$$

The equation \eqref{ddv-eq} has been solved numerically for various values of 
the parameters $a_1,a_2,a_3$ and $p_1,p_2,p_3$. Using the obtained
numerical values of $\epsilon_i(\theta)$ we calculated \eqref{apspaospa} for $n=1,2$ and \eqref{spspsao} for $n=1$
and
checked that
they are in an excellent agreement  with
Eqs.\eqref{sopsopsaspaa},\,  \eqref{sjusyu} and 
the analytical formula   for $x^{(i)}_{\sigma,1}|_{L=0}$  from  Footnote\ \ref{footn9}.  Also
we numerically checked Eq.\eqref{opssasap},
where the l.h.s.
is  given by \eqref{sisiossaio} with $g^{(0)}_i=g^{(0)}_j=1$.

\section{\label{sec6}Hidden algebraic  structures  (continuation)}

In a view of identification \eqref{isussospsaps}, the formal power series  $B^{(k)}(\lambda)$
in the asymptotic  formula \eqref{ystaspoaskasususys}
can be understood as    eigenvalues
of the formal operator
\bea\label{ysstsaopsspas}
\mathbb{B}^{(k)}(\lambda)=\exp\bigg(\, -\sum_{n=1}^\infty\
\frac{\Gamma(1-( n-\frac{1}{2})\, a_{k-1})\Gamma(1-( n-\frac{1}{2})\, a_{k+1})}
{\Gamma(( n-\frac{1}{2})\, a_{k})}\ \ \ \frac{\Gamma(n-\frac{1}{2})}{
2^{2n-2}\, n!\, \sqrt{\pi}}\
\ \mathbb{I}_{2n-1}\, \lambda^{1-2 n} \bigg) 
\eea
in the Fock space ${\cal F}_{\bf P}$ with $P_i=\frac{2p_i}{\sqrt{a_i}}$. 
(Here and below, we always assume that the parameters $\alpha_i$ and  $a_i$ are related as in \eqref{paaossasai}.)
In fact,  all other terms in  \eqref{ystaspoaskasususys} can be also understood as eigenvalues
of certain operators commuting with the local IM.

\subsection{\label{secnn}Corner-brane $W$-algebra and reflection operators}

Here we argue that the factor $\big(g^{(L,\alpha)}_j\big)^{\sigma'}\
 \big(g^{(L,\alpha)}_i\big)^{\sigma}$ in \eqref{sisiossaio} can be identified  with an  eigenvalue of certain 
$\lambda$-independent  operator 
$\mathbb{R}^{(k)}_{\sigma'\sigma}$ acting in the Fock ${\cal F}_{\bf P}$ space and commuting the local IM 
$\mathbb{I}_{2m-1}$:
\bea\label{aopsospospa}
\mathbb{R}^{(k)}_{\sigma'\sigma}\ |\,L,\alpha\,\rangle=\big(g^{(L,\alpha)}_j\big)^{\sigma'}\ 
\big(g^{(L,\alpha)}_i\big)^{\sigma}\, |\,L,\alpha\,\rangle\ .
\eea

The  operators  $\mathbb{R}^{(k)}_{\sigma'\sigma}$  are similar to the reflection operator from Ref.\cite{Zamolodchikov:1995aa}.
The main part in the  construction
belong to a $W$-algebra whose  r${\hat{\rm o}}$le is analogous to that
of the Virasoro algebra in the
quantum  Liouville theory. This $W$-algebra was introduced in
Ref.\cite{fate} and studied in Ref.\cite{Feigin:2001yq}. 
Bellow we closely follow the consideration from Ref.\cite{Lukyanov:2012wq}, 
where this $W$-algebra was called  ``corner-brane'' $W$-algebra.

Let us introduce  four  vectors
\bea\label{sospsopsa}
{\boldsymbol \alpha}_1&=&\ri\ \big(+\alpha_1,+\alpha_2,+\alpha_3\,\big)\nonumber\\
{\boldsymbol \alpha}_2&=&\ri\ \big(+\alpha_1,-\alpha_2,-\alpha_3\,\big)\nonumber\\
{\boldsymbol \alpha}_3&=&\ri\ \big(-\alpha_1,+\alpha_2,-\alpha_3\,\big)\\
{\boldsymbol \alpha}_4&=&\ri\  \big(-\alpha_1,-\alpha_2,+\alpha_3\,\big)\ ,\nonumber
\eea
and define the   exponential vertex operators  
\bea\label{sopsassaop}
V_{A}(u)=\re^{2 \boldsymbol \alpha_A\cdot {\boldsymbol \phi}}(u)\ \ \ \ \ \  (A=1,2,3,4)\ .
\eea
Here  $ {\boldsymbol \phi}=(\phi_1,\phi_2,\phi_3)$ is the
three-component chiral Bose field \eqref{sakshs} and the dot product stands for  ${\boldsymbol x}\cdot {\boldsymbol y}=
\sum_{i=1}^3 x_i y_i$.
We now   choose  the first three vectors ${\boldsymbol \alpha}_1$, ${\boldsymbol \alpha}_2$
and ${\boldsymbol \alpha}_3$
from the set \eqref{sospsopsa}
and
define the algebra ${\cal W}^{(1,2,3)}$ as an algebra generated by the holomorphic currents
$W_s$ of spin $s$ characterized by the condition that they commute with three ``screening charges''
\bea\label{soipssposa}
\oint_u\rd v\  W_s(u)\,V_{A}(v)=0\ \ \ \ \ \ \ \ \ \ \ (A=1,2,3)\ .
\eea
The integration here is taken over a small contour around the point $u$.
For small $s$ the condition \eqref{soipssposa} can be straightforwardly analyzed.
In particular,  one can show that  spin-1  currents satisfying \eqref{soipssposa} are absent,
but there is one (up to an overall multiplier)  spin-2 current  
\bea\label{usyopspsa}
W_2=\partial{\boldsymbol\phi}\cdot\partial{\boldsymbol\phi}+{\boldsymbol \rho}\cdot \partial^2{\boldsymbol\phi}\ ,
\eea
with
\bea\label{sopospaos}
{\boldsymbol \rho}=\frac{\ri}{2}\ \Big(\frac{1}{\alpha_1}, \frac{1}{\alpha_2}, \frac{1}{\alpha_3}\,\Big)\ ,
\eea
which generate the Virasoro subalgebra with the central charge
\bea\label{asopsopa}
c=3-6\,\sum_{i=1}^3\frac{1}{a_i}\ .
\eea
Furthermore, there are no non-trivial 
spin-3 currents since the spin-3 fields satisfying
\eqref{soipssposa} turns to be  the derivative $ \partial W_2$.
For spin-4  there are three fields -- two ``descendent'' currents $ \partial^2 W_2$ and $(W_2)^2$,  but
also one  new current $W_4$. Explicit form of $W_4$ is somewhat cumbersome and can be found in 
Appendix A of Ref.\cite{Lukyanov:2012wq}.
For $s>4$, the calculations based on definition \eqref{soipssposa} become very complicated. However one can
argue (see Ref.\cite{Lukyanov:2012wq}) 
that  there is exactly one independent current $W_{2n}$ at each even spin $s=2n$, having the form
\bea\label{saopasospaos}
W_{2n}=W^{(\rm sym)}_{2n}+\partial V_{2n-1}\ ,
\eea
where the non-derivative term $W^{(\rm sym)}_{2n}$  (but not $V_{2n-1}$) is symmetric with respect to all
$180^o$ rotation around the coordinate axes of the $(\phi_1,\phi_2,\phi_3)$ space:
\bea
\label{aaiiioas}
(\phi_1,\phi_2,\phi_3)\mapsto (\phi_1,-\phi_2,-\phi_3),\, \ \  (-\phi_1,\phi_2,-\phi_3),\, \ \  (-\phi_1,-\phi_2,\phi_3)\ .
\eea

The above construction can be repeated  for any choice of three vectors 
${\boldsymbol \alpha}_A$, ${\boldsymbol \alpha}_B$ and ${\boldsymbol \alpha}_C$ from the set  \eqref{sospsopsa}
to yield four corner-brane
$W$-algebras which are labeled by are triple integers $(A,B,C)$:\footnote{These $W$-algebras are 
naturally associated with four  corners of the pillow-brane from Ref.\cite{Lukyanov:2012wq}.
The notations $X^i$ and $\alpha_i$ from  Ref.\cite{Lukyanov:2012wq}
coincides with ours $\phi_i$ and $2\ri\, \alpha_i$, respectively.}
\bea\label{asopsosapaso}
{\cal W}^{(1,2,3)}\ ,\ \ \  {\cal W}^{(2,3,4)}\ ,\ \ \ {\cal W}^{(3,4,1)}\, ,\ \ \ \   {\cal W}^{(4,1,2)}\ .
\eea
To simplify formulae, 
bellow we will  use the   shortcut  notations
\bea\label{asopssapas}
{\cal W}^{(A)}\equiv {\cal W}^{(B,C,D)}\ ,\ \ \ \ \ {\rm where}\ \ \  \ \ \ (A,B,C,D)={\tt perm}(1,2,3,4)\ .
\eea

Of course,  all algebras ${\cal W}^{(A)}$ are isomorphic to  ${\cal
  W}^{(4)}\equiv{\cal W}^{(1,2,3)}$, differing from it only by
the way they are embedded in the Heisenberg algebra \eqref{alksu}. 
To be more precise, it is expected that for generic values of the
parameters there exist  
twelve   invertible linear operators
\bea\label{sapossasa}
{\mathbb R}^{(A,B)}\ :\ \ \ \  {\cal F}_{\bf P}^{(L)}\mapsto {\cal F}_{\bf P}^{(L)}\ \ \ \
\big(\,A,B=1,\ldots 4\, ,\ \ A\not=B\,\big)
\eea
satisfying the condition:
\bea\label{aisiosaissaio}
W^{(B)}_s(u)=
{\mathbb R}^{(A,B)}\ W^{(A)}_s(u)\ \big[{\mathbb R}^{(A,B)}\big]^{-1}
 \ \ \ \ \ \ \ \ (s=2,4,\ldots)\ .
\eea
It is also  expected that  the  whole ${\cal W}^{(A)}$-algebra is generated by
the spin-4 current $W^{(A)}_4(u)$, so that relations
\eqref{aisiosaissaio} for any $s$ follow from $s=4$ case. 
The operators \eqref{sapossasa} will be referred to bellow  as {\it
  reflection operators}. 

A rigorous proof of existence of the reflection operators is absent.
However, assuming that they are exist,
it is not difficult to  describe the   procedure which allows one to 
construct  them explicitly.

Let us denote the Fourier coefficients of the   $W_4^{(A)}$-currents
by ${\tilde W}^{(A)}_4(n)$
$(n\in\mathbb{Z}, \ A=1,2,3,4)$.
For generic values of the parameters the Fock space   possesses a natural structure
of  the highest weight irreducible  representation of 
the ${\cal W}^{(A)}$-algebra. It is expected that,
for a given $A$, each level subspace ${\cal F}^{(L)}_{\bf P}$ is spanned      on the  vectors  
\bea\label{aspoaspaosp}
{\tilde W}^{(A)}_4(n_1)\ldots {\tilde W}^{(A)}_4(n_M)\, |\,{\bf P}\,\rangle\ ,\ \ \ \ \ \ \ \ \ L=-\sum_{i=1}^Mn_i\ 
\ \ \ \ \ (n_i\in\mathbb{Z})\ ,
\eea
and
one can chose ${\cal N}_L$ linear independent vectors 
of the form \eqref{aspoaspaosp} to build the basis
in  ${\cal F}^{(L)}_{\bf P}$:
\bea\label{sossasapo}
\big\{w^{(A)}_{b}\big\}_{b=1}^{{\cal N}_L}\ :\ \ \  
w^{(A)}_{b}=
{\tilde W}^{(A)}_4(n_1)\ldots {\tilde W}^{(A)}_4(n_M)\, |\,{\bf P}\,\rangle\ .
\eea
(Here we use subscript $b$ to enumerate the basis vectors.) 
The  choice of the monomials ${\tilde W}^{(A)}_4(n_1)\ldots$ ${\tilde W}^{(A)}_4(n_M)$ in \eqref{sossasapo}
is not  particularly  important for us here. What's important is that for the given choice of monomials  one can build four
different bases in  ${\cal F}^{(L)}_{\bf P}$ corresponding to different values of $A=1,2,3,4$ and therefore,
using these bases,  one  can introduce  the set  of  linear  operators according to the rule
\bea\label{rsaopsosap}
{\mathbb R}^{( A,B)}\ :\ \ \ \ \ \ \ {\mathbb R}^{(A,B)}\, w^{(A)}_{b}= w^{(B)}_{b}\ .
\eea
Let 
\bea\label{saospsospa}
\{e_\beta\}_{\beta=1}^{{\cal N}_L}\ :\ \ 
e_\beta=a_{i_1}(-m_1)\ldots a_{i_M}(-m_N)\, |\,{\bf P}\,\rangle\ ,\ \ \ \ L=\sum_{i=1}^N m_{i}\ \ \ \ \ \ (m_i=1,2,\ldots)\ ,
\eea
be the basis in the level subspace ${\cal F}_{\bf P}^{(L)}$.
Then the $W$-basis \eqref{sossasapo} can be expressed in terms of the Heisenberg  states \eqref{saospsospa}:
\bea\label{asopssaop}
 w^{(A)}_{b}=\big(R^{(A)}\big)_b^\beta\ e_\beta\ .
\eea
The matrix of the linear operator ${\mathbb R}^{( A,B)}$
in the Heisenberg 
basis is given by
\bea\label{aspospsaopy}
{\mathbb R}^{(A,B)}\  e_\beta=\big(\big[R^{(A)}\big]^{-1}\big)^b_\beta\ \big(R^{(B)}\big)_b^{\beta'}\ e_{\beta'}\ .
\eea

In a view of Eq.\eqref{saopasospaos},  the operators
${\tilde W}^{(A)}_{2n}(0)=\int_0^{2\pi}\frac{{\rm d} u}{2\pi}\ W^{(A)}_{2n}(u)$ are elements of all  $W$-algebras
${\cal W}^{(A)}$. We introduce  special notations for these elements:
\bea\label{ksaoasais}
\big\{\,{\mathbb I}_{2n-1}\,\big\}_{n=1}^\infty\ :\ \ \ \ \ \
{\mathbb I}_{2n-1}:={\tilde W}^{(A)}_{2n}(0)=\int_0^{2\pi}\frac{\rd u}{2\pi}\ W^{\rm (sym)}_{2n}(u)\ .
\eea
Each operator from this set   is written  in the form  of integral over
the local density and commute with the reflection operators
\bea\label{sapoasaio}
[\,{\mathbb R}^{(A,B)},\,{\mathbb I}_{2n-1}\,]=0\ .
\eea
To prove the last relation, one should  rewrite 
Eq.\eqref{aisiosaissaio} in the form
\bea\label{aisiissaio}
W^{(B)}_{2n}(u)\ {\mathbb R}^{(A,B)}=
{\mathbb R}^{(A,B)}\ W^{(A)}_{2n}(u)\ ,
\eea
and then integrate both sides of the obtained relation
over the period.
Much of this work is 
based on the assumption that the operators ${\mathbb I}_{2n-1}$
form a maximal commuting set, despite that at the moment a rigorous
proof of  mutual commutativity 
of  ${\mathbb I}_{2n-1}$ defined by Eq.\eqref{ksaoasais} is lacking.

Let us illustrate the construction above by  the simplest $L=1$ case.
At the first level, there are three linear independent states and the Heisenberg  basis  is generated by the vectors
\bea\label{saospsapsaoaa}
e_i=a_{i}(-1)\, |\,{\bf P}\,\rangle\ \ \ \ \ \ \ \ \ \ \ \ (i=1,2,3)\ .
\eea
As for  $W^{(4)}$-basis, one can use  the following three linear independent vectors
\bea\label{saopsospa}
w_1={\tilde W}^{(4)}_4(-1)\, |\,{\bf P}\,\rangle\ ,\ \ \ \ \
w_2={\tilde W}^{(4)}_4(0)\, {\tilde W}^{(4)}_4(-1)\, |\,{\bf P}\,\rangle\ ,\ \ \ \
w_3={\tilde W}^{(4)}_4(1)\, {\tilde W}^{(4)}_4(-2)\, |\,{\bf P}\,\rangle\ .
\eea
An explicit form of  the $W_4$-current  for ${\cal W}^{(1,2,3)}\equiv{\cal W}^{(4)}$ algebra  is given
by formulae (A.1)-(A.5) in   Appendix A of Ref.\cite{Lukyanov:2012wq}.
Using  those formulae    one can calculate $(3\times 3)$-matrix
$\big(R^{(4)}\big)_b^\beta$, defined in
\eqref{asopssaop}.\footnote{We are grateful to A.V. Litvinov for
  writing a 
computer code for this calculation.} Having at hand an explicit expression for this matrix,  other matrixes 
$(R^{(A)})_b^\beta$ can be obtained by means of the formal substitutions
\bea\label{sopsaop}
(R^{(1)})_b^\beta=(R^{(4)})_b^\beta|_{{\alpha_1}\mapsto-{\alpha_1}\atop
{\alpha_2}\mapsto-{\alpha_2}}\ ,\ \ \ (R^{(2)})_b^\beta=(R^{(4)})_b^\beta|_{{\alpha_1}\mapsto-{\alpha_1}\atop
{\alpha_3}\mapsto-{\alpha_3}}\ ,\ \ \ (R^{(3)})_b^\beta=(R^{(4)})_b^\beta|_{{\alpha_2}\mapsto-{\alpha_2}\atop
{\alpha_3}\mapsto-{\alpha_3}}\ .
\eea
Then Eq.\eqref{aspospsaopy} allows one to construct  twelve  $3\times 3$-matrices $\big({\mathbb R}^{(A,B)}\big)_\beta^{\beta'}$
and then check the commutativity condition \eqref{sapoasaio}.

Returning to  general properties of the reflection operators, 
it can be easily seen from Eqs.\eqref{aspospsaopy}, \eqref{sapoasaio} that they are mutually commute 
\bea\label{asospsasaop}
{\mathbb R}^{(A,B)}\,{\mathbb R}^{(C,D)}={\mathbb R}^{(C,D)}\,{\mathbb R}^{(A,B)}
\eea
and satisfy the relations
\bea\label{aopssopa}
{\mathbb R}^{(A,B)}\,{\mathbb R}^{(B,A)}=1\ ,\ \ \  \ 
{\mathbb R}^{(A,C)}={\mathbb R}^{(A,B)} {\mathbb R}^{(B,C)}\ ,\ \ \ \  
{\mathbb R}^{(1,2)}{\mathbb R}^{(2,3)}{\mathbb R}^{(2,3)}{\mathbb R}^{(3,4)}=1\ .
\eea
For our purposes it is useful to enumerate twelve  reflections operators
in slightly different manner than in the definition \eqref{aisiosaissaio}.
Namely we define
\bea\label{sospsap}
{\mathbb R}^{(k)}_{\sigma'\sigma}:={\mathbb R}^{(A,B)}\ \ \ \ \ \ \ \ \ \ \ 
 \ (\sigma,\sigma'=\pm 1\ ,\ \ \ k=1,\,2,3)\ ,
\eea
by means of the following relations
\bea\label{asiosaisao}
&&{\mathbb R}^{(k)}_{++}={\mathbb R}^{(4,4-k)}\ \ \ \ \ (k=1,2,3)\nonumber\\
&&{\mathbb R}^{(1)}_{+-}={\mathbb R}^{(1,2)}\ ,\ \ \ {\mathbb R}^{(2)}_{+-}={\mathbb R}^{(3,1)}\ ,\ \ \ 
{\mathbb R}^{(3)}_{+-}={\mathbb R}^{(2,3)}\\
&&{\mathbb R}^{(k)}_{-\sigma',-\sigma}=
{\mathbb R}^{(B,A)}\ .\nonumber
\eea
Then,  Eqs.\eqref{sapoasaio},\,\eqref{asospsasaop} and \eqref{aopssopa} imply that  eigenvalues of 
the operators ${\mathbb R}^{(k)}_{\sigma'\sigma}$ in the
level subspace ${\cal F}_{\bf P}^{(L)}$ have  the form \eqref{aopsospospa}, where $g^{(L,\alpha)}_i$ $(i=1,2,3)$
stand for  some constants. 
Furthermore, explicit calculations in the case $L=1$ shows that 
$g^{(1,\alpha)}_i$ $(i=1,2,3;\,\alpha=1,2,3)$ are the same as those 
quoted in Footnote\,\ref{footn8}.
Notice  that  the square of  $\Lambda_i(\lambda)$ 
\eqref{ossaaosp} can be identified with eigenvalues of the following reflection operators acting in the 
level subspace ${\cal F}_{\bf P}^{(L)}$: 
\bea\label{ospsopsa}
&&\Big\{\,\big(\,\Lambda^{(\alpha)}_1(\lambda)\,\big)^2\,\Big\}_{\alpha=1}^{{\cal N}_L}= {\rm Spect}_{{\cal F}_{\bf P}^{(L)}}\Big[
\,\big(\, S(  p_1 |p_2+p_3)\,S(  p_1|p_2-p_3)\,\big)^{\frac{1}{2}}\ 
{\mathbb R}^{(4,1)}\,{\mathbb R}^{(3,2)}\,\Big]\nonumber\\
&&\Big\{\,\big(\,\Lambda^{(\alpha)}_2(\lambda)\,\big)^2\,\Big\}_{\alpha=1}^{{\cal N}_L}=
 {\rm Spect}_{{\cal F}_{\bf P}^{(L)}}\Big[\,\big(\, S(  p_2 |p_3+p_1)\,S(  p_2|p_3-p_1)\,\big)^{\frac{1}{2}}
\ {\mathbb R}^{(4,1)}\,{\mathbb R}^{(2,3)}\,\Big]\\
&&\Big\{\,\big(\,\Lambda^{(\alpha)}_3(\lambda)\,\big)^2\,\Big\}_{\alpha=1}^{{\cal N}_L}=
 {\rm Spect}_{{\cal F}_{\bf P}^{(L)}}\Big[\,\big(\, S(  p_3 |p_1+p_2)\,S(  p_3|p_1-p_2)\,\big)^{\frac{1}{2}}
\ {\mathbb R}^{(4,2)}\,{\mathbb R}^{(1,3)}\,\Big]\ ,\nonumber  
\eea
where $S(p_i|q)$ is given by \eqref{asosaosapao} and $P_i$ related to $p_i$ as in
\eqref{osososap}, i.e.,
$P_i=\frac{2p_i}{\sqrt{a_i}}$.

\subsection{Large-$\lambda$ asymptotic expansion and  dual non-local IM}

Here we discuss  the formal asymptotic series $X_\sigma^{(i)}(\lambda)$ \eqref{sospospsoa} which appears in the
large-$\lambda$ asymptotic expansion of the connection coefficients $A_{\sigma'\sigma}^{(k)}(\ri\lambda)$
\eqref{ystaspoaskasususys}.

In the previous section we have described the characteristic property of the local IM -- they 
are integrals over the local densities $ W_{2n}^{({\rm sym})}(u)$ 
satisfying the conditions
\bea\label{isuoipssposa}
\oint_u\rd v\  W_{2n}^{({\rm sym})}(u)\,V_{A}(v)=\partial_u F^{(A)}_{2n-1}
\eea
for $A=1,2,3,4$, where  the
vertex operators  $V_A$ given by \eqref{sopsassaop}
and $ F^{(A)}_{2n-1}$ are some local fields.
In fact, there exists another set of vertex operators satisfying
similar conditions
\cite{fate,Feigin:2001yq}. 
Namely,
consider six vertex operators
\bea\label{isuasopsopsosasp}
V^{(i)}_\pm (u)= \big(\,  \alpha_{k}\, \partial\phi_{k}
\pm \alpha_{i}\, \partial\phi_{i} 
\pm \alpha_{j}\, 
\partial\phi_{j}\, \big)\ \re^{\pm \frac{\ri\phi_i}{\alpha_i}}(u)\ 
\eea
$\big(\,(i,j,k)={\tt cyclic\ perm}(1,2,3)\,\big),$
then using  explicit formulae for the first two $W$-currents, it is
straightforward to check that for $m=1$ and $m=2$ 
\bea\label{drisuoipssposa}
\oint_u\rd v\  W_{2n}^{({\rm sym})}(u)\,V^{(i)}_{\pm}(v)=\partial_u {\tilde F}^{(i,\pm)}_{2n-1}\ .
\eea
We expect, that both Eqs.\eqref{isuoipssposa} and
\eqref{drisuoipssposa} hold for any $m=1,2\ldots\infty$. 
Let us  introduce  the following notations for the integrals of the vertex operators \eqref{isuasopsopsosasp}:
\bea\label{sopsaosap}
{\tilde x}^{(i)}_{0}=\int_0^{2\pi} \rd u\ V^{(i)}_- (u)\ ,\ \ \ \ \ \ \ {\tilde x}^{(i)}_{1}=
\int_0^{2\pi} \rd u\ V^{(i)}_+ (u)\ .
\eea
Repeating the calculations from Ref.\cite{Bazhanov:1998dq},
one can show that 
$\big({\tilde x}^{(i)}_{0},\,{\tilde x}^{(i)}_{1}\big)$ satisfy the
Serre relations for  the quantum Kac-Moody algebra
$U_{{\tilde q}_i}\big(\widehat{{sl}}(2)\big)$: 
\bea\label{saosopsapas}
\big({\tilde x}^{(i)}_a\big)^3\,
{\tilde x}^{(i)}_b-[3]_{{\tilde q}_i}\ 
\big({\tilde x}^{(i)}_a\big)^2\,
{\tilde x}^{(i)}_b\, {\tilde x}^{(i)}_a  
+[3]_{{\tilde q}_i}\  {\tilde x}^{(i)}_a\, {\tilde x}^{(i)}_b\, \big({\tilde x}^{(i)}_a\big)^2-
{\tilde x}^{(i)}_b\, \big({\tilde x}^{(i)}_a\big)^3=0
\ \ \ \ (a,b=0,1)\ ,
\eea
where
\bea\label{susyasopsosap}
{\tilde q}_i=\re^{\ri\pi(1+\frac{1}{a_i})}
\eea
and  the conventional notation  $[n]_q=(q^n-q^{-n})/(q-q^{-1})$ is applied.
We may now employ  the whole machinery developed in the work \cite{Bazhanov:1998dq}, to construct 
families  of  mutual commuting   operators
which are also commute with the local IM.
In particular,  let us introduce
the operators
\bea\label{aspaopaossposa}
\mathbb{X}^{(i)}_\pm (\lambda_i)&=&
Z^{-1}_\pm( P_i)\ {\rm Tr}_{\rho_\pm}\bigg[\, \re^{\pm \frac{\pi\ri P_i}{2\alpha_i}\, {\cal H}^{(i)}}\times\\
&&\ 
{\cal P}\,\exp\Big({\lambda}_i\int_0^{2\pi}\rd u\ \big(\,V^{(i)}_-(u)\ 
{\tilde q}_i^{\pm\frac{{\cal H}^{(i)}}{2}}{\cal E}^{(i)}_\pm+
V^{(i)}_+(u)\ {\tilde q}_i^{\mp\frac{{\cal H}^{(i)}}{2}}{\cal E}^{(i)}_\mp\,\big)\Big)\,\bigg]\  .\nonumber
\eea
Here $\rho_\pm$ are representations of the so-called $q$-oscillator algebra generated by the elements
${\cal H}^{(i)}$, ${\cal E}^{(i)}_+$, ${\cal E}^{(i)}_-$ subject to the relations
\bea\label{aasaosao}
{\tilde q}_i\, {\cal E}^{(i)}_+\,{\cal E}^{(i)}_- -
{\tilde q}^{-1}_i\, 
{\cal E}_-^{(i)}\,{\cal E}_+^{(i)}=
\frac{1}{{\tilde q}_i-{\tilde q}_i^{-1}}\ ,
\ \ \ [{\cal H}^{(i)},\,{\cal E}^{(i)}_\pm]=\pm 2\, {\cal E}^{(i)}_{\pm}\ ,
\eea
and such that the traces
\bea\label{aspoapsaosaasp}
Z_\pm( P_i)={\rm Tr}_{\rho_\pm}\bigg[\, \re^{\pm \frac{\pi\ri P_i}{2\alpha_i}\, {\cal H}^{(i)}}\,\bigg]
\eea 
exist and do not vanish for complex $P_i$ belonging to
the lower half plane  $\Im m (P_i)<0$.
The operator  $\mathbb{X}^{(i)}_\pm(\lambda_i)$ can be understood as the series expansion in $({\lambda}_i)^2$
\bea\label{saoipsasppsaposa}
\mathbb{X}^{(i)}_\sigma(\lambda_i)=1+\sum_{m=0} \mathbb{X}^{(i)}_{ \sigma,n}\ ({\lambda}^2_i)^{n}\ \ \ \ \ \ \ \ \ \ 
\ (\sigma=\pm)\  , 
\eea
and, as it follows from the result of Ref.\cite{Bazhanov:1998dq},
each of the expansion coefficient commute with the local IM
\bea\label{apospaopsaoap}
\mathbb{X}^{(i)}_{ \sigma,n}\ \ \ :\ \ \ \ {\cal F}_{\bf P}\mapsto {\cal F}_{\bf P}\ ,\ \ \ \ \ \
[\mathbb{X}^{(i)}_{ \sigma,n},\,\mathbb{I}_{2m-1}]=0\ .
\eea
In fact, the definition \eqref{aspaopaossposa} and the series expansion \eqref{saoipsasppsaposa}
can be applied literally only  within the domain
\bea\label{saopsasasaop}
-2<\Re e(a_i)<-1\ .
\eea
In this case all the matrix elements of $\mathbb{X}^{(i)}_{ \sigma,n}$
are  represented by  convergent  $2n$-fold ${\cal P}$-ordered
integrals.
Furthermore,  all the  matrix elements are  entire functions of $(\lambda_i)^2$ in this case. 
Following the approach developed in Ref.\cite{Bazhanov:1998dq}, one can show  that  ${\cal P}$-ordered integrals
$\mathbb{X}^{(i)}_{ \sigma,n}$ can be always rewritten as   contour integrals. The such representation
allows one to define  the operators $\mathbb{X}^{(i)}_{ \sigma,n}$ outside the domain \eqref{saopsasasaop}
through the analytical continuation. In particular,  the action of $\mathbb{X}^{(i)}_{ \sigma,n}$ 
can be defined in the  domain
of our  current interest, i.e.  for $0<a_i<2$. 
Following the terminology of Ref.\cite{Bazhanov:1996dr} we will referred to
 $\mathbb{X}^{(i)}_{ \sigma,n}$  in the domain  $0<a_i<2$
as {\it dual non-local} integrals of motion.
Notice that, the possibility of  analytical continuation of the coefficients in the expansion \eqref{saoipsasppsaposa}
does not  necessarily imply the convergency of the series.
As $0<a_i<2$, Eq.\eqref{saoipsasppsaposa} should be understood as a formal series expansion with
zero 
radius of convergence.\footnote{
In the domain  $ a_3<-2$   the operators 
$\mathbb{X}^{(3)}_{ \sigma}$  \eqref{aspaopaossposa} were studied in  Ref.\cite{Lukyanov:2006gv}.}

The analytical  calculation of the spectrum of  dual non-local IM 
is a complicated unsolved problem. An explicit result can be obtained
only for the vacuum eigenvalue 
of the first IM, $\mathbb{X}^{(i)}_{ \sigma,1}$. It suggests that, in all likelihood,
the formal asymptotic series $X_\sigma^{(i)}(\lambda)$ \eqref{sospospsoa} coincides with 
the eigenvalue of the
formal operator \eqref{saoipsasppsaposa} provided the  following relation between the expansion parameters
holds
\bea\label{sopssposap}
{\lambda}_i=\frac{1}{\Gamma(-\frac{1}{a_i})}\ 
 \Big(\frac{\lambda}{a_i}\Big)^{-\frac{1}{a_i}}\ .
\eea

\subsection{Relation to  quantum superalgebra 
$U_q\big(\widehat{D}(2,1;\alpha)\big)$}

The appearance of the exponential fields $V_A(u)$ and the relation 
\eqref{isuoipssposa} suggests a strong connection of our problem with
the quantized exceptional affine superalgebra
$U_q\big(\widehat{D}(2,1;\alpha)\big)$. This algebra is generated
by twelve elements
$h_1,h_2,h_3,h_4$, $x_1,x_2,x_3,x_4$ and 
$y_1,y_2,y_3,y_4$. An unusual feature of this superalgebra 
is that its Cartan matrix 
\beq
\Vert C_{A,B}\Vert=\left(\begin{array}{cccc}
0&1&\alpha&-1-\alpha\\[.2cm]
1&0&-1-\alpha&\alpha\\[.2cm]
\alpha&-1-\alpha&0&1\\[.2cm]
-1-\alpha&\alpha&1&0\end{array}\right)
\eeq
contains an arbitrary (complex) parameter $\alpha$. 
So, that together with the ``deformation'' parameter $q$ the algebra   
$U_q\big(\widehat{D}(2,1;\alpha)\big)$ has two continuous parameters.
For our purposes it is convenient to connect these parameters to 
our constants $a_1$, $a_2$ and $a_3$, used before in \eqref{qi-def}, 
\beq
q=\re^{\ri\pi a_1}\,,\qquad q^\alpha=\re^{\ri\pi a_2}\,,\qquad q^{-1-\alpha}=
\re^{\ri\pi a_3}\,,
\eeq
and introduce additional notations 
\beq
q_{AB}=-\exp\big({2 \pi\ri\, {\boldsymbol \alpha}_A\cdot {\boldsymbol
  \alpha}_B}\big)\ .
\eeq
The generating elements of the algebra satisfy the following
commutation relations
\cite{Feigin:2001yq,Yamane} 
\bea
&&[h_A,h_B]=0\,,\qquad [h_A, x_B] = C_{AB}\, x_B\,,\qquad [h_A, y_B] =-
C_{AB}\, y_B\nonumber\\
&&x_A \, y_B + y_B  \, x_A=\delta_{AB}\ 
\,\frac{q^{h_A}-q^{-h_A}}{q-q^{-1}}\, \qquad\ \ \ \ (A,B=1,2,3,4)\ ,  \label{xy-comm}  
\eea
and also the Serre relations 
\beq
x_A^2=y_A^2=0
\label{ferm}
\eeq
and for any triple $(A,B,C)$ with $A,B,C$ all
different\footnote{
If any two of indices in \eqref{ser-1}
coincide the relation trivially reduces to \eqref{ferm}.} 
\bea
&&[q_{AB}](x_A x_C x_B- x_B x_C x_A)
+[q_{BC}](x_B x_A x_C- x_C x_A x_B)
+[q_{CA}](x_C x_B x_A- x_A x_B x_C)=0\,,\nonumber\\
[.3cm]
&&[q_{AB}](y_A \,y_C\, y_B- y_B \,y_C\, y_A)
+[q_{BC}](y_B\, y_A\, y_C- y_C \,y_A\, y_B)
+[q_{CA}](y_C\, y_B\, y_A- y_A\, y_B\, y_C)=0\,,\nonumber
\\
\label{ser-1}
\eea
where
\be
[q]=q-1/q\ .
\ee
The last two relations are totally symmetric under any permutations of
$A,B,C$, so there only four pairs different relations \eqref{ser-1} with
$(A,B,C)=(1,2,3),\ (1,2,4),\ (1,3,4)$, or $ (2,3,4)$. 

Remarkably, as shown in \cite{Feigin:2001yq}, the integrals of the
vertex operators 
\beq
x_A=\int_0^{2\pi}\rd u\,  V_A(u)
\eeq
satisfy the above Serre relations \eqref{ferm},
\eqref{ser-1}. Therefore, one could again execute the program of
the work \cite{Bazhanov:1998dq}
$\big($now based on the quantum affine algebra
$U_q(\widehat{D}(2,1;\alpha))$\,$\big)$  
and define families of commuting transfer matrices, which are entire
functions of the variable $\lambda^2$ and  act directly in
the Fock  spaces discussed in Section\,\ref{jassysa}. Moreover, in
view of the relation \eqref{isuoipssposa}, these transfer matrices
will commute with all local integrals of motion
${\mathbb I}_{2n-1}$. 
This direction,
however, would requires a lot of additional work, since, to our
knowledge, the representation
theory of $U_q(\widehat{D}(2,1;\alpha))$ is not sufficiently studied. 
Nevertheless, we expect that the values of the Wilson loop ${\mathcal
  W}(\lambda)$ \eqref{osposoappsa} for various
PGHO's will be given by eigenvalues of the transfer matrix obtained as
a trace over some finite-dimensional representation of
$U_q(\widehat{D}(2,1;\alpha))$. Moreover, we expect that 
the corresponding values of the
connection coefficients $A^{(k)}_{\sigma'\sigma}(\lambda)$ \eqref{sospsosapo} will be given
by eigenvalues of appropriate analogs of Baxter ${ Q}$-operators,
obtained as traces over some special oscillator-type representations, first
introduced in \cite{Bazhanov:1996dr} in the context of $U_q(\widehat{sl}(2))$. 

As a
justification of the above picture, 
consider the value ${\mathcal W}_0$, given by
\eqref{opasospaos}. This expression looks like a diagonal character
      ${\rm Tr}\big[\, \exp(\sum_A\beta_A h_A)\,\big]$ of 
an 18-dimensional representation (indeed, it contains 18 exponential
terms, where some exponents vanish). Quite excitingly, Zengo Tsuboi
has pointed out that his calculations with analytic Bethe Ansatz 
suggest \cite{Tsuboi} 
that the algebra $U_q(\widehat{D}(2,1;\alpha))$ does
indeed have an 18-dimensional representation. Note, that the
  corresponding non-affine algebra has only 17-dimensional representation,
but there is no an ``evaluation map'', so that dimensions of
representations of the affine and
non-affine algebras should not necessarily coincide in this case 
\cite{Tsuboi}. We hope to return to this interesting 
question in the future, 
 as well as to
  other question relevant to an algebraic construction of commuting transfer  
matrices in this case.

\section{\label{sec7}MShG equation and  the auxiliary  linear problem}

\subsection{Complex solutions of MShG}

We now  turn to further development of the concept of PGHO, where 
the central r${\hat{\rm o}}$le is played by the 
modified sinh-Gordon (MShG) equation
\bea\label{oaisoaoso}
\partial_z\partial_{\bar z}\eta-\re^{2\eta}+ \rho^4\ {\cal P}(z){\bar {\cal P}}({\bar z})\, \re^{-2\eta}=0\ ,
\eea
where ${\cal P}(z)$ is still  given by \eqref{oasioaq}, i.e.,
$${\cal P}(z)=\frac{(z_3-z_2)^{a_1}\,(z_1-z_3)^{a_2}\,(z_2-z_1)^{a_3}}
{(z-z_1)^{2-a_1}(z-z_2)^{2-a_2}(z-z_3)^{2-a_3}}\ ,
$$
${\bar {\cal P}}({\bar z})$ stands for complex conjugate 
of ${\cal P}(z)$ and $\rho$ is an arbitrary constant.
For $\rho=0$,  this partial differential equation  reduces to the Liouville equation,  $\partial_z\partial_{\bar z}\eta-\re^{2\eta}=0$.
In what follows,  the field $\eta$ is understood as a solution
of the MShG equation, rather than the  Liouville equation.
 
The subject of our interest are  solutions of \eqref{oaisoaoso}, which can be thought as
``$\rho$-deformation''
of  the  complex solutions of  the Liouville equation from Section\,\ref{Liouv}.
To describe their  properties  it is still convenient to employ 
the function $\re^{-\eta}$\ (see Eq.\eqref{apoososa}).
As before, we  assume that $\re^{-\eta}$ is a   smooth, single valued complex function without zeroes  
on the sphere with $3+L+{\bar L}$ punctures.
Since  $z=\infty$ is a regular point on the Riemann sphere,  $\re^{-\eta}$ satisfy the
asymptotic condition  $\re^{-\eta}\sim |z|^{2}$  as $z\to\infty$.
The asymptotic  behavior  at the punctures  $z=z_1,z_2,z_3$ are given  by the  same  formulae as  \eqref{aospsao},
i.e.,
$\re^{-\eta}\sim |z-z_i|^{-2m_i}$ as $|z- z_i|\to 0$. 
Notice that as $m_i<-\frac{1}{4}\,(2-a_i)$ the first term in the r.h.s. of \eqref{oaisoaoso}
dominates as $|z-z_i|\to  0$. Therefore, the term $\propto  \re^{-2\eta}$
can be neglected for  sufficiently small  $|z-z_i|$ and we return to the Liouville equation.
From the 
other hand, it is easy to see that in the case of the Liouville equation,  
the parameters $m_i$ should be bounded from  below.
For this reason we assume that the constraints
$-\frac{1}{2}< m_i<-\frac{1}{4}\,(2-a_i)$ are enforced. 

In the case of the Liouville field the behavior at the punctures 
$z=x_a$ $(a=1,\ldots L)$ and ${\bar z}={\bar y}_b$ $(b=1,\ldots {\bar L})$ 
are given by \eqref{sopossaposa}.  This  singular behavior
consistent with the Liouville dynamics, however
the   term $\propto \re^{-2\eta}$ in  the MShG equation \eqref{oaisoaoso} can not be treated as a small perturbation 
in the vicinity 
$z=x_a$  and ${\bar z}={\bar y}_b$ ---
it essentially modifies  the  singular behavior at  these points.
A brief analysis of \eqref{oaisoaoso}  suggests to replace \eqref{sopossaposa} 
by
$\re^{-\eta}\sim \frac{{\bar z}-{\bar
    x}_a}{z-x_a}$ and
$\re^{-\eta}\sim \frac{{ z}-{ y}_b}{{\bar z}-{\bar
    y}_b}$, i.e. the  asymptotic formulae \eqref{sopasopspoa} from the introduction.

We also  impose   certain ``monodromy-free'' constraints on
positions of the punctures \eqref{sopasopspoa}.
For this purpose,  let us recall that  the MShG equation
constitute the flatness condition 
for $\mathfrak{sl}(2)$ connection \eqref{ystopsso}. 
Suppose $ {\boldsymbol \Psi}$ is  a  general  solution of the associated linear
problem $(\partial_z-{\boldsymbol { A}}_z){\boldsymbol \Psi}=(\partial_{\bar z}-{\boldsymbol { A}}_{\bar z}){\boldsymbol \Psi}=0$.
The  monodromy-free constraints mean  that $ \re^{\pm \frac{1}{2}\eta\sigma_3}\ {\boldsymbol \Psi}$
is single-valued in the neighborhood of the points
 $z=x_a$ $(a=1,\ldots L)$ and ${\bar z}={\bar y}_b$ $(b=1,\ldots {\bar L})$.
(Notice  that, since  $\re^{-\eta}$
is a single valued function,  the factor $\re^{\pm \frac{1}{2}\eta\sigma_3}$
does  not essentially affect 
the monodromy properties -- its r${\hat {\rm o}}$le is clarified by the forthcoming consideration.)

It is not difficult to reformulate the  monodromy-free constraints as
local conditions at the punctures  imposed on the MShG 
field $\eta$ \cite{Feigin:2007mr}. For this purpose,  it is useful to
rewrite the matrix differential operators in \eqref{asopssaopasopq} in the form
\bea\label{aopssapso}
\partial_z-{\boldsymbol { A}}_z=
\lambda^{-\frac{1}{2}\sigma_3}\ 
\re^{\frac{1}{2}\eta\sigma_3}\  {\boldsymbol {\cal D}}\ \re^{-\frac{1}{2}\eta\sigma_3}\,
\lambda^{\frac{1}{2}\sigma_3} \ ,\ \ \ \ \ \ \ 
\partial_{\bar z}-{\boldsymbol { A}}_{\bar z}={\bar \lambda}^{\frac{1}{2}\sigma_3}\
\re^{-\frac{1}{2}\eta\sigma_3}\ {\bar  {\boldsymbol {\cal D}}}\ \re^{\frac{1}{2}\eta\sigma_3}\,
{\bar \lambda}^{-\frac{1}{2}\sigma_3}\ ,
\eea
where
\bea\label{sopsso}
{\boldsymbol {\cal D}}&=&
\partial_z+\partial_z\eta\,\sigma_3-\lambda\ \big(\,\sigma_++\sigma_-\ {\cal P}(z)\,\big)\\
\bar{\boldsymbol {\cal D}}&=&
\partial_{\bar z}-\partial_{\bar z}\eta\,\sigma_3-{\bar \lambda}\ \big(\,\sigma_-+\sigma_+\ {\bar {\cal P}}({\bar z})\,\big)\  .\nonumber
\eea
Let us focus on the differential operator ${\boldsymbol {\cal D}}$ in the vicinity of the puncture  $z=x_a$, where
\bea\label{apoaapsoa}
\partial_z\eta\to \frac{1}{z-x_a}+f_a\ \ \ \ \ \ (f_a={\rm const})\ .
\eea
It can be easily seen  that as $z\to x_a$
\bea\label{ospssspaoussyyst}
{\boldsymbol C}^{-1}_a\, {\boldsymbol  {\cal D}}\, {\boldsymbol C}_a=
\partial_z+\, \big(\lambda {\cal P}(x_a)\big)^{-1}\ \frac{2 f_a-\gamma(x_a)}{z-x_a}\ \sigma_++O(1)\ .
\eea
Here we use the notation
\bea\label{osasao}
\gamma(z)=\partial_z\log {\cal P}(z)\ ,
\eea
and
the gauge transformation is performed by the  singular, but single-valued at $z=x_a$, matrix
\bea\label{sospsao}
{\boldsymbol C}_a=
\begin{pmatrix}
1&(\lambda {\cal P}(x_a))^{-1}\ \frac{1}{z-x_a}\\
0&1
\end{pmatrix}\ .
\eea
Hence for
\bea\label{sopsaospa}
f_a=\frac{1}{2}\ \gamma(x_a)\ ,
\eea
${ \boldsymbol {\cal  D}}$ is gauge equivalent to  a  nonsingular at $z=x_a$ differential operator. 
Similarly in the case
\bea\label{usysapoaapsoa}
\partial_{\bar z}\eta\to-\frac{1}{{\bar z}-{\bar x}_a}+{\bar g}_a\ \ \ \ \ \ ({\bar g}_a={\rm const})\ ,
\eea
one can consider the gauge transformation
\bea\label{usysospssspao}
{\boldsymbol {\bar C}}_{a}^{-1}\
{\boldsymbol {\bar {\cal D}}}\ {\boldsymbol {\bar C}}_{a}=
\partial_{\bar z}- {\bar \lambda}^{-1}\ \frac{2\, {\bar g}_a}{ {\bar z}-{\bar x}_a}\ \sigma_++O(1)
\eea
with
\bea\label{aospsa}
 {\boldsymbol {\bar C}}_{a}=
\begin{pmatrix}
1&
\frac{{\bar \lambda}^{-1}}{{\bar z}-{\bar x}_a}\\
0&  1
\end{pmatrix} \ .
\eea
Therefore, as
\bea\label{siospsai}
{\bar g}_a=0\ ,
\eea
${\boldsymbol {\bar {\cal  D}}}$ is gauge equivalent to a
regular  at ${\bar z}={\bar x}_a$ differential operator.
An immediate consequence of our analysis is that 
Eqs.\eqref{sopsaospa} and \eqref{siospsai} constitute the single-valuedness  condition
for
$  \re^{\pm \frac{1}{2}\eta\sigma_3}\ {\boldsymbol\Psi}$ at $z=x_a$.
Of course, similar consideration can be done for   the  punctures
at ${\bar z}={\bar y}_b$.

The partial differential equation \eqref{oaisoaoso} is  invariant with respect of  to 
$\mathbb{PSL}(2,\mathbb{C})$ transformation. One can use this symmetry to sent
the  punctures $(z_1,z_2,z_3)$ to any positions. Then 
we expect that, for a given   triple ${\bf m}=(m_1,m_2,m_3)$ \eqref{asioasiso} and
pair  $(L,{\bar L})$,
the MShG equation possesses a  finite  set  ${\cal A}^{(L,{\bar L})}_{\bf m}$ of solutions
such that  $\re^{-\eta}$ is a  smooth, single valued complex function without zeroes 
on the punctured Riemann sphere, whereas 
$\eta$
satisfy Eqs.\eqref{sospsaosapyst}-\eqref{sopasopspoa},\,\eqref{isuposksi},\,\eqref{saopsosap}.

\subsection{Conserved charges for MShG on the punctured sphere}

As it was already explained in the introduction, the elements of  ${\cal A}^{(L,{\bar L})}_{\bf m}$
can be characterized by means of the   set of  conserved charges $\{{\mathfrak q}_{2n-1},{\bar {\mathfrak q}}_{2n-1}\}_{n=1}^\infty$
generated by the asymptotic expansions \eqref{aaosioasao}  of the Wilson loop.
These conserved quantity  are given by the integrals 
\bea\label{osapsospa}
{\mathfrak q}_{2n-1}=\oint_{\gamma_P}\omega_{2n}\ ,\ \ \ \ \ \ \ \ \ \
{\bar {\mathfrak q}}_{2n-1}=\oint_{{\bar \gamma}_P}{\bar \omega}_{2n}\ ,
\eea
where $\{\omega_{2n-1},{\bar {\omega}}_{2n-1}\}_{n=1}^\infty$ 
constitute an infinite  hierarchy of  one-forms, which are closed by virtue of the MShG equation only,
\bea\label{spsapsaos}
\rd \omega_{2n}=
\rd {\bar \omega}_{2n}=0\  .
\eea
Explicit formulae for $\{\omega_{2n-1},{\bar {\omega}}_{2n-1}\}_{n=1}^\infty$ are not particular important for us.
They can be found in Ref.\cite{Lukyanov:2013wra} (Here
we closely follow notations  from  this paper.) It is useful to mention
that $\omega_{2n}$
are usually normalized by
the condition
\bea\label{aopssap}
\omega_{2n}=\rho^{1-2n}\
\Big(\, \big({\cal P}(z)\big)^{\frac{1}{2}-n}\,(\partial_z\eta)^{2n}+\ldots\,\Big)\, \rd z+
\big(\,\ldots\,\big)\ \rd {\bar z}\ ,
\eea
where dots in the first bracket involves terms with  higher derivatives of $\partial_z\eta$  and/or ${\cal P}(z)$.
Similarly
\bea\label{aopssapissu}
{\bar \omega}_{2n}=
\rho^{1-2n}\ \Big(\, \big({\bar {\cal P}}({\bar z})\big)^{\frac{1}{2}-n}\,(\partial_{\bar z}\eta)^{2n}+
\ldots\,\Big)\, \rd {\bar z}+
\big(\,\ldots\,\big)\ \rd { z}\ .
\eea
The one-forms  are not single-valued on the punctured sphere
due to the presence of  the multivalued functions  ${{\cal P}(z)},\,  {{\bar {\cal P}}({\bar z})}$.
However,  the restriction of $\omega_{2n}$ to the Pochhammer  loop
depicted in   Fig.\ref{fig1av}
are  single-valued and  the integrals \eqref{osapsospa}
are  not sensitive to continuous deformations of the contour.
(The second  integral in \eqref{osapsospa} is taken over the contour
${\bar \gamma}_P$ which is complex conjugate $\gamma_P$.)

\subsection{\label{alksals}Relation to PGHO}

Here we describe a relation  between  the linear problem associated with the
complex solutions from the set ${\cal A}^{(L,{\bar L})}_{\bf m}$ 
and the  Perturbed Generalized Hypergeometric Opers.

It is well known, the matrix linear problem 
$(\partial_z-{\boldsymbol { A}}_z){\boldsymbol \Psi}=(\partial_{\bar z}-{\boldsymbol { A}}_{\bar z}){\boldsymbol \Psi}=0$
can be reduced to second order linear differential operators.
The relations \eqref{aopssapso},\,\eqref{sopsso} allows one to write its general solution 
as
\bea\label{ilslsa}
{\boldsymbol \Psi}=
\begin{pmatrix}
 \re^{\frac{\eta}{ 2}}\ \psi\\
 \re^{-\frac{\eta}{ 2}}\ 
(\partial_z+\partial_z\eta)\, \psi
\end{pmatrix}=
\begin{pmatrix}
  \re^{-\frac{\eta}{ 2}}\   ( \partial_{\zb}+
\partial_{\zb} \eta)\,   {\bar \psi}\\ 
\re^{\frac{\eta}{ 2}}\ 
{\bar\psi}
\end{pmatrix}\ ,
\eea
where $\psi$ and ${\bar \psi}$ solve the equations
\bea\label{sksaksayst}
&&\big[\, \partial_{z}^2-u(z,{\bar z})-\lambda^2\ \ {\cal P}(z)\, \big]\ \psi=0\\
&&
\big[\, \partial^2_{\zb}-{\bar u}(z,{\bar z})-{\bar \lambda}^2\ \ 
{\bar  {\cal P}}({\bar z})\, \big]\ {\bar \psi}=0\ ,\nonumber
\eea
with
\bea\label{aspospsaop}
u(z,{\bar z})=(\partial_z\eta)^2-\partial^2_z\eta\ ,\ \ \ \ \ \ \ \ 
{\bar u}(z,{\bar z})=(\partial_{\bar z}\eta)^2-\partial^2_{\bar z}\eta\ .
\eea
In the vicinity of the monodromy-free puncture  $z=x_a$
\bea\label{sososap}
u(z,{\bar z})=\frac{2}{(z-x_a)^2}+\frac{\gamma_a}{z-x_a}+O(1)\ ,
\eea
whereas  the monodromy-free conditions \eqref{isuposksi} imply that
\bea\label{sosossa}
\gamma_a\ \big(\,\gamma^2_a-4\,u^{(0)}_{a}\,\big)+4\,u^{(1)}_{a}=0\ ,\ \ \ \ \ \ a=1,\ldots L\ ,
\eea
where $u^{(0)}_{a}$ and $u^{(1)}_{a}$ are defined through the expansion
\bea\label{sosospsa}
u(z,{\bar z})=\frac{2}{(z-x_a)^2}+\frac{\gamma_a}{z-x_a}+u_a^{(0)}+
u_a^{(1)}\ (z-x_a)+O\big((z-x_a)^2\big)\ .
\eea
Notice that $u(z,{\bar z})$ remains finite at the monodromy-free punctures at
$z=y_b$ $(b=1,\ldots {\bar L})$.
Similarly the field ${\bar u}(z,{\bar z})$ is nonsingular at $z=x_a$ $(a=1,\ldots { L})$, whereas
\bea\label{sossusap}
{\bar u}(z,{\bar z})=\frac{2}{({\bar z}-{\bar y}_b)^2}+\frac{{\bar \gamma}_b}{{\bar z}-{\bar y_b}}+
{\bar u}_b^{(0)}+
{\bar u}_b^{(1)}\ ({\bar z}-{\bar y}_b)+O\big(({\bar z}-{\bar y}_b)^2\big)
\ \ \  \ \ {\rm as}\ \ \ {\bar z}\to {\bar y}_b
\eea
and
\bea\label{isusosossa}
{\bar \gamma}_b\ \big(\,
{\bar \gamma}^2_b-4\,{\bar u}_b^{(0)}\,\big)+4\,{\bar u}_b^{(1)}=0\ ,\ \ \ \ \ \ b=1,\ldots {\bar L}\ .
\eea

We may now consider the limit $\rho\to 0$. Contrary to the MShG field, the composite fields  $u(z,{\bar z})$
and ${\bar u}(z,{\bar z})$
admit  small-$\rho$ perturbative  expansion even in the vicinity of the monodromy-free punctures.
It can be easily seen  that, with the identification
\bea\label{aospsoap}
p_i=m_i+{\textstyle\frac{1}{2}}\ ,
\eea
the limiting form  of  $u(z,{\bar z})$  coincides with
holomorphic potential $T_L(z)$  \eqref{oaposaosa}:
\bea\label{apospaopsa}
\lim_{\rho\to 0}u(z,{\bar z})=T_{L}(z)\ ,
\eea
and the monodromy-free equations \eqref{sosossa}
become identical to the system of equations \eqref{hsgspossaopa},\eqref{saopsoposa}.
Similarly the field  ${\bar u}(z,{\bar z})$  turns to be  ${\bar T}_{\bar L}({\bar z})$ -- an obvious
antiholomorphic  counterpart of  $T_L(z)$.
Notice that 
\bea\label{saospsopsaos}
q^{(L)}_{2n-1}=\lim_{\rho\to 0}\big(\, \rho^{2n-1}\,{\mathfrak q}_{2n-1}\,\big)\  ,\ \ \ \ \ \ \ \
{\bar q}^{({\bar L})}_{2n-1}=\lim_{\rho\to 0}\big(\, \rho^{2n-1}\,{\bar {\mathfrak q}}_{2n-1}\,\big)\ ,
\eea
where $q^{(L)}_{2n-1}$ are defined by Eq.\eqref{opsoposaaps} and
${\bar q}^{({\bar L})}_{2n-1}$ stand for
their antiholomorphic  counterparts.

\section{\label{sec8}Local IM versus MShG conserved charges}

In the case without  the monodromy-free punctures (i.e. $L={\bar L}=0$),
the  values of
conserved charges $ {\mathfrak q}_{2n-1}$ and $ {\bar {\mathfrak q}}_{2n-1}$ coincide:
\bea\label{sopsaosaaops}
 {\mathfrak q}_{2n-1}|_{L={\bar L}=0}={\bar {\mathfrak q}}_{2n-1}|_{L={\bar L}=0}\ .
\eea
In the recent work \cite{Lukyanov:2013wra},  a relation between 
these quantities and  vacuum eigenvalues of
local IM for the Fateev model was  proposed. 
In this section we discuss a natural generalization of that relation to
the excited states spectrum.

The definition and some basic properties of the Fateev model \cite{Fateev:1996ea} have been already  described 
in the introduction.
The following clarifying    remark on the decomposition  \eqref{saossasap} is in order at this stage.
A brief inspection of the Lagrangian  \eqref{aposoasio}  reveals
that the  model can be understood within the Conformal Perturbation Theory -- the Gaussian theory of three-component
Bose field  perturbed by the relevant operator.
As $\mu=0$ the general solution of the equation of motion can be written in the form,
\bea\label{sopsaossa}
{\textstyle \frac{1}{2}}\ \varphi(x,t)=\phi\big({\textstyle\frac{2\pi}{R}}\, (x-t)\big)-
{\bar \phi}\big({\textstyle\frac{2\pi}{R}}\, (x+t)\big)\ ,
\eea
where we use the ``right-moving'' chiral Bose field  \eqref{sakshs}. The ``left-moving'' chiral Bose field
${\bar \phi}({\bar  u})$ is defined by similar formulae, in particular,
\bea\label{sakshsisu}
{\bar \phi}_i({\bar u})=  \frac{1}{2}\ ( \mathbb{ \bar Q}_i-  \mathbb{\bar P}_i\, {\bar u})-\ri\
\sum_{n\not=0} \frac{{\bar a}_i(-n)}{n}\
\re^{-\ri  n {\bar u}}\, \ \ \ \ \ \ \ \ \ \ \ (i=1,2,3)\ .
\eea
As it follows from the periodic boundary condition the zero-modes eigenvalues  satisfy the
condition ${\bf P}+{\bar {\bf P}}=0$,
and hence 
the Hamiltonian of the Gaussian theory acts irreducibly  in the tensor product
\bea
{\cal F}_{\bf P}\otimes {\bar {\cal F}}_{-{\bf P}}\ .
\eea
Here ${\cal F}_{\bf P}$ is the Fock space defined in Section\,\ref{jassysa} and  ${\bar {\cal F}}_{-{\bf P}}$ 
is  similar space generated by the ``left-moving'' chiral Bose field.
The Hamiltonian  of the model   \eqref{aposoasio} with $\mu\not=0$
acts in the space 
\bea\label{sossapsapo}
\oplus_{{\boldsymbol\alpha} }\, \big(\, {\cal F}_{{\bf P}+{\boldsymbol\alpha}}\otimes {\bar {\cal F}}_{-{\bf P}-{\boldsymbol\alpha}}\,\big)\ ,
\eea
where 
\bea\label{soospsosaop}
{\textstyle\frac{1}{2}}\, {\bf P}=
(\alpha_1\,k_1,\alpha_2\,k_2,\alpha_3\,k_3)\ ,\ \ \ \ \ \ \  \ \ -{\textstyle \frac{1}{2}}<k_i\leq{\textstyle \frac{1}{2}}\ ,
\eea
and sum is taken over the vectors of the form 
\bea\label{aiosoiaos}
{\boldsymbol \alpha}=(n_1\,\alpha_1,\, n_2\,\alpha_2,\,
n_3\,\alpha_3)\ \ \  \ \ \ (n_i\in{\mathbb Z})\ .
\eea
The component ${\cal H}^{(0)}_{\bf k}:={\cal H}^{(0,0,0)}_{\bf k}$ in the decomposition  \eqref{saossasap}
can be realized as  a certain subspace of \eqref{sossapsapo}
spanned by the stationary states    such that
\bea\label{saospsopsao}
|\,\Psi^{(0)}_{\bf k}\,\rangle\in {\cal H}^{(0)}_{\bf k}\ :\ \ \ \ \ \ \ 
\lim_{\mu\to 0}\,|\,\Psi^{(0)}_{\bf k}\,\rangle\in {\cal F}_{\bf P}\otimes {\bar {\cal F}}_{-{\bf P}}\ ,
\eea
where ${\bf P}=(P_1,P_2,P_3)$ related to  ${\bf k}=(k_1,k_2,k_3)$ as in Eq.\eqref{soospsosaop}.
In this work we do not consider the stationary states  corresponding to  the higher Brillouin zones, i.e.,
the subspaces  ${\cal H}^{(n_1,n_2,n_3)}_{\bf k}$ of \eqref{sossapsapo}
whose CFT limit is described in terms of the tensor product   ${\cal F}_{{\bf P}+{\boldsymbol\alpha}}\otimes 
{\cal F}_{-{\bf P}-{\boldsymbol\alpha}}$ with  a given vector 
${\boldsymbol \alpha}\not= 0$  of the form \eqref{aiosoiaos}. 
Also we do not consider  ``charged'' sectors of the model associated with quasiperiodic boundary conditions
\bea\label{aospaossaops}
\varphi_i(x+R,t)=\varphi_i(x,t)+2\pi\, l_i/\alpha_i\ ,
\ \ \  \ \ \ l_i\in{\mathbb Z}\ .
\eea

The QFT \eqref{aposoasio} possesses an infinite set 
of commuting local integrals of motion \eqref{isusospsasopas}. 
In the CFT limit the operators $\mathbb{I}^{(+ )}_{2n-1}$ becomes  chiral local IM
discussed in Section\,\ref{jassysa}
\bea\label{sopsaospspao}
\lim_{\mu\to 0} \mathbb{I}^{(+)}_{2n-1}=\Big(\frac{2\pi}{R}\Big)^{2n-1}\  \mathbb{I}_{2n-1}\ .
\eea
Of course,  similar relations hold for $\mathbb{I}^{(- )}_{2n-1}$ whose  CFT limit is defined    by
${ \mathbb{{\bar I}}}_{2n-1}$ -- the  antiholomorphic counterpart of $\mathbb{I}_{2n-1}$:
\bea\label{usyssopsaospspao}
\lim_{\mu\to 0} \mathbb{I}^{(-)}_{2n-1}=\Big(\frac{2\pi}{R}\Big)^{2n-1}\  \mathbb{\bar I}_{2n-1}\ .
\eea
Let  $|\,\Psi^{(A)}_{\bf k}\,\rangle\in {\cal H}^{(0)}_{\bf k}$ be joint  eigenvectors of the
operators $\mathbb{I}^{(\pm)}_{2n-1}$ and $A$  is some  multi-index   labeling 
different eigenvectors
\bea\label{aopsasaosp}
\mathbb{I}^{(\pm)}_{2n-1}\ |\,\Psi^{(A)}_{\bf k}\,\rangle=
I^{(\pm,A)}_{2n-1}\ |\,\Psi^{(A)}_{\bf k}\,\rangle\ .
\eea
Recall that in Section\,\ref{jassysa} we considered   joint
eigenvectors $|\,L,\,\alpha\,\rangle$ for the commuting family of chiral IM 
$\{\mathbb{I}_{2n-1}\}_{n=1}^\infty$  (see Eq.\eqref{sopssaop}). 
Let  $|\,{\bar L},\,{\bar \alpha}\,\rangle$ be
their antiholomorphic analog.
Then our  consideration suggests that 
\bea\label{asopopassp}
\lim_{\mu\to 0} \, |\,\Psi^{(A)}_{\bf k}\,\rangle=|\,L,\alpha\,\rangle\otimes |\,{\bar L},{\bar \alpha}\,\rangle
\in {\cal F}_{\bf P}\otimes  {\bar {\cal F}}_{-{\bf P}}\ ,
\eea
where ${\bf P}=(P_1,P_2,P_3)$ related to  $(k_1,k_2,k_3)$ as in Eq.\eqref{soospsosaop}.

In Ref.\cite{Lukyanov:2013wra}, the $k$-vacuum
eigenvalues
were considered, 
\bea\label{akssasuaiassusys}
I^{(\rm vac)}_{2n-1}=I^{(+)}_{2n-1}(\{k_i\}\,|\,R)={ I}^{(-)}_{2n-1}(\{k_i\}\,|\,R)\ .
\eea
They are correspond to the vacuum states from  ${\cal H}^{(0)}_{\bf k}$, i.e., the states with
the lowest value of  energy $E^{(A)}=I^{(+,A)}_{1}+I^{(-,A)}_{1}$.
In the large-$R$ limit all vacuum  eigenvalues $I^{(\rm vac)}_{2n-1}$
vanish except $I^{(\rm vac)}_1$. 
The vacuum energy is composed of an extensive part proportional to the length of
the system,
\bea\label{saospsaos}
E^{(\rm vac)}=R\,{\cal E}_0+o(1)\ \ \ \ \ \ \ \ \ {\rm at }\ \ \ R\to \infty\ ,
\eea
where ${\cal E}_0$ stands for the specific  bulk energy \eqref{asososa}\ \cite{Fateev:1996ea}.

The main  observation of Ref.\cite{Lukyanov:2013wra} is  that  the  vacuum eigenvalues \eqref{akssasuaiassusys}
can be expressed in terms of  the classical  conserved charges \eqref{sopsaosaaops}. The relations are described by 
Eqs.\eqref{aspspsapo}-\eqref{saossaops}.
One of the main objectives 
of this work is to promote Eqs.\eqref{aspspsapo}-\eqref{saossaops}
to more general relations between the  joint spectrum of
the local IM \eqref{aopsasaosp} for the eigenstates $|\,\Psi^{(A)}_{\bf k}\,\rangle\in {\cal H}_{\bf k}^{(0)}$
\eqref{asopopassp}  and the conserved charges associated with
complex solutions  the MShG equation from the finite set ${\cal A}_{\bf m}^{(L,{\bar L})}$ described in  the previous section.
Notice that, in a view of relations \eqref{saospsopsaos},\,\eqref{sopsaospspao} and \eqref{saossaops},
the formulae \eqref{aspspsapo} and \eqref{apospaosp} 
reduce to Eqs.\eqref{isussospsaps} in the CFT limit.

\section{\label{ninesec}Non-linear integral  equations for the Fateev model}

The usual approach for studying off-shell physics is the
Thermodynamic Bethe Ansatz (TBA).  Its key input is factorizable
scattering theory underlying integrable QFT.  In principle, TBA is
a mathematically well defined method for evaluating thermodynamic
quantities by solving a set of coupled integral equations.  However,
in the case of a non-diagonal scattering this method 
requires many {\em ad hoc} assumptions (such as ``string
hypotheses'') and, therefore, is not very practical 
for complicated theories. 
The model described by the
Lagrangian \eqref{aposoasio} in the regime where all the couplings
$\alpha_i$ are real, seems to be a good illustration of this statement.  
Even though the corresponding factorizable scattering theory has been 
proposed  quite a while ago \cite{Fateev:2004un}, the derivation 
TBA equations for this complicated theory (to the best of our
knowledge) has never appeared in the literature.

The purpose of this section is to demonstrate  that
the correspondence between classical and quantum integrable
systems proposed in the 
previous section, provides  an alternative powerful tool for deriving
{\em functional and Bethe Ansatz type equations} which determine
the full spectrum of  local  IM in the massive QFT. 
For the vacuum sector of the Fateev model, we convert our
functional equations into the  non-linear
integral equations  and numerically study their
solutions, extending the similar analysis 
from  Section~\ref{nl-cft}.
Note, that a system of integral equations corresponding to  the ground state
was independently proposed by Fateev \cite{fate}.


\subsection{Connection matrices for MShG linear problem}

Let us consider  to the  axillary linear  problem \eqref{apsosaospa}
associated with   some element of  the finite set ${\cal A}^{(L,{\bar L})}_{\bf m}$.
We introduce  three  matrix solutions
\bea\label{asosopsa}
{\boldsymbol  \Psi}^{(i)}=\big({\boldsymbol  \Psi}^{(i)}_{-  }, {\boldsymbol  \Psi}^{(i)}_{+  }\,\big)\in
\mathbb{SL}(2,{\mathbb C})\ \ \ \ \ \ (i=1,2,3)\ .
\eea
For given $i$,  ${\boldsymbol  \Psi}^{(i)}$
solves the linear problem and satisfies  the
following asymptotic  condition
\bea\label{apsopaosaps}
{\boldsymbol  \Psi}^{(i)}
&\to&
\bigg(\,\re^{\frac{4\theta}{ a_i}}\
\frac{{ z}-{ z}_i} {{\bar z}-{\bar z}_i}\,\bigg)^{\frac{1}{4}\, (1-2p_i)\,\sigma_3}\ \ \re^{\ri \beta_i\sigma_3} \ \ \ \ \ \ \ \ \ 
{\rm as}\ \ \ \ z-z_i\to 0\ .
\eea
Here $p_i=m_i+\frac{1}{2}$ whereas  $\beta_i$ stands for arbitrary constant which will be fixed later (see Eq.\eqref{aopsaspsa} bellow).
The connection  matrices
are   defined  as follows:
\bea\label{saospsao}
{\boldsymbol  \Psi}^{(i)}={\boldsymbol  \Psi}^{(j)}\ {\boldsymbol S}^{(j, i)}(\theta)\ .
\eea
At this stage we will treat them as matrix  functions of the spectral parameter $\theta$ \eqref{osapsoasp};  that is
indicated explicitly in \eqref{saospsao}.
They are
entire  functions of $\theta$
satisfying  the  equations identical  to \eqref{Slambda}:
\bea\label{NS-rel2}
\det\big({\boldsymbol S}^{(j,i)}(\theta)\big)=1\ ,\ \ \
{\boldsymbol S}^{(i,j)}(\theta)\,{\boldsymbol S}^{(j,i)}(\theta)={\boldsymbol I}\ ,\ \ \
{\boldsymbol S}^{(i,k)}(\theta)\
{\boldsymbol S}^{(k,j)}(\theta)\
{\boldsymbol S}^{(j,i)}(\theta)={\boldsymbol I}\ .
\eea

The  axillary linear problem  is invariant with respect to the symmetries
analogous to  \eqref{O-def}
\bea\label{O-defn}
\wh\Omega_i:\qquad z\mapsto \gamma_i\circ z\,,\qquad
\, \ \ {\bar z}\mapsto {\bar \gamma}_i\circ {\bar z}\,,\qquad\ \ \ \ \ \ \ \  \theta\mapsto \theta-\ri\pi a_i\
\ \ \  \ \ \ (i=1,2,3)\ .
\eea
(Note that  $\wh\Omega_i$ involves now
the translation of  the variable ${\bar z}$ along the
complex conjugate contour ${\bar \gamma}_i$.)
These symmetries  act as linear transformations in the space of solutions and
in the basis ${\boldsymbol\Psi}^{(i)}$ they read
\bea\label{N123yt}
\wh\Omega_i\big({\boldsymbol\Psi}^{(i)}\big)&=&
  {\boldsymbol\Psi}^{(i)}\nonumber\\
\wh\Omega_j\big({\boldsymbol\Psi}^{(i)}\big)
&=&{\boldsymbol\Psi}^{(i)}\,
{\boldsymbol S}^{(i,j)}(\theta)\,
{\boldsymbol S}^{(j,i)}(\theta-\ri\pi a_j)
 \label{N123bn}\\
\wh\Omega_k\big({\boldsymbol\Psi}^{(i)}\big)
&=&{\boldsymbol\Psi}^{(i)}\,{\boldsymbol
  S}^{(i,k)}(\theta)\,
{\boldsymbol S}^{(k,i)}(\theta-\ri\pi a_k)\,
\ .\nonumber
\eea
Similar to derivations of Eqs.\eqref{S-rel2} and \eqref{S-rel3},
the  symmetry transformations \eqref{N123yt} allow one to
obtain the relation
\bea\label{NS-rel2h}
{\boldsymbol S}^{(i,k)}(\theta)\
{\boldsymbol S}^{(k,j)}(\theta-\ri\pi a_k)\
{\boldsymbol S}^{(j,i)}(\theta+\ri\pi a_i)=
{\boldsymbol I}
\eea
and express the  Wilson loop \eqref{sospsaosapo} in terms of the connection matrices
\beq\label{SW-rel3}
W={\rm Tr}\Big[\,\boldsymbol{S}^{(i,k)}(\theta-\ri\pi a _k)\,
\boldsymbol{S}^{(k,j)}(\theta)\, \boldsymbol{S}^{(j,i)}(\theta+\ri\pi a_j)\,\Big]\ .
\eeq

Another easily established symmetry of the  axillary linear problem \eqref{apsosaospa} involves the operation
\bea\label{aoipaspsoa}
&&{\wh \Pi}\ :\ \ \ \theta\mapsto\theta-\ri\pi\ ,\\
&&{\wh \Pi}\big[\,\partial_z-{\boldsymbol A}_z\,\big]=
\partial_z-{\boldsymbol A}_z\ ,\ \ \ \ \ \
{\wh \Pi}\big[\,\partial_{\bar z}-{\boldsymbol A}_{\bar z}\,\big]=
\partial_z-{\boldsymbol A}_z\ .\nonumber
\eea
Using this symmetry it is easy to show that ${\boldsymbol  S}^{(j,i)}$ are
quasiperiodic matrix functions of the spectral parameter $\theta$:
\bea\label{aopaopas}
{\boldsymbol  S}^{(j,i)}(\theta+\ri\pi)=
\re^{\frac{\ri\pi}{a_j} (2p_{j}-1)  \, \sigma_3}\
{\boldsymbol  S}^{(j,i)}(\theta)\ \re^{-\frac{\ri\pi}{a_i}  (2p_{i}-1)\, \sigma_3}\ .
\eea
To describe  properties of ${\boldsymbol S}^{(j, i)}(\theta)$
it is  convenient to use the matrices
$ {\boldsymbol  Q}^{(k)}(\theta)$   defined through the relation
\bea\label{usyisusaospsao}
{\boldsymbol  S}^{(j, i)}(\theta)=\frac{1}{\sqrt{
4
s\big({\textstyle \frac{2 p_{i}}{a_{i}}}\big)
s\big({\textstyle \frac{2 p_{j}}{a_{j}}}\big)}}\ \
 \re^{-\frac{\theta}{a_{j}}\, \sigma_3}\ \Big[\, -
\sigma_2\ {\boldsymbol  Q}^{(k)}(\theta+\ri\pi b_k)\ \Big]
\ \re^{\frac{\theta}{a_i}\, \sigma_3}\ ,
\eea
where $b_i$ stand for the constants given by Eq.\eqref{usystsaospsopsao}.
In terms of the matrix
 ${\boldsymbol  Q}^{(k)}(\theta)$ the quasiperiodicity condition \eqref{aopaopas}
looks somewhat simpler:
\bea\label{aopaopasysst}
{\boldsymbol  Q}^{(k)}(\theta+\ri\pi)=
\re^{-\frac{2\ri\pi p_j}{a_j}  \, \sigma_3}\
{\boldsymbol  Q}^{(k)}(\theta)\ \re^{-\frac{2\ri\pi p_i}{a_i}  \, \sigma_3}\ .
\eea
The shift of the argument in the definition \eqref{usyisusaospsao}
makes simpler the relation between the matrix elements
$Q^{(k)}_{\sigma' \sigma }(\theta)$,
\bea\label{usysusyisusaospsao}
{\boldsymbol  Q}^{(k)}=
\begin{pmatrix}
Q^{(k)}_{--}&Q^{(k)}_{-+}\\
Q^{(k)}_{+-}&Q^{(k)}_{++}
\end{pmatrix}\ ,
\eea
and
connection coefficients $A^{(k)}_{\sigma'\sigma}(\lambda)$  \eqref{sospsosapo}.
To describe this relation we note that $Q^{(k)}_{\sigma' \sigma }(\theta)$
can be written in the form
\bea\label{aopaospasp}
Q^{(k)}_{\sigma' \sigma }(\theta+\ri\pi b_k)=\ri
\ \sqrt{
4
s\big({\textstyle \frac{2 p_{i}}{a_{i}}}\big)
s\big({\textstyle \frac{2 p_{j}}{a_{j}}}\big)}\
\ \re^{\frac{\theta}{ a_{i}}\,\sigma+\frac{\theta}{ a_{j}}\,\sigma'}\
\  \det\big(  {\boldsymbol  \Psi}^{(j)}_{\sigma' },
{\boldsymbol \Psi}^{(i)}_{\sigma  }\big)\  .
\eea
Then
using
the relation between the  axillary problem \eqref{apsosaospa} and PGHO,
one can
show that
\bea\label{ospsaspsopaospspsa}
\lim_{\rho\to 0,\,\Re e(\theta)\to+\infty\atop
\lambda= \rho {\rm e}^{\theta}-{\rm fixed}}
\bigg(\, \re^{-\big(\frac{2 p_i}{a_i}\, \sigma+\frac{2 p_{j}}{a_{j}}\, \sigma'\big)\,\theta}\
Q^{(k)}_{\sigma'\sigma}(\theta)\,\bigg)&=&
\sqrt{{ f}^{(j)}_{\sigma'}\, { f}^{(i)}_{\sigma}}\ \ A^{(k)}_{\sigma'\sigma}(\ri \lambda)\nonumber\\
\lim_{\rho\to 0,\,\Re e(\theta)\to-\infty\atop
{\bar \lambda}= \rho {\rm e}^{-\theta}-{\rm fixed}}
\bigg(\, \re^{-\big(\frac{2 p_i}{a_i}\, \sigma+\frac{2 p_{k}}{a_{k}}\, \sigma'\big)\,\theta}\
Q^{(k)}_{\sigma'\sigma}(\theta)\,\bigg)&=&
\sqrt{{ f}^{(j)}_{\sigma'}\ { f}^{(i)}_{\sigma}}\ {\bar A}^{(k)}_{-\sigma',-\sigma}(\ri {\bar \lambda}) \ ,
\eea
where
\bea\label{ystspsospsa}
{ f}^{(i)}_{\sigma}=
2\,
s\big( {\textstyle\frac{2 p_{i}}{a_i}}\big)\ \
\re^{({\bar \omega}_i-\omega_i-2\beta_i)\,\sigma}\ ,
\eea
and  ${\bar A}^{(k)}_{\sigma'\sigma}({\bar \lambda})$ is an antiholomorphic counterpart of
$A^{(k)}_{\sigma'\sigma}(\lambda)$ \eqref{sospsosapo}.

Let us fix  the  value of  constant $\beta_i$ in Eqs.\eqref{apsopaosaps} and \eqref{ystspsospsa}:
\bea\label{aopsaspsa}
\re^{\ri \beta_i}=
\bigg(\frac{z_{ji}z_{ik}}{z_{jk}}\ \frac{{\bar z}_{jk}}{{\bar z}_{ji}{\bar z}_{ik}}\bigg)^{\frac{p_i}{2}}\ .
\eea
Then, by virtue of a  WKB analysis similar to one employed in  derivation Eq.\eqref{aspoaskasus},
the following asymptotic formulae   within the strip $\big|\Im m(\theta)\big|<\frac{\pi}{2}$ can be obtained:
\bea\label{ospsahssgst}
Q_{\sigma'\sigma}^{(k)}(\theta)\to\
\begin{cases}
\big({\mathfrak S}_{j}\big)^{\frac{\sigma'}{4}}\
\big({\mathfrak S}_i\big)^{\frac{\sigma}{4}}\
 \exp\big(\,   r_{k}\ \rho\,\re^{\theta }\,\big)\   \ \  &{\rm as}\ \
\Re e(\theta)\to+\infty\\
\big({\mathfrak { S}}_{j}\big)^{-\frac{\sigma'}{4}}\
\big({\mathfrak { S}}_i\big)^{-\frac{\sigma}{4}}\
\exp\big(\,  r_{k}\  \rho\,\re^{-\theta }\,\big)\   \ \  &{\rm as}\ \
\Re e(\theta)\to-\infty
\end{cases}\  .
\eea
Here $r_k$  stand for the constants given  by Eq.\eqref{ospspsa} and
\bea\label{opasasos}
\big({\mathfrak { S}}_{i}\big)^{\frac{1}{2}}=
\Big(\frac{\rho}{a_i}\Big)^{-\frac{4 p_i}{a_i}}\ \
\frac{\Gamma(1+\frac{2p_i}{a_i})}
{\Gamma(1-\frac{2p_i}{a_i})}\ \ \frac{\exp({\eta^{(\rm reg)}_i})}{2\,p_i}\ \ \Big|\frac{z_{jk}}{z_{ji}z_{ik}}\Big|^{-2p_i},
\eea
where $\eta^{(\rm reg)}_i$ is  the regularized value of the MShG field at the puncture $z_i$ $(i=1,2,3)$, i.e.,
$\eta=(2p_i-1)\ \log|z-z_i|+\eta_i^{(\rm reg)}+o(1)$.

\subsection{\label{sec9.2}Reconstruction of the connection matrices for  $L={\bar L}=0$}

The quasiperiodic entire  function $Q^{(k)}_{\sigma' \sigma }(\theta)$ is  completely determined by its zeroes 
in the strip $|\Im m(\theta)|\leq \frac{\pi}{2}$ and 
the leading asymptotic behavior given by  \eqref{ospsahssgst}. 
On the other hand,  positions of the zeroes are restricted by the
Bethe Ansatz equations similar to \eqref{BAE}.
In fact, the problem of reconstruction of  the connection matrices ${\boldsymbol S}^{(ij)}(\theta)$
is almost identical to that is studied in Section\,\ref{seccon}.
Thus, we will not repeat the analysis in detail  but quote
the  non-linear integral equation determining   the connection matrices
in  the case without monodromy-free punctures.
In this case, using the 
arguments similar to those given  in Section 3.2 from Ref.\cite{Lukyanov:2010rn}, one can argue that
all the roots of  $Q^{(k)}_{\sigma' \sigma }(\theta)$ 
are simple and located at the lines
$\Im m(\theta)=\ri\  \big( n+\frac{1}{2}\big)\ \pi\ (n\in \mathbb{Z})$.
After that  the derivation becomes straightforward and yields the system of 
non-linear integral equations
which differs from \eqref{ddv-eq}  in the source terms only:
\beq\label{Mddv-eq}
\epsilon_i(\theta)=4\rho r_i\, \sinh(\theta) -\pi\, \big(
\,\sigma'\, {\textstyle \frac{2p_j}{a_j}}+\sigma''\ {\textstyle \frac{2p_k}{a_k}}\,\big)+
 \sum_{l=1}^3
\int_{-\infty}^\infty\frac{\rd\theta'}{\pi}\  G_{il}(\theta-\theta') \,{\Im m}
\Big[\log\big(1+\re^{-\ri \epsilon_l(\theta'-\ri 0)}\,\big)\Big]\ .
\eeq
Once the numerical data for $\epsilon_i(\theta)$ are available, $Q^{(k)}_{\sigma' \sigma }(\theta)$ can be computed
by means of the relation
\bea\label{pospsosaop}
\log Q^{(k)}_{\sigma' \sigma }(\theta)=2\rho\, r_k\, \cosh(\theta)+
\sum_{l=1}^3\int_{-\infty}^\infty\frac{\rd\theta'}{\pi}\ F_{kl}(\theta-\theta')\ 
\Im m \Big[\log\big(1+\re^{-\ri \epsilon_l(\theta'-\ri 0)}\,\big)\Big]\, ,
\eea
where
\bea\label{aopsosapas}
F_{kl}(\theta)=\int_{0}^\infty\rd\nu\ \frac{\Phi_{kl}(\nu)}{\sin(\frac{\pi\nu}{2})}\ 
\sin(\nu\theta)\ ,
\eea
and
$\Phi_{kl}(\nu)$ are defined by \eqref{akasissau}.

Notice that,
in the case $L={\bar L}=0$ the   
conserved charges ${\mathfrak q}_{2n-1}$ and ${\bar {\mathfrak q}}_{2n-1}$  have the same 
value which is given by
\bea\label{integ}
{\mathfrak q}_{2n-1}=
\frac{8\, n!\,\sqrt{\pi}}{\Gamma(n-\frac{1}{2})}\ 
\sum_{l=1}^3 \sin\big(\,\pi (n-{\textstyle\frac{1}{2}})\,a_l\,\big)
\ \int_{-\infty}^{\infty} \frac{\rd\theta}{\pi} \ \re^{(2n-1)
  \theta}\
\Im m\Big[\log\big(\,1+\re^{-\ri\epsilon_l(\theta-\ri 0)}\,\big)\Big]\ .
\eea
Also the subleading term in the asymptotic  \eqref{ospsahssgst} is given by
\bea\label{asopaso}
\big({\mathfrak S}_{j}\big)^{\frac{\sigma'}{4}}\
\big({\mathfrak S}_i\big)^{\frac{\sigma}{4}}=\exp\bigg(
\sum_{l=1}^3 \Phi_{kl}(0)\  \int_{-\infty}^{\infty}\frac{ \rd\theta}{\pi} \ 
\Im m\Big[\log\big(\,1+\re^{-\ri\epsilon_l(\theta-\ri 0)}\,\big)\Big]\,\bigg)\ ,
\eea 
where
\bea\label{siosasaiop}
 \Phi_{ki}(0)=\frac{1}{2 a_j}\ ,\ \ \ \  \Phi_{kj}(0)=\frac{1}{2 a_i}\ ,\ \ \ \ 
\Phi_{kk}(0)=\frac{1}{2 a_i}+\frac{1}{2 a_j}\ .
\eea

We solved the integral equation \eqref{Mddv-eq} numerically 
for various values of
the parameters $a_1,a_2,a_3$ and $p_1,p_2,p_3$ and then calculated the
values the conserved charge ${\mathfrak q}_{1}$ 
from the formula \eqref{integ}. The
results are in an excellent agreement with 
the expression for the vacuum energy given by Eqs.(3.25)-(3.28) of
Ref.\cite{Lukyanov:2013wra}:
\bea\label{ysfasosposa}
{\mathfrak q}_1=\frac{2\pi^2}{\prod_{i=1}^3\Gamma(\frac{a_i}{2})}\, 
\bigg[-\frac{1}{6\rho}\,\sum_{i=1}^3\Big(1-\frac{24 }{ a_i}\ p_i^2\, \Big)+
4\, \rho\, \prod_{i=1}^3
\gamma\Big({\frac{a_i}{2}}\Big)-\frac{4}{\pi}\, \rho^3 \int
\rd^2 z\,{\cal P}(z){\bar {\cal P}}({\bar z})\, \re^{-2\eta}\,\bigg] ,
\eea
where 
$\gamma(x):=\frac{\Gamma(x)}{\Gamma(1-x)}$.
Note that the third term in this expression involves  
the solution of the MShG equation without  monodromy-free punctures
(its contribution is essential for large values of $\rho$). The MShG
equation has been solved numerically to find the value of the
integral, entering \eqref{ysfasosposa}, and calculate the constants
$\eta_i^{\rm (reg)}$, entering \eqref{opasasos}. 
Using the latter, we have verified the numerical agreement between
 \eqref{opasasos} and \eqref{asopaso}.

\subsection{$k$-vacuum eigenvalues of local IM in the Fateev model}

First of all let us recall some
facts concerning  the factorizable  scattering  theory associated with   QFT \eqref{aposoasio}.
All the details can be found in Appendix F  in
Ref.\cite{Fateev:2004un}. 

The spectrum consists of three quadruplets of fundamental particles 
\bea\label{asosopsapsa}
Z_{\epsilon\epsilon'}^{(i)}\ , \ \ \ \ \ \ \ \ \ \epsilon,\,\epsilon'=\pm\,,\ \ \ \  i=1,\,2,\,3\ ,
\eea
with the masses
\bea\label{apsaisaosao}
M_i=M_0\ \sin\Big(\frac{\pi a_i}{2}\Big)\ ,\ \ \ \ M_0= \frac{2\mu}{\pi}\ \prod_{i=1}^3\Gamma\Big(\frac{a_i}{2}\Big)\ 
\eea
and their bound states. (Here the relation  $a_i=4\alpha_i^2$ is assumed to hold.) 
The Zamolodchikov-Faddeev commutation  relations for the fundamental particles
read
\bea\label{sopspospas}
&&Z_{\epsilon_1\epsilon'_1}^{(i)}(\theta_1)Z_{\epsilon_2\epsilon'_2}^{(i)}(\theta_2)=
-\sum_{\epsilon_3\,\epsilon'_3\atop
\epsilon_4\,\epsilon'_4}
\big[\,S_{a_j}(\theta_1-\theta_2)\,\big]^{\epsilon_3\epsilon_4}_{\epsilon_1\epsilon_2}\
\big[\,S_{a_k}(\theta_1-\theta_2)\,\big]^{\epsilon'_3\epsilon'_4}_{\epsilon'_1\epsilon'_2}\
Z_{\epsilon_4\epsilon'_4}^{(i)}(\theta_2)Z_{\epsilon_3\epsilon'_3}^{(i)}(\theta_1)\nonumber\\
&&Z_{\epsilon\epsilon'_1}^{(i)}(\theta_1)Z_{\epsilon'_2\epsilon''}^{(j)}(\theta_2)=\epsilon\,\epsilon''
\sum_{\epsilon_3\,\epsilon'_4}
\big[\,{\hat S}_{a_k}(\theta_1-\theta_2)\,\big]^{\epsilon'_3\epsilon'_4}_{\epsilon'_1\epsilon'_2}\
Z_{\epsilon_4\epsilon''}^{(j)}(\theta_2)Z_{\epsilon\epsilon'_3}^{(i)}(\theta_1)\ ,
\eea
where $(i,j,k)={\tt cyclic\  perm}(1,2,3)$ and
\bea\label{aspopsao}
{\hat S}_{a}(\theta)=\ri\ \tanh\big({\textstyle\frac{\theta}{2}}+
\ri\, {\textstyle\frac{\pi a}{4}}\,\big)\ S_a\big(\theta+
\ri\, {\textstyle\frac{\pi a}{2}}\,\big)\ .
\eea
Also $S_a(\theta)$ stands for 
the conventional $S$-matrix in the quantum sine-Gordon theory \cite{Zamolodchikov:1978xm}
with the renormalized coupling constant  $a$, 
related to  the Coleman coupling $\beta^2_{C}$ \cite{Coleman:1974bu}
as follows
\bea\label{sosposa}
a=\frac{\beta^2_C}{8\pi-\beta_C^2}\ .
\eea
In particular, the sine-Gordon soliton-soliton scattering amplitude reads
explicitly as
\bea\label{usypaospaos}
s_a(\theta):=[S_a(\theta)]_{++}^{++}
=-\exp\bigg(-\ri\ \int_{0}^\infty\frac{\rd\nu}{\nu}\ \frac{\sinh(\frac{\pi\nu}{2}\,(1-a))}{
\cosh(\frac{\pi\nu}{2})\,\sinh(\frac{\pi\nu }{2}\, a)}\ \sin(\nu\theta)\,\bigg)\ .
\eea

In a view  of our previous discussion it is expected that  the non-linear integral equations \eqref{Mddv-eq}
solves the problem of calculation of  the   $k$-vacuum eigenvalues \eqref{akssasuaiassusys}
in the Fateev model in the finite volume.
To make the  link more  explicit let
us note 
the kernels \eqref{uasysa} in the equations \eqref{Mddv-eq}  are simply related to the amplitude \eqref{usypaospaos}
and
\bea\label{aoisisiosa}
{\hat s}_{a}(\theta):=
[{\hat S}_a(\theta)]_{++}^{++}
=\exp\bigg(-\ri \int_{0}^\infty\frac{\rd\nu}{\nu}\ \frac{\sinh(\frac{\pi\nu}{2})}{
\cosh(\frac{\pi\nu}{2})\sinh(\frac{\pi\nu }{2}\, a)}\ \sin(\nu\theta)\,\bigg)\  .
\eea
Namely, it is easy to see that
\bea\label{GS-rel}
G_{ii}(\theta)= \ri\ \partial_\theta\log\big(\,s_{a_j}(\theta)\,s_{a_k}(\theta)\,\big)\,,\qquad
G_{ij}(\theta)= \ri\  \partial_\theta\log\big({\hat s}_{a_k}(\theta)\big)
\  .
\eea
These formulae  confirm an
empirical rule that the kernels of the non-linear integral equations
coincide with the logarithmic derivative of some diagonal elements of
the $S$-matrix, which has been previously observed for some other models
(e.g., the sine-Gordon model).
The $\mu-\rho$ relation \eqref{saossaops} combined with  the definition of $r_i$ \eqref{ospspsa},
implies that  the combination  $4\rho r_i$, which appears in the source term of
the integral equation \eqref{Mddv-eq}, is simply expressed in terms of the particle mass $M_i$ \eqref{apsaisaosao}:
\bea\label{ospsaopsa}
4\,\rho\ r_i=M_i\, R\ .
\eea
Also, as it follows from \eqref{sssaopsa}, 
the parameters $2\,p_j/a_j$  should be identified with the quasimomentum magnitudes  $|k_j|$.
Finally the $k$-vacuum 
eigenvalues \eqref{akssasuaiassusys} can be calculated using   Eqs.\eqref{aspspsapo},\,\eqref{apospaosp}
and  \eqref{integ}.

%
Unfortunately, at the moment there is no independent derivation of
the results of Section~\ref{sec9.2} from the field theory side --
%
the Fateev model does not have any known
lattice analog, and neither it has any known 
coordinate or algebraic Bethe Ansatz
solutions.\footnote{The limiting case $\alpha_3=0$ of the Fateev model can be 
reduced to the Bukhvostov-Lipatov
model, where the non-linear integral equations were derived
from the coordinate Bethe Ansatz in Ref.\cite{Saleur:1998wa}.} 
 
\section{Concluding remarks}

In this paper we have described the relation between the MShG equation,
on one hand and the Fateev model on the other. 
We belive that  the outlined results
open a new general way of approaching integrable QFT.

As an immediate (but perhaps not entirely straightforward) application
one could consider various Toda QFT's.  This would involve 
differential operators of higher orders (the  $\mathfrak{g}$-opers \cite{Beilinson}) 
and the  classical modified Toda equations.
Some  basic  ingredients, required for this 
development have already been  revealed.
Among them
a classification of
third order differential operators with monodromy-free singular
points, corresponding to stationary states in CFT's with the extended
$W_3$-symmetry \cite{Bazhanov2013xxx}, and the relation 
between
the vacuum sector  in the ${\hat A}_2^{(2)}$ Toda QFT and
the modified Bullough-Dodd equation \cite{Dorey:2012bx}.
However, perhaps the most important potential outcome of our approach 
is related to the problem of  non-perturbative quantization of classically integrable  non-linear sigma models.
Here, we are motivated by the following consideration.

This work  has been focused on the ``symmetric'' regime
of the Fateev  model where
all the couplings   $\alpha_i$ in  \eqref{aposoasio} are real, so that
the Lagrangian  is completely symmetric under simultaneous
permutations of the real fields  $\varphi_i$ and the real couplings $\alpha_i$.
The theory   is apparently non-unitary in this case.
In the most interesting regime
one of the couplings, say $\alpha_3$,
is pure imaginary
\bea\label{aisaisaosa}
\alpha_1^2>0\ ,\ \ \ \ \alpha_2^2>0\ ,\ \ \ \ \alpha_3^2:=-b^2<0 \ ,
\eea
and the theory is governed by  the real Lagrangian
\bea\label{usyaposoasio}
{\cal L}&=& \frac{1}{16\pi}\ \sum_{i=1}^3
\big(\, (\partial_t\varphi_i)^2-(\partial_x\varphi_i)^2\,\big)\\
&-&2\mu\
\big(\, \re^{b\varphi_3}\ \cos(\alpha_1\varphi_1+\alpha_2\varphi_2)+\re^{-b\varphi_3}\
\cos(\alpha_1\varphi_1-\alpha_2\varphi_2)\,\big) \ ,
\nonumber
\eea
where
\bea\label{isusaposapoas}
\alpha_1^2+\alpha_2^2-b^2=\frac{1}{2}\ .
\eea
The physical content in the unitary regime 
is   different from  the symmetric one.
However, assuming the same periodic boundary conditions  for each field $\varphi_i\  (i=1,2,3)$, 
we can  use the same symbols  ${\cal H}$ and ${\cal H}_{\bf k}$ 
to denote  the   spaces of states and their certain  linear subspaces
in the both cases. 
Just remember that because of the lack
of periodicity in $\varphi_3$-direction in the unitary regime,
Eq.\eqref{sapsapo} can be applied for $i=1,2$ only. Therefore
${\bf k}$ should be regarded as a pair of
quasimomenta, ${\bf k}=(k_1,k_2)$, and
Eq.\eqref{saossasap} should be now substituted by
\bea\label{isusaossasap}
{\cal H}_{\bf k}=\oplus_{n_1,n_2\in\mathbb{Z}}\,{\cal H}^{(n_1,n_2)}_{\bf k}\ .
\eea
Again, it makes sense to focus on the component ${\cal H}^{(0)}_{\bf k}:={\cal H}^{(0,0)}_{\bf k}$
corresponding to the first Brillouin zone.
We would like to emphasize that the fact of existence of the local IM 
and their  form
are not sensitive to the choice of the regime.
In particular, with the formal
substitution $\alpha_3\to-\ri\, b$,   Eqs.\eqref{isusospsasopas} and \eqref{saosopsaosa} can be applied
to the unitary case.
The eigenstates in  ${\cal H}^{(0)}_{\bf k}$ are again specified 
by the joint spectra of local IM.

Having in mind   relations between ${\cal H}^{(0)}_{\bf k}$ and ${\cal A}_{\bf m}$ in the symmetric regime, 
let us consider the MShG equation in the regime $a_1,\, a_2>0,\,  a_3<0$
(the constraint $a_1+a_2+a_3=2$ is still assumed). 
A brief inspection shows that  set of requirements \eqref{sospsaosapyst}-\eqref{saopsosap} imposed on the MShG field
looks  quite  meaningful in this case. 
Only the formulae \eqref{sospsaosap} and \eqref{asioasiso}
which describe
the  behavior  of the solution in the vicinity of the third puncture
$z_3$, call for a special attention. 
As  $a_i>0$ we had a freedom to control 
the asymptotic behavior of $\eta$ as $z\to z_i$, with the free
parameter $m_i$. If  
$a_3<0$, the situation is different -- the leading asymptotic behavior
of the solution  at $z=z_3$  is fixed by the  
MShG equation itself \cite{Lukyanov:2010rn}:
\bea\label{osapsaosaasps}
\re^{-\eta}\sim \big|{\cal P}(z)\big|^{-\frac{1}{2}}\propto |z-z_3|^{\frac{a_3}{2}-1}\ .
\eea
With this modification, we expect Eq.\eqref{oaioasoas} 
to remain a meaningful definition of the moduli space ${\cal A}_{\bf
  m}$, provided ${\bf m}$ is understood now as a pair $(m_1,m_2)$. 
Using  the intuition gained from the study of the sine and sinh-Gordon
models \cite{Lukyanov:2010rn}, we expect that 
the relations \eqref{aspspsapo}-\eqref{saossaops} remain valid for the
case $a_3=-b^2/4<0$. 
The only exemption is  the second formula  in Eq.\eqref{sssaopsa} for $i=3$ -- 
evidently it cannot be applied literally.
Notice also that the definition of 
the set of the conserved charges
$\{{\mathfrak q}_{2n-1},\, {\bar {\mathfrak q}}_{2n-1}\}_{n=1}^\infty$ 
remains unchanged. 
We expect that, with  these simple modifications 
the relation between the subspace ${\cal H}_{\bf k}^{(0)}$ and
the moduli space ${\cal A}_{\bf m}$ holds for the case $a_3=-b^2/4<0$.

The Fateev model in the unitary regime  
admits a dual description in terms of the action
$\int\rd^2x\ G_{\mu\nu}(X)\ \partial_a X^\mu\partial_a X^\nu$,
where
$G_{\mu\nu}$ is a certain two-parameter families
of metric on the topological three-sphere which possesses two $U(1)$
Killing vector fields \cite{Fateev:1996ea}. The sigma-model
description is especially useful in 
the strong coupling  limit
($\alpha_i^2,\ b^2\to \infty$ with  $\alpha_i^2/b^2$ kept fixed), which can be regarded as
the classical limit. Notice that  the classical  integrability of the theory 
was established only  recently in  Ref.\cite{Lukyanov:2012zt}.

\section*{Acknowledgments}

The authors are grateful to B.~A. Dubrovin, A.~V. Litvinov, F.~A. Smirnov,
Z. Tsuboi and  A.~B. Zamolodchikov for fruitful discussions.
Part of this work was done during the visit of the  second author to IPhT at
CEA Saclay in May-June 2013. SL would like to express his sincere gratitude
to members of the laboratory and especially Didina  Serban and Ivan Kostov
for their kind hospitality and interesting discussions. The research of VB was
partially supported by the Australian Research Council.

\appendix
\section{GHO with $L=1$\label{app1}}

The   system  of  algebraic equations \eqref{hsgspossaopa}-\eqref{osaospa} looks rather cumbersome.
Here  we   discuss the  simplest case $L=1$ in some details. 

First, using M\"obius transformation
one can move the first three punctures 
to the  standard positions, $(z_1,z_2,z_2)= (0,1,\infty)$.
The coordinate of the forth puncture $x$ (which is the only monodromy-free
puncture for $L=1$) will now coincides with the projective invariant $X_1$ \eqref{aoapospaspo},
while the corresponding accessary parameter will be denoted as $C$.
Eqs.\eqref{osaospa} allows one to express
the accessory parameters
$c_1,\,c_2,\,c_3$ in terms of $C$ and $x$, and, thereby,  
to reduce Eq.\eqref{hsgspossaopa} to a single algebraic
equation. The later  
can be brought to the form
\bea\label{soiasisisa}
P_3(x,\,y )=0\ , \ \ \ \ \ \ \ \ \ {\rm where}\ \ \ y=1-2\,x-x\,(1-x)\, C\, ,
\eea
and
\bea\label{sisisoaois}
P_3(x,y)&=&
y^3+ (1-2 x)\, y^2+
\big(\, 4\, \delta_1-1+ 4\, (\delta_2-\delta_1-\delta_3\,\big)\,  x+4\, 
\delta_3\, x^2\,\big)\,  y\nonumber\\
&+&
  (4\, \delta_1-1)\, (1-2x)+4\, (\delta_1-\delta_2)\, x^2\ .
\eea
For generic  values of $\delta_i$, \eqref{soiasisisa} considered
as a cubic equation for $C$,
has three different roots. We will label them 
by an integer $\epsilon=0,\ \pm 1 $.
For small $x$ the roots
admit Laurent expansions, 
which can be related to
the series  expansions
for the classical conformal blocks   depicted in Fig.\,\ref{fig2aa}:
\begin{figure}
\centering
\psfrag{z1}{$(0,\delta_1)$}
\psfrag{z2}{$(x,-2)$}
\psfrag{z3}{$\delta(p_1+\epsilon)$}
\psfrag{z4}{$(1,\delta_2)$}
\psfrag{z5}{$(\infty,\delta_3)$}
\includegraphics[width=10  cm]{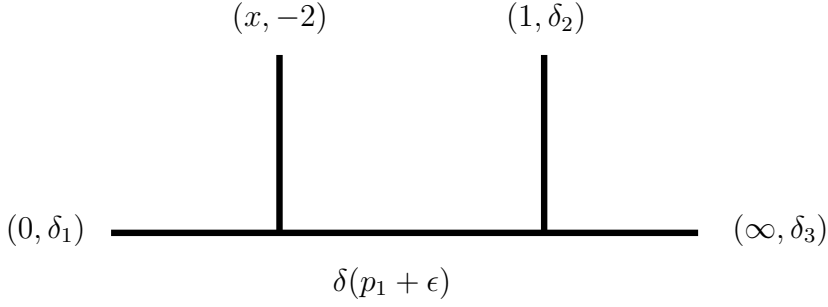}
\caption{Dual diagram for the classical conformal block from \eqref{oopsspa}. Here
$\epsilon=0,\,\pm 1$.}
\label{fig2aa}
\end{figure}
\bea\label{oopsspa}
C^{(\epsilon)}=\frac{\partial}{\partial x}\, 
f_{\delta(p_1+\epsilon)}\big[\,_{\ \delta_1,\, \delta_3}^{-2,\,\delta_2}\,\big](x
)\ \ \ \ \ \ \ \ \   \ 
(\epsilon=0,\ \pm 1)\ .
\eea
Here we 
use the standard notation for
the general  4-point classical conformal block 
\bea\label{uisais}
f_{\delta}\big[\,_{\delta_1,\,\delta_4}^{\delta_2,\delta_3}\big](x)
=(\delta-\delta_1-\delta_2)\ \log(x)+
\frac{(\delta-\delta_1+\delta_2)(\delta+\delta_3-\delta_4)}{2\ \delta}\ x+O(x^2)
\ .
\eea
In fact, \eqref{oopsspa}
is the simplest illustration of 
the general relation \eqref{aopsaps}.

The roots $C^{(\epsilon)}$  corresponds to different branches of 
a multivalued function, which has algebraic singularities
in the complex plane of the variable $x$.
For sufficiently small positive $p_i$ all branch points
lie outside of the real axis.
In this case  the branches $C^{(\epsilon)}$ can  be  unambiguously defined
for all real  $x$ through the analytic continuation of
the  series expansions \eqref{oopsspa},\,\eqref{uisais}
along the real axis. 
The real functions
$C^{(\epsilon)}$, defined in this way, have the following expansion 
in the vicinity of points $x=0,\,1$ and $\infty$
\begin{subequations}
\label{tewsiosioaw}
\bea\label{tewsiosioa}
C^{(0)}(x)&=&
\begin{cases}
&\frac{2}{x}
+O(1)\ \ \ \ \ \  \ \ \ \ \ \ \ \ \ \ \ \ \ \ \ \ \ 
{\rm as}\ \ x\to 0\\
&\frac{1-2 p_2}{x-1}
+O(1)\ \ \ \ \
\ \ \  \ \ \ \ \ \  \ \ \ \  {\rm as}\ \ x\to 1
\\
&\frac{2}{x}+
O(x^{-2}) \ \ \ \ \ \ \ \ \ \ \ \ \ \ \ \ \ \ \ \ {\rm as}\ \ x\to\infty
\end{cases}
\ ,
\\[.4cm] 
\label{tewsiosioayst}
C^{(+)}(x)&=&
\begin{cases}
&\frac{1-2 p_1}{x}
+O(1)\ \ \ \ \ \ \ \ \ \ \
\ \ \ \ \ \ \ \   {\rm as}\ \ x\to 0\\
&\frac{2}{x-1}+O(1)
\ \ \ \ \
\ \ \ \ \ \ \ \  \ \ \ \ \ \ \  \  {\rm as}\ \ x\to 1
\\
&\frac{3+2 p_3}{x}
+O(x^{-2}) \ \ \ \ \ \ \ \ \ \ \ \ \ \ \ \  {\rm as}\ \ x\to\infty
\end{cases}
\ ,
\\[.4cm] 
\label{hshsiosioa}
C^{(-)}&=&
\begin{cases}
&\frac{1+2 p_1}{x}+O(1)
\ \ \ \ \ \ \ \ \ \ \ 
\ \ \ \ \ \ \ \ \  {\rm as}\ \ x\to 0\\
&\frac{1+2 p_2}{x-1}+O(1)
\ \ \ \ \ 
\ \ \  \ \ \ \ \ \   \ \ \ \ \ \   {\rm as}\ \ x\to 1 
\\
&\frac{3-2 p_3}{x}+O(x^{-2})\ \ \ \ \ \ \ \ 
\ \ \ \ \ \ \ \ \  {\rm as}\ \ x\to\infty 
\end{cases}
\ .
\eea
\end{subequations}
To simplify the notations, we shall
denote by $f(x)$ the classical conformal block 
associated to the ``principle'' branch $C^{(-)}$ \eqref{hshsiosioa}:
\bea\label{soosasa}
f(x)=(1+2\,p_1)\ \log(x)+
\int_0^x\rd x\ \Big(\, C^{(-)}-\frac{1+2p_1}{x}\Big)\ .
\eea
This  defines $f(x)$ unambiguously in the neighborhood of $x=0$.
In general, the integral
depends on a integration contour connecting $x$ to the origin, and
the classical  conformal blocks corresponding to
$\epsilon=0$ and $\epsilon=+1$ are just different 
branches of the multivalued function \eqref{soosasa}.

To describe  global properties of the
multivalued functions $C(x)$ and $f(x)$
we  use  the  fact that any nondegenerate  cubic is 
homeomorphic
to an elliptic curve.
In the case under consideration the corresponding  
elliptic modulus (denoted by $k$ bellow)  can be chosen  as
\bea\label{sosioaois}
k^2=\frac{8\ p_1 p_2 p_3}
{(\frac{1}{2} +  p_1 +  p_2 +  p_3)
 (\frac{1}{2} +  p_1 -  p_2 -  p_3)
 (\frac{1}{2} -  p_1 +  p_2 -  p_3) 
 (\frac{1}{2} -  p_1 -  p_2 +  p_3)}\ .
\eea
(Recall that $k^2$ is defined up to modular transformations
$k^2\mapsto 1-k^2, 1/k^2$).
Bellow we use the  nome $q$ which is related to the elliptic modulus as
\bea\label{iaosaoos}
k^2=\frac{\vartheta^4_2(0,q)}{\vartheta^4_3(0,q)}\ .
\eea
For the purpose of uniformization of the cubic \eqref{soiasisisa},
it is useful to introduce three parameters $u_1,u_2,u_3$ such that
\bea\label{osooppos}
p_i=\frac{1}{2}\ \rho(u_j-u_i,q)\,\rho(u_k-u_i,q)\ .
\eea
Here
$\rho(u,q)$ stands for the double periodic function
\bea\label{asiossaoi}
\rho(u,q)=\frac{\vartheta_3(u,q)\vartheta_4(u,q)}
{\vartheta_1(u,q)\vartheta_2(u,q)}=\frac{\vartheta_4(2u,q^2)}{\vartheta_1(2u,q^2)}\ ,
\eea
and $(i,j,k)$ is an arbitrary permutation of $(1,2,3)$.
Notice that the above relations define $u_i$ up to the overall shift
$u_i\to u_i+const$. 
For  real  $p_i$ restricted as in Eq.\eqref{aopsaospap}, the elliptic nome 
is real and 
$0<q<1$, whereas the parameters $u_i$ can be chosen in the form
\bea\label{oospsop}
u_i=u_0+\ri\ v_i\ \ \ \ 0<v_i<-\log (q)\ .
\eea
In terms of an uniformizing variable
\bea\label{soosapsao}
u\ :\ \ \ \ \ \ u\sim u+N\pi+\ri\ M\log q\ \ \ \ \ (N,M\in \mathbb{Z})\ ,
\eea
the cubic  \eqref{soiasisisa}
is  described  as follows
\bea\label{osospsa}
x&=&-\frac{\vartheta_3(u-u_3-u_2+u_1,q)\vartheta_1(u-u_1,q)\vartheta_2(u-u_1,q)}
{\vartheta_3(u+u_3-u_2-u_1,q)\vartheta_1(u-u_3,q)\vartheta_2(u-u_3,q)}
\ \frac{\vartheta_1(u_3-u_2,q)\vartheta_2(u_3-u_2,q)}
{\vartheta_1(u_2-u_1,q)\vartheta_2(u_2-u_1,q)}\nonumber\\
y&=&\rho(u-u_3,q)\,\rho(u_2-u_1,q)\ ,
\eea
whereas   the accessory parameter is given by
\bea\label{oosaps}
C=\frac{1}{x}+\frac{1}{x-1}+\frac{y}{x(x-1)}\ .
\eea
These  equations imply that $C$ is a single-valued doubly-periodic  function
with simple poles located at
\bea\label{oapopsapsa}
u\in\big\{\, u_i,\, u_i+
{\textstyle\frac{1}{2}}\,\pi\, ,\, u_j+u_k-u_i+{\textstyle \frac{1}{2}}\ 
(\pi+\ri \log q)\,\big\}
\eea
corresponding to $x=0,1$ and $\infty$ at  the three sheets of the Riemann surface.
Since
the classical conformal block 
\eqref{soosasa} has  the logarithmic branching   at these points, it
is not a single-valued  function on the two-torus.
Note that the residues of $C$
at  $u=u_j+u_k-u_i+\frac{1}{2}\, (\pi+\ri \log q)$ do not depend on
$p_i$, whereas all the residues of $\partial_{p_i} C$ equals to $\pm 2$ 
(see \eqref{tewsiosioaw}).
Using this observation 
one can show that
\bea\label{sisoasaoi}
\exp\bigg(\,\frac{1}{2}\ \frac{\partial f}
{\partial p_i}\,\bigg)=\xi_i\ \ \frac{\vartheta_1(u-u_i, q)}{\vartheta_2(u-u_i,q)}\ ,
\eea
i.e., it is a double periodic function as well as the accessary parameter itself.
The constant $\xi_i$ depends on the normalization prescription
for the classical conformal block. For our assignment \eqref{soosasa},
it reads explicitly as 
\bea\label{sosaosa}
\xi_i=
\frac{\vartheta_3(u_{ji}+u_{ki},q)\, 
\vartheta_1(u_{kj},q)\, \vartheta^2_2(0,q)}
{\vartheta_1(u_{ji},q)\,\vartheta_2(u_{ji},q)\,
\vartheta_1(u_{ki},q)\vartheta_2(u_{ki},q) }\ \ \ \ \ \ \ (u_{ji}=u_j-u_i)\ .
\eea

\section{\label{app2}Some explicit formulae for GHO with $L=0$ and $L=1$ }

For the ordinary hypergeometric oper (i.e., without any monodromy free
punctures)  the solutions $\chi^{(i)}_\sigma$ \eqref{opaosap}  are expressed 
in terms of the hypergeometric functions (see e.g. Ref.\cite{Zamolodchikov:1995aa}):
\bea\label{osiaspspspo}
\chi^{(i)}_\sigma&=& 
\frac{1}{\sqrt{2p_i}}\ 
 (z-z_i)^{\frac{1}{2}+\sigma p_i}\ \
\Big(\frac{z-z_j}{z_i-z_j}\Big)^{-\sigma (p_i+p_k)}\
\Big(\frac{z-z_k}{z_i-z_k}\Big)^{\frac{1}{2}+\sigma p_k}\\
&\times& {}_2F_{1}\bigg(
{\frac{1}{2}}+\sigma (p_i-p_j+p_k),{ \frac{1}{2}}+\sigma( p_i+p_j+p_k),1+ 2\sigma p_i; 
\frac{(z-z_i) z_{jk}}{(z-z_j) z_{ik}}\, \bigg)\ .\nonumber
\eea
In this case, the combination\ \eqref{asospasoaso} reads explicitly
\bea\label{saopopsa}
\exp(\omega_i)\ \Big(\frac{z_{jk}}{z_{ji} z_{ik}}\Big)^{-p_i}
=\big(\,{\Omega}(p_i,p_j+p_k)\, 
{\Omega}(p_i,p_j-p_k)\,\big)^{\frac{1}{4}}\ \ \ \ \ \ \ \ \ (L=0)\ ,
\eea
where
\bea\label{asopsaapo}
{\Omega}(p,p')=\frac{\Gamma(\frac{1}{2}+p-p')\, \Gamma(\frac{1}{2}+p+p')}
{ \Gamma(\frac{1}{2}-p-p')\, \Gamma(\frac{1}{2}-p+p')}\ \frac{\Gamma(1-2 p)}{\Gamma(1+2p)}\ .
\eea

In the  case $L=1$, one can show that
\bea\label{soaoposap}
\exp(\omega_i)\ \Big(\frac{z_{jk}}{z_{ji} z_{ik}}\Big)^{-p_i}=
\ri\
\frac{\vartheta_2(u-u_i,q)}{\vartheta_1(u-u_i,q)}\ \frac{\vartheta_3(u_j-u_k,q)}{\vartheta_4(u_j-u_k,q)}\ 
\big(\,{\Omega}(p_i,p_j+p_k)\,
{\Omega}(p_i,p_j-p_k)\,\big)^{\frac{1}{4}}
\ ,
\eea
where  the notations are inherited from Appendix A.
The derivation  is based on the general facts \eqref{sopasapsissu}
and \eqref{apossosp} specialized for GHO with $L=1$. Combining these relations
with the result  \eqref{sisoasaoi} from Appendix A,
one obtains $\omega_i$  up to an additive coordinate-independent constant
$\partial_{p_i}F_0$. We may now consider the limit when the 
monodromy-free puncture  approaches to $z_i$.
At this limit the differential equations \eqref{sopspa}
becomes the  hypergeometric one,
and the limiting behavior of $\omega_i$
can be analyzed explicitly. This fix the value of $\partial_{p_i}F_0$ for $L=1$
and yields formula \eqref{soaoposap}.

\end{document}